\documentclass[a4paper, 12pt]{article}

\pdfoutput = 1

\usepackage{styleBuding}
\usepackage{bm}
\usepackage{color,listings}
\usepackage[usenames,dvipsnames,svgnames,table]{xcolor}
\usepackage{amsthm, amsmath}
\usepackage{amssymb}
\usepackage{mathrsfs}
\usepackage{graphicx}
\usepackage{slashed}
\usepackage{soul}
\usepackage{multirow}
\usepackage{subfigure}
\usepackage{diagbox}
\usepackage{siunitx} 
\usepackage{url} 
\usepackage[utf8]{inputenc}
\usepackage[figuresright]{rotating}
\definecolor{back}{HTML}{F8F8F8}
\usepackage{epstopdf}
\usepackage{lipsum}
\newcommand\blfootnote[1]{%
	\begingroup
	\renewcommand\thefootnote{}\footnote{#1}%
	\addtocounter{footnote}{-1}%
	\endgroup
}

\newcommand{\rom}[1]{\uppercase\expandafter{\romannumeral #1\relax}}

\allowdisplaybreaks
\DeclareUnicodeCharacter{2212}{-}

\title{\boldmath Dark Matter Physics in General NMSSM}

\author{Lei Meng$^{b}$, Junjie Cao$^{a,c,*}$\blfootnote{*Corresponding author.}, Fei Li$^{c}$, and Shenshen Yang$^{c}$}
\affiliation{$^a$ School of Physics, Zhengzhou University, Zhengzhou 450000, China}
\affiliation{$^b$ School of Physics and Electrical Engineering, Anyang Normal University, Anyang 455000, China}
\affiliation{$^c$ School of Physics, Henan Normal University, Xinxiang 453007, China}

\emailAdd{mel18@foxmail.com}
\emailAdd{junjiec@alumni.itp.ac.cn}
\emailAdd{hnufeili@163.com}
\emailAdd{yangshenshen@stu.htu.edu.cn}

\abstract{
In the General Next-to-Minimal Supersymmetric Standard Model (GNMSSM), singlet particles may form a secluded sector of dark matter (DM), in which Singlino-like DM could achieve the observed relic abundance through various channels such as $\tilde{\chi}_1^0 \tilde{\chi}_1^0 \to h_s h_s, A_s A_s, h_s A_s$, where $h_s$ and $A_s$ represent singlet-dominated CP-even and CP-odd Higgs bosons. We provide analytical formulas for both the spin-independent and spin-dependent cross sections of Singlino DM scattering with nucleons, illustrating their dependence on the model's parameters in a clear manner. We also present analytic expressions for the annihilation cross sections of these three important channels. Based on these preparations, we conducted Bayesian analyses of the GNMSSM and concluded that the theory significantly favored Singlino-dominated DM over Bino-like DM across a much broader range of parameters. The combined results from our numerical analyses and the formulas distinctly highlight crucial aspects of DM physics within the GNMSSM.
}

\begin{document}
\maketitle
\flushbottom

\section{Introduction}
\label{sec:intro}

Although the presence of non-baryonic dark matter (DM) in the cosmos is compelling from the perspective of astrophysics and cosmological observations and constitutes a pivotal constituent of the prevailing standard model of cosmology, its precise nature remains undetermined.  Weakly Interacting Massive Particles (WIMPs) are taken as the favored Cold Dark Matter (CDM) candidates in the standard $\Lambda$CDM model. These large-mass, stable particles are favored because they can be produced in thermal equilibrium conditions, and as a consequence, the DM relic density measured by the Planck experiment can be naturally predicted, referred to as the 'WIMP Miracle'~\cite{Planck:2015fie,Planck:2018vyg,Griest:2000kj,Bertone:2004pz}. The WIMPs are typically assumed to couple to Standard Model (SM) particles through weak interactions, resulting in spin-independent (SI) and spin-dependent (SD) scattering cross sections with nucleons around $10^{-45}$ and $10^{-39}\ {\rm cm}^2$, respectively~\cite{Baum:2017enm}. This coincidence has spurred extensive direct~\cite{XENON:2018voc,PandaX-II:2020oim,PandaX-II:2017hlx,LZ:2022lsv}, indirect~\cite{PAMELA:2010kea,Fermi-LAT:2011baq,AMS01:2007rrn}, and collider searches for WIMPs~\cite{Goodman:2010yf,Fox:2011pm}. In particular, recent direct detection experiments like LUX-ZEPLIN (LZ) have achieved unprecedented precision in probing both SI and SD DM-nucleon scatterings but found no evidence of DM, restricting the scattering cross sections to be less than the order of $10^{-47}$ and $10^{-42}\ {\rm cm}^2$, respectively~\cite{LZ:2022lsv}. This implies that the interaction between DM and nucleons is at most feeble, posing significant experimental challenges to simple WIMP theories~\cite{Cao:2019qng}. Consequently, there is growing interest in exploring scenarios that go beyond the traditional thermal WIMP paradigm.

WIMPs can naturally arise in the supersymmetric extension of the SM in particle physics. Furthermore, supersymmetric theories possess numerous intriguing features, such as the absence of quadratic divergences, gauge coupling unification, and distinctive signatures of supersymmetric particles produced in present and future colliders. In the framework of the Minimal Supersymmetric Standard Model (MSSM), recognized as one of the most economical supersymmetric theories, the neutral electroweakino sector encompasses the mixing of Bino (denoted as $\tilde{B}$), Wino (referred to as $\tilde{W}$), and neutral Higgsinos (designated as $\tilde{H}_u$ and $\tilde{H}_d$), giving rise to the formation of four distinct neutralinos~\cite{Gunion:1984yn,Haber:1984rc,Djouadi:2005gj}. Among these, the lightest mass eigenstate, denoted as $\tilde{\chi}_1^0$, usually corresponds to the lightest supersymmetric particle (LSP) and acts as a feasible candidate of DM. It prefers to be $\tilde{B}$-dominated and co-annihilate with the $\tilde{W}$-like electroweakinos to achieve the measured relic abundance. As indicated by the research in Ref.~\cite{He:2023lgi}, the LZ experiment alone imposes a lower bound on the Higgsino mass, $\mu \gtrsim 380$ GeV, in the absence of strong cancellations between different contributions to the SI scattering. This bound can be enhanced by several tens of GeV after incorporating the radiative correction to the scattering~\cite{Bisal:2023fgb,Bisal:2023iip}.  Although such a large $\mu$
may be naturally generated by the well-known Giudice--Masiero mechanism in the gravity-mediated supersymmetry breaking scenario~\cite{Giudice:1988yz}, it leads to severe fine-tuning problems in precisely predicting the mass of the $Z$ boson, after considering the Large Hadron Collider (LHC) Higgs discovery and non-observations of any DM and supersymmetric particle signals in experiments when the theory runs down from an ultraviolet high-energy scale to the electroweak
scale~\cite{Arvanitaki:2013yja,Evans:2013jna,Baer:2014ica}.

The dilemma faced by the MSSM inspired us to extend this model. The Next-to-Minimal Supersymmetric Standard Model with a $\mathbb{Z}_3$ symmetry ($\mathbb{Z}_3$-NMSSM) introduces a gauge singlet superfield $\hat{S}$ on top of the MSSM~\cite{Ellwanger:2009dp,Maniatis:2009re}.
In this framework, the $\mu$-parameter of the MSSM is dynamically generated once the scalar component of the superfield $\hat{S}$ acquires a vacuum expectation value (vev) below approximately 1 TeV. This places it naturally at the electroweak scale.
The neutralino sector includes an additional fermionic partner of $\hat{S}$, known as the Singlino. Both the Bino-dominated (which is the case in most physical scenarios)
and Singlino-dominated neutralinos are viable DM candidates~\cite{Cao:2016nix,Ellwanger:2016sur,Xiang:2016ndq,Baum:2017enm,Cao:2018rix,Ellwanger:2018zxt,Domingo:2018ykx,Baum:2019uzg,vanBeekveld:2019tqp, Abdallah:2019znp,Cao:2019qng,Guchait:2020wqn,Abdallah:2020yag,Das:2012rr,Ellwanger:2014hia,Chatterjee:2022pxf,Cao:2022htd,Datta:2022bvg,Cao:2023juc,Roy:2024yoh,Bisal:2023mgz}. Primarily, the Bino-dominated DM candidate differs from the MSSM  given the former's ability to co-annihilate with a Singlino-dominated neutralino in achieving the measured abundance. Nonetheless, this scenario is confined to an exceedingly limited parameter space defined by specific conditions: $|2\kappa \mu/\lambda|\simeq|M_1|$, moderately large $\lambda$ and $\kappa$, and $\mu$ values exceeding 300 GeV~\cite{Baum:2017enm,Cao:2019qng}. This space predicts a significant mixing of the SM Higgs field with the singlet Higgs field and is thus disfavored by the Higgs property measurements at the LHC. The characteristics of Singlino-dominated DM are contingent upon the values of $\lambda$, $\tan\beta$, $m_{\tilde{\chi}_1^0}$, and $\mu_{\rm eff} \equiv \lambda v_s/\sqrt{2}$ ~\cite{Zhou:2021pit}, where $\lambda$ denotes the singlet--doublet Higgs Yukawa coupling within the superpotential and $m_{\tilde{\chi}_1^0} \simeq 2 \kappa \mu/\lambda$ requires $2 |\kappa|/\lambda < 1$. Considering that both SI and SD DM-nucleon scattering cross sections scale with $\lambda^4$ in the regime of heavy singlet Higgs bosons and consequently, the LZ experiment generally sets an upper limit of $\lambda \lesssim 0.1$, the DM is likely to achieve the observed abundance through one of two main mechanisms~\cite{Cao:2018rix}: co-annihilating with Higgsino-dominated electroweakinos or undergoing resonant annihilations facilitated by singlet-dominated CP-even or CP-odd Higgs bosons. The former scenario is only viable within a narrow parameter space characterized by $2|\kappa|/\lambda \simeq 1$, $\lambda < 0.1$, and $\mu<400$ GeV, where $\kappa$ represents the self-coupling coefficient of the singlet Higgs field~\cite{Zhou:2021pit}. To achieve the measured density, the latter scenario requires $2|m_{\tilde{\chi}_1^0}|$ to be close to the scalar mass. Thus, all these scenarios correspond to an exceptionally constrained parameter space, resulting in a significant suppression of Bayesian evidence\footnote{Bayesian evidence is a fundamental concept in Bayesian statistics~\cite{MR1647885}. It is typically denoted as $Z(D|M)\equiv\int{P(D|O(M,\Theta))P(\Theta|M)\prod d \Theta_i}$, where $P(\Theta|M)$ represents the prior probability density function of input parameters $\Theta = (\Theta_1,\Theta_2,\cdots)$ in model $M$, and $P(D|O(M, \Theta))\equiv \mathcal{L}(\Theta)$ is the likelihood function for observed values $O$. The Bayesian evidence considers both the theoretical predictions $O(M, \Theta)$ and the experimental data $D$, and  its computation involves marginalizing the model's parameters, which requires integrating the likelihood function and prior probability over all possible parameter values. In Bayesian statistics, evidence is used to compare different models and determine which one is more plausible for explaining the observed data. While a higher value of $Z$ suggests that the respective model is more likely to agree with the data,  an extremely small $Z$ means that the theory requires its parameters to be fine-tuned for the same.}.

The situation becomes different in the General NMSSM (GNMSSM), where both $\lambda$ and $\kappa$ can play distinct roles in DM physics~\cite{Cao:2021ljw,Cao:2022ovk}. Specifically, the relationship that holds for Singlino DM in the $\mathbb{Z}_3$-NMSSM, i.e., $2 |\kappa|/\lambda < 1$, no longer applies, implying that $|\kappa|$ can be significantly larger than $\lambda$. In this case, particles predominantly comprising the singlet sector, including the Singlino-dominated DM and singlet-dominated Higgs bosons, can form an isolated DM sector~\cite{Pospelov:2007mp}, where the DM attains the correct relic abundance by tuning the value of $\kappa$ and through processes such as $\tilde{\chi}_1^0 \tilde{\chi}_1^0 \to h_s A_s, h_s h_s, A_s A_s$. Here, $h_s$ and $A_s$ denote singlet-dominated CP-even and CP-odd Higgs bosons, respectively. Similar to the situation of $\mathbb{Z}_3$-NMSSM, the SI DM-nucleon scattering rate is proportional to $\lambda^2 \kappa^2$ for moderately light $h_s$ and $\lambda^4$ for very massive $h_s$. It is thus suppressed to satisfy the LZ restrictions for a small $\lambda$. As a result, the vacuum of the scalar potential in the GNMSSM becomes more stable than that of the MSSM~\cite{Hollik:2018yek,Hollik:2018wrr}.
Additionally, given the singlet nature of the DM and the complex mass hierarchy, the decay chains of heavy supersymmetric particles
in the GNMSSM are lengthened compared to the prediction of the MSSM, which causes  their detection at the LHC to be rather tricky. These characteristics allow a more extensive parameter space of the GNMSSM to be consistent with current experimental results.

Based on the challenges faced by the MSSM and $\mathbb{Z}_3$-NMSSM, an extensive investigation of the DM physics in the GNMSSM was performed in this study. The rest of this paper is organized as follows. Section~\ref{sec:theory} outlines the properties of the GNMSSM, Section~\ref{sec:NR} describes the research strategy and numerical results to elucidate the characteristics of DM physics, and Section \ref{sec:sum} provides a summary of the research findings.

\section{Theoretical preliminaries}
\label{sec:theory}

The GNMSSM augments the MSSM by a gauge singlet superfield $\hat{S}$ that does not carry any leptonic or baryonic number. Thus, its Higgs sector contains $\hat{S}$ and two $SU(2)_L$ doublet superfields, $\hat{H}_u=(\hat{H}_u^+,\hat{H}_u^0)$ and $\hat{H}_d=(\hat{H}_d^0,\hat{H}_d^-)$. The general form of its superpotential is given by~\cite{ Ellwanger:2009dp}
\begin{eqnarray}
W_{\rm GNMSSM} = W_{\text{Yukawa} }+ \lambda \hat{S}\hat{H_u} \cdot \hat{H_d} + \frac{\kappa}{3}\hat{S}^3 + \mu \hat{H_u} \cdot \hat{H_d}  + \frac{1}{2} \mu^{\prime} \hat{S}^2 + \xi\hat{S}, \label{Superpotential}
\end{eqnarray}
where $W_{\rm Yukawa}$ is the MSSM superpotential containing quark and lepton Yukawa couplings, and the dimensionless coupling coefficients $\lambda$ and $\kappa$ parameterize the interactions between the Higgs fields, the same as those of the $\mathbb{Z}_3$-NMSSM. The bilinear mass parameters $\mu$ and $\mu^\prime$ and the singlet tadpole parameter $\xi$ depict $\mathbb{Z}_3$-symmetry-violating effects, which are advantageous for solving the tadpole problem~\cite{Ellwanger:1983mg, Ellwanger:2009dp} and the cosmological domain-wall problem of the $\mathbb{Z}_3$-NMSSM~\cite{Abel:1996cr, Kolda:1998rm, Panagiotakopoulos:1998yw}. Given that one of these parameters can be eliminated by shifting the $\hat{S}$ field by a particular constant and redefining the other parameters~\cite{Ross:2011xv}, we set $\xi$ to be zero without losing the generality of this study. As suggested by previous studies \cite{Abel:1996cr,Lee:2010gv,Lee:2011dya,Ross:2011xv,Ross:2012nr}, the natural values of the electroweak order for $\mu$ and $\mu^{\prime}$ may arise from breaking the fundamental discrete $R$-symmetry, $\mathbb{Z}^R_4$ or $\mathbb{Z}^R_8$, at high energy scales. They can significantly change the properties of neutral Higgs bosons and neutralinos and predict richer phenomenology than the $\mathbb{Z}_3$-NMSSM, which is the focus of this study.

\subsection{Higgs sector of GNMSSM}

The soft supersymmetry-breaking Lagrangian for the Higgs fields in the GNMSSM is given by
\begin{eqnarray}
 -\mathcal{L}_{soft} = &\Bigg[\lambda A_{\lambda} S H_u \cdot H_d + \frac{1}{3} \kappa A_{\kappa} S^3+ m_3^2 H_u\cdot H_d + \frac{1}{2} {m_S^{\prime}}^2 S^2 + \xi^\prime S + h.c.\Bigg]  \nonumber \\
& + m^2_{H_u}|H_u|^2 + m^2_{H_d}|H_d|^2 + m^2_{S}|S|^2 , \label{Soft-terms}
  \end{eqnarray}
where $H_u$, $H_d$, and $S$ denote the scalar components of the Higgs superfields, and $m^2_{H_u}$, $m^2_{H_d}$, and $m^2_{S}$ are their supersymmetry-breaking masses.
After the electroweak symmetry breaking, the neutral Higgs fields acquire non-zero vevs:
\begin{eqnarray}
\left\langle H_u^0 \right\rangle = v_u/\sqrt{2}, ~~\left\langle H_d^0 \right\rangle = v_d/\sqrt{2}, ~~\left\langle S \right\rangle = v_s/\sqrt{2},
\end{eqnarray}
with $v = \sqrt{v_u^2+v_d^2}\simeq 246~\mathrm{GeV}$, and the three masses are determined in solving the conditional equations to minimize the scalar potential.
The Higgs sector is then described by eleven free parameters:
\begin{eqnarray}
    \tan \beta,~\lambda,~\kappa,~v_s,~A_{\lambda},~A_{\kappa},~\mu,~\mu^{\prime},~m_3^2,~{m_{\text{S}}^{\prime}}^2,~\xi^\prime,
\end{eqnarray}
where $\tan \beta$ is defined by $\tan \beta \equiv v_u/v_d$.

In revealing the properties of Higgs physics, it is customary to work with the field combinations of $H_{\rm SM} \equiv \sin\beta {\rm Re}(H_u^0) + \cos\beta {\rm Re} (H_d^0)$, $H_{\rm NSM} \equiv \cos\beta {\rm Re}(H_u^0) - \sin\beta {\rm Re}(H_d^0)$, and $A_{\rm NSM} \equiv \cos\beta {\rm Im}(H_u^0) - \sin\beta  {\rm Im}(H_d^0)$, where $H_{\rm SM}$ denotes the SM Higgs field, and $H_{\rm NSM}$ and $A_{\rm NSM}$ represent the extra doublet Higgs fields~\cite{Cao:2012fz,Miller:2003ay}.
In the bases $\left(H_{\rm NSM}, H_{\rm SM}, {\rm Re}[S]\right)$, the mass matrix of the CP-even Higgs fields takes the following form~\cite{Ellwanger:2009dp,Miller:2003ay}
\begin{eqnarray}
  {\cal M}^2_{S, 11}&=& \frac{ \lambda v_s (\sqrt{2} A_\lambda + \kappa v_s + \sqrt{2} \mu^\prime ) + 2 m_3^2  }{\sin 2 \beta} + \frac{1}{2} (2 m_Z^2- \lambda^2v^2)\sin^22\beta,  \label{Mass-CP-even-Higgs} \\
  {\cal M}^2_{S, 12}&=&-\frac{1}{4}(2 m_Z^2-\lambda^2v^2)\sin4\beta, \quad {\cal M}^2_{S, 13} = -\frac{\lambda v}{\sqrt{2}} ( A_\lambda + \sqrt{2} \kappa v_s + \mu^\prime ) \cos 2 \beta, \nonumber \\
  {\cal M}^2_{S, 22}&=&m_Z^2\cos^22\beta+ \frac{1}{2} \lambda^2v^2\sin^22\beta,\nonumber  \\
  {\cal M}^2_{S, 23}&=& \frac{\lambda v}{\sqrt{2}} \left[(\sqrt{2} \lambda v_s + 2 \mu) - (A_\lambda + \sqrt{2} \kappa v_s + \mu^\prime ) \sin2\beta \right], \nonumber \\
  {\cal M}^2_{S, 33}&=& \frac{(A_\lambda + \mu^\prime) \sin 2 \beta}{2 \sqrt{2} v_s} \lambda v^2   + \frac{\kappa v_s}{\sqrt{2}} (A_\kappa +  2 \sqrt{2} \kappa v_s + 3 \mu^\prime ) - \frac{\mu}{\sqrt{2} v_s} \lambda v^2 - \frac{\sqrt{2}}{v_s} \xi^\prime. \nonumber
\end{eqnarray}
In the bases $\left( A_{\rm NSM}, {\rm Im}(S)\right)$, the elements of the CP-odd Higgs matrix are given by
\begin{eqnarray}
{\cal M}^2_{P,11}&=& \frac{ \lambda v_s (\sqrt{2} A_\lambda + \kappa v_s + \sqrt{2} \mu^\prime ) + 2 m_3^2  }{\sin 2 \beta}, \quad {\cal M}^2_{P,12} = \frac{\lambda v}{\sqrt{2}} ( A_\lambda - \sqrt{2} \kappa v_s - \mu^\prime ), \nonumber  \\
{\cal M}^2_{P,22}&=& \frac{(A_\lambda + 2 \sqrt{2} \kappa v_s + \mu^\prime ) \sin 2 \beta }{2 \sqrt{2} v_s} \lambda v^2  - \frac{\kappa v_s}{\sqrt{2}} (3 A_\kappa + \mu^\prime) \nonumber \\ & & - \frac{\mu}{\sqrt{2} v_s} \lambda v^2 - 2 m_S^{\prime\ 2} - \frac{\sqrt{2}}{v_s} \xi^\prime. \quad \quad \label{Mass-CP-odd-Higgs}
\end{eqnarray}
Three CP-even mass eigenstates $h_i=\{h, H, h_s\}$ and two CP-odd mass eigenstates $a_i=\{A_H, A_s\}$ are acquired through unitary rotations of $V$ and $V_P$ to diagonalize ${\cal{M}}_S^2$ and ${\cal{M}}_P^2$, respectively, leading to
\begin{eqnarray} \label{Mass-eigenstates}
    h_i & = & V_{h_i}^{\rm NSM} H_{\rm NSM}+V_{h_i}^{\rm SM} H_{\rm SM}+V_{h_i}^{\rm S} Re[S], \nonumber \\
    a_i & = &  V_{P, a_i}^{\rm NSM} A_{\rm NSM}+ V_{P, a_i}^{\rm S} Im [S].
\end{eqnarray}
Among these states, $h$ corresponds to the scalar discovered at the LHC, $H$ and $A_H$ represent heavy doublet-dominated Higgs bosons, and $h_s$ and $A_s$ denote singlet-dominated states. These states are also labeled as $h_i$ (i=1,2,3) and $A_j$ (j=1,2) in ascending mass orders for convenience, i.e., $m_{h_1} < m_{h_2} < m_{h_3}$ and $m_{A_1} < m_{A_2}$ in this study. Therefore, $h \equiv h_1$ and $m_{h_s} > m_h$ for the $h_1$ scenario, and $h \equiv h_2$ and $m_h > m_{h_s}$ for the $h_2$ scenario.
The model also predicts a pair of charged Higgs bosons, $H^\pm = \cos \beta H_u^\pm + \sin \beta H_d^\pm$, with their masses given by
\begin{eqnarray}
    m^2_{H^{\pm}} &=&  \frac{ \lambda v_s (\sqrt{2} A_\lambda + \kappa v_s + \sqrt{2} \mu^\prime ) + 2 m_3^2  }{\sin 2 \beta} + m^2_W - \frac{1}{2}\lambda^2 v^2. \label{Charged Hisggs Mass}
  \end{eqnarray}
Note that charged Higgs bosons $H^\pm$ degenerate with the CP-even doublet scalar $H$ and the CP-odd scalar $A_H$ in mass.

So far, the LHC experiments have performed intensive searches for the additional Higgs bosons $H$, $A_H$, $H^{\pm}$, $h_s$, and $A_s$
and set limits on their properties, such as their masses and couplings~\cite{ATLAS:2020zms,ATLAS:2021upq}. We note that the parameters
$\mu$, $\mu^\prime$, $m_3^2$, $m_S^{\prime\ 2}$, and $\xi^\prime$ are not directly related to experimental measurements. This
motivated us to use the masses of CP-odd heavy doublet Higgs field, $CP$-even and -odd singlet Higgs fields, and Higgsino and Singlino fields, denoted as $m_A \equiv \sqrt{{\cal M}^2_{P,11}}$, $m_B \equiv \sqrt{{\cal M}^2_{S,33}}$, $m_C \equiv \sqrt{{\cal M}^2_{P,22}}$, $\mu_{\text{tot}} \equiv \mu_{\text{eff}} + \mu$, and $m_N \equiv \frac{2 \kappa}{\lambda} \mu_{\rm eff} + \mu^{\prime}$, respectively, as theoretical inputs. The former set of parameters are then given by
\begin{eqnarray}
\mu &= & \mu_{tot} - \frac{\lambda}{\sqrt{2}} v_s, \quad \mu^\prime = m_N - \sqrt{2} \kappa v_s, \quad  m^2_3 = \frac{m^2_A \sin{2\beta}}{2} - \lambda v_s (\frac{\kappa v_s}{2} + \frac{\mu^\prime}{\sqrt{2}} + \frac{A_\lambda}{\sqrt{2}}), \nonumber \\
\xi^\prime &=& \frac{v_s}{\sqrt{2}} \left [ \frac{(A_\lambda + \mu^\prime) \sin 2 \beta}{2 \sqrt{2} v_s} \lambda v^2   + \frac{\kappa v_s}{\sqrt{2}} (A_\kappa +  2 \sqrt{2} \kappa v_s + 3 \mu^\prime ) - \frac{\mu}{\sqrt{2} v_s} \lambda v^2 - m_B^2 \right ],  \nonumber\\
m_S^{\prime 2} &=& \frac{1}{2} \left[ m_B^2 - m^2_C + \lambda \kappa \sin 2\beta v^2 - 2 \sqrt{2} \kappa v_s ( A_\kappa + \frac{\kappa}{\sqrt{2}} v_s + \mu^\prime) \right],   \label{Simplify-1}
\end{eqnarray}
and Eqs.~(\ref{Mass-CP-even-Higgs}), (\ref{Mass-CP-odd-Higgs}), and (\ref{Charged Hisggs Mass}) take the following simplified forms:
\begin{eqnarray}
 {\cal M}^2_{S, 11}&=& m_A^2 + \frac{1}{2} (2 m_Z^2- \lambda^2v^2)\sin^22\beta, \quad {\cal M}^2_{S, 12}=-\frac{1}{4}(2 m_Z^2-\lambda^2v^2)\sin4\beta, \nonumber \\
  {\cal M}^2_{S, 13} &=& -\frac{\lambda v}{\sqrt{2}} ( A_\lambda + m_N ) \cos 2 \beta, \quad {\cal M}^2_{S, 22} = m_Z^2\cos^22\beta+ \frac{1}{2} \lambda^2v^2\sin^22\beta, \nonumber  \\
  {\cal M}^2_{S, 23}&=& \frac{\lambda v}{\sqrt{2}} \left[ 2 \mu_{tot} - (A_\lambda + m_N ) \sin2\beta \right], \quad {\cal M}^2_{S, 33} = m_B^2, \quad {\cal M}^2_{P,11} = m_A^2,  \nonumber \\
 \quad {\cal M}^2_{P,22} &=& m_C^2, \quad {\cal M}^2_{P,12} = \frac{\lambda v}{\sqrt{2}} ( A_\lambda - m_N ), \quad  m^2_{H^{\pm}} = m_A^2 + m^2_W -  \frac{1}{2}\lambda^2 v^2.  \label{New-mass-matrix}
\end{eqnarray}
These formulas indicate that the Higgs mass matrices are determined by just eight out of the eleven parameters: $\tan \beta$, $\lambda$, $A_\lambda$, $m_A$, $m_B$, $m_C$, $m_N$, and $\mu_{tot}$. As will be shown in Eqs.~(\ref{eq:hihshs}-\ref{Coupling-size}), the remaining three parameters, $\kappa$, $A_\kappa$, and $v_s$, specifically influence triple Higgs coupling strengths.

In the case of very massive charged Higgs bosons, the following approximations hold~\cite{Baum:2017enm}:
\begin{eqnarray}
m_{h_s}^2 & \simeq & m_B^2 - \frac{{\cal M}^4_{S, 13}}{m_A^2 - m_B^2}, \quad m_{A_s}^2 \simeq m_C^2 - \frac{{\cal M}^4_{P, 12}}{m_A^2 - m_C^2}, \quad \frac{V_{P,A_s}^{\rm NSM}}{V_{P,A_s}^{\rm S}} = \frac{{\cal M}^2_{P, 12}}{m_{A_s}^2 - m_A^2} \simeq 0, \nonumber \\
\frac{V_{h}^{\rm S}}{V_h^{\rm SM}} & \simeq &  \frac{{\cal M}^2_{S, 23}}{m_h^2 - m_B^2}, \quad V_{h}^{\rm NSM} \sim 0, \quad V_h^{\rm SM} \simeq \sqrt{1 - \left ( \frac{V_{h}^{\rm S}}{V_h^{\rm SM}} \right )^2}  \sim 1, \nonumber  \\
\frac{V_{h_s}^{\rm SM}}{V_{h_s}^{\rm S}} & \simeq &  \frac{{\cal M}^2_{S, 23}}{m_{h_s}^2 - m_h^2}, \quad V_{h_s}^{\rm NSM} \sim 0, \quad V_{h_s}^{\rm S} \simeq \sqrt{1 - \left ( \frac{V_{h_s}^{\rm SM}}{V_{h_s}^{\rm S}} \right )^2 } \sim 1.   \label{Approximations}
\end{eqnarray}
Evidently, the singlet fields will decouple from the doublet Higgs fields in the limit of $\lambda \to 0$, and the field masses $m_B$, $m_C$, and $m_N$ can be regarded as physical particle masses to a good approximation. Additionally, the singlet masses are independent and may all take small values since they are weakly constrained by current experiments.

\subsection{Neutralino sector of GNMSSM}

The mixing between fermionic partners for neutral Higgs bosons and gauginos results in five neutralinos and two charginos, denoted as $\tilde{\chi}_i^0$ ($i=1,\dots,5$) and $\tilde{\chi}_i^\pm$ ($i=1,2$), respectively.
In the gauge eigenstate bases $\psi^0 = \left(-i \tilde{B}, -i\tilde{W}, \tilde{H}_d^0, \tilde{H}_u^0, \tilde{S}\right)$, the symmetric neutralino mass matrix is given by~\cite{Ellwanger:2009dp}
\begin{equation}
    M_{\tilde{\chi}^0} = \left(
    \begin{array}{ccccc}
    M_1 & 0 & -m_Z \sin \theta_W \cos \beta & m_Z \sin \theta_W \sin \beta & 0 \\
      & M_2 & m_Z \cos \theta_W \cos \beta & - m_Z \cos \theta_W \sin \beta &0 \\
    & & 0 & -\mu_{\text{tot}} & - \frac{1}{\sqrt{2}} \lambda v \sin \beta \\
    & & & 0 & -\frac{1}{\sqrt{2}} \lambda v \cos \beta \\
    & & & & m_{\text{N}}
    \end{array}
    \right), \label{eq:mmn}
\end{equation}
where $M_1$ and $M_2$ are gaugino soft-breaking masses, $s_W = \sin \theta_W$, and $c_w = \cos \theta_W$.
We use the rotation matrix $N$ to diagonalize this mass matrix and label the resulting mass eigenstates by an ascending mass order
\begin{eqnarray}  \label{Mass-eigenstate-neutralino}
    \tilde{\chi}_i^0 = N_{i1} \psi^0_1 +   N_{i2} \psi^0_2 +   N_{i3} \psi^0_3 +   N_{i4} \psi^0_4 +   N_{i5} \psi^0_5,
\end{eqnarray}
where the matrix elements $N_{i3}$ and $N_{i4}$ characterize the $\tilde{H}_d^0$ and $\tilde{H}_u^0$ components in $\tilde{\chi}_i^0$, and $N_{i5}$ denotes the Singlino component. We say $\tilde{\chi}_1^0$ is Singlino-dominated if $N_{15}^2 > 0.5$.

In the case that the gauginos are very massive and $\mu_{\text{tot}}^2 - m_N^2 \gg \lambda^2 v^2$, the following approximations for the Singlino-dominated $m_{\tilde{\chi}_1^0}$ hold~\cite{Badziak:2017uto,Badziak:2015nrb,Cheung:2014lqa}:
\begin{eqnarray} \label{Approximation-neutralinos}
    m_{\tilde{\chi}_1^0} & \simeq & m_N + \frac{1}{2} \frac{\lambda^2 v^2 ( m_{\tilde{\chi}_1^0} - \mu_{\text{tot}} \sin 2 \beta )}{m_{\tilde{\chi}_1^0}^2 - \mu_{\text{tot}}^2} \simeq m_N, \quad N_{11} \sim 0, \quad N_{12} \sim 0, \label{Neutralino-Mixing}  \\
    \frac{N_{13}}{N_{15}} &= & \frac{\lambda v}{\sqrt{2} \mu_{\rm tot}} \frac{(m_{\tilde{\chi}_1^0}/\mu_{\text{tot}})\sin\beta-\cos\beta} {1-\left(m_{\tilde{\chi}_1^0}/\mu_{\rm tot}\right)^2}, \quad \quad  \frac{N_{14}}{N_{15}} =  \frac{\lambda v}{\sqrt{2} \mu_{\text{tot}}} \frac{(m_{\tilde{\chi}_1^0}/\mu_{\text{tot}})\cos\beta-\sin\beta} {1-\left(m_{\tilde{\chi}_1^0}/\mu_{\text{tot}}\right)^2}, \nonumber \\
    N_{15}^2 & \simeq & \left(1+ \frac{N^2_{13}}{N^2_{15}}+\frac{N^2_{14}}{N^2_{15}}\right)^{-1} \nonumber \\
    &= & \frac{\left[1-(m_{\tilde{\chi}_1^0}/\mu_{\text{tot}})^2\right]^2}{\left[(m_{\tilde{\chi}_1^0}/{\text{tot}})^2 -2(m_{\tilde{\chi}_1^0}/\mu_{\text{tot}})\sin2\beta+1 \right]\left(\frac{\lambda v}{\sqrt{2}\mu_{\text{tot}}}\right)^2
    +\left[1-(m_{\tilde{\chi}_1^0}/\mu_{\text{tot}})^2\right]^2}. \nonumber
\end{eqnarray}
The ratio of the Higgsino fraction in $\tilde{\chi}_1^0$, denoted as $Z_h \equiv N_{13}^2+N_{14}^2$, to the Singlino fractions $Z_s \equiv N_{15}^2$ is acquired by
\begin{eqnarray}
    \frac{Z_h}{Z_s} &= & \left(\frac{\lambda v}{\sqrt{2} \mu_{\rm tot}}\right)^{\!\!2} \frac{\left(m_{\tilde{\chi}_1^0}/\mu_{\rm tot}\right)^2-2{(m_{\tilde{\chi}_1^0}}/{\mu_{\rm tot}})\sin2\beta+1}
{\left[1-\left({m_{\tilde{\chi}_1^0}}/{\mu_{\rm tot}}\right)^2\right]^2}.
\end{eqnarray}

These approximations reveal that the characteristics of $\tilde{\chi}_1^0$ are represented by five independent parameters: $\tan \beta$, $\lambda$, $\kappa$, $\mu_{\text{tot}}$, and $m_{\tilde{\chi}_1^0}$, where $\kappa$ quantifies the self-interactions of the singlet fields. They reflect that both $\lambda$ and $\mu_{\text{tot}}$ play a significant role in determining $Z_h$; specifically, a smaller $\lambda$ or a larger $\mu_{\rm tot}$ can suppress $Z_h$. Notably, if one chooses $m_N$, or equivalently $m_{\tilde{\chi}_1^0}$, as a theoretical input, $\kappa$ does not explicitly affect $m_{\tilde{\chi}_1^0}$, and it is independent of $\lambda$ in influencing the other properties of $\tilde{\chi}_1^0$. This feature distinguishes it from the Singlino-dominated $\tilde{\chi}_1^0$ in the $Z_3$-NMSSM, whose properties are determined by four parameters, i.e.,  $\tan\beta$, $\lambda$, $\mu_{\text{eff}}$, and $m_{\tilde{\chi}_1^0} \simeq 2 \kappa \mu_{\text{eff}}/\lambda$, with $\kappa$ satisfying $2 |\kappa| < \lambda$~\cite{Zhou:2021pit}.

We point out the difference between the GNMSSM and $\mu$-extended NMSSM discussed in Ref.~\cite{Cao:2021ljw}: in the former, the singlet masses $m_B$, $m_C$, and $m_N$ and the Higgsino mass $\mu_{\rm tot}$ are independent, while in the latter, they are correlated.

\subsection{DM physics of GNMSSM} \label{DM-physics}

In the GNMSSM, the following interactions are pertinent to the DM physics~\cite{Ellwanger:2009dp}:
\begin{eqnarray}
C_{\tilde {\chi}^0_1 \tilde {\chi}^0_1 h_i }  & = &
V_{h_i}^{\rm SM} C_{\tilde {\chi}^0_1 \tilde {\chi}^0_1 H_{\rm SM} }+V_{h_i}^{\rm NSM} C_{\tilde {\chi}^0_1 \tilde {\chi}^0_1 H_{\rm NSM} }+V_{h_i}^{\rm S} C_{\tilde {\chi}^0_1 \tilde {\chi}^0_1 Re[S] } \nonumber \\
& \simeq & \sqrt{2}\lambda \left[ V_{h_i}^{\rm SM} N_{15}\left(N_{13}\sin\beta+N_{14}\cos\beta\right)
+V_{h_i}^{\rm NSM}N_{15}\left(N_{13}\cos\beta- N_{14}\sin\beta\right)\right] \nonumber \\
& & + \sqrt{2} V_{h_i}^{\rm S}
\left(\lambda{N_{13}}{N_{14}}-\kappa N_{15}^2\right)\,,
\label{C-hichi01chi01_S}
\\
C_{\tilde {\chi}^0_1 \tilde {\chi}^0_1 a_i }
& = & V_{P,a_i}^{\rm NSM}C_{\tilde {\chi}^0_1 \tilde {\chi}^0_1 A_{\rm NSM} }+ V_{P,a_i}^{\rm S}C_{\tilde {\chi}^0_1 \tilde {\chi}^0_1 Im[S] } \nonumber \\
& \simeq & - \sqrt{2} V_{P,a_i}^{\rm NSM}\lambda N_{15} ( N_{13}\cos\beta +  N_{14}\sin\beta) + \sqrt{2} V_{P,a_i}^{\rm S} (\lambda
N_{13}N_{14} - \kappa N_{15}^2), \label{C-aichi01chi01_S} \\
C_{\tilde {\chi}^0_1 \tilde {\chi}^0_1 Z } & = & \frac{m_Z}{v} ( N_{13}^2 - N_{14}^2), \quad \quad C_{\tilde {\chi}^0_1 \tilde {\chi}^0_1 G^0 } \simeq - \sqrt{2} \lambda N_{15} (N_{13} \sin \beta - N_{14}\cos \beta ), \label{C-Zchi01chi01} \\
C_{h_i h_s h_s} &=& \lambda v\, V_{h_i}^{\rm SM} V_{h_s}^{\rm S} V_{h_s}^{\rm S} (\lambda - \kappa \sin2\beta) - \lambda \kappa v V_{h_i}^{\rm NSM} V_{h_s}^{\rm S} V_{h_s}^{\rm S} \cos2\beta \nonumber \\
& & + \sqrt{2} \kappa V_{h_i}^{\rm S} V_{h_s}^{\rm S} V_{h_s}^{\rm S} ( 3 m_N + A_\kappa ) + C^\prime_{h_s} (\lambda, \kappa, \tan \beta, v_s, A_\lambda, m_N), \label{eq:hihshs} \\
C_{h_i A_s A_s} &=& \lambda v\, V_{h_i}^{\rm SM} (\lambda + \kappa \sin2\beta) + \lambda \kappa v V_{h_i}^{\rm NSM} \cos2\beta  \nonumber \\
& & + \sqrt{2} \kappa V_{h_i}^{S}( m_N - A_\kappa ) + C^\prime_{A_s} (\lambda, \kappa, \tan \beta, v_s, A_\lambda, m_N), \label{eq:hiasas}
\end{eqnarray}
where the last contributions in $C_{h_i h_s h_s}$ and $C_{h_i A_s A_s}$ are suppressed by the Higgs mixings and they are functions of $\lambda$, $\kappa$, $\tan \beta$, $v_s$, $A_\lambda$, and $m_N$. In the small $\lambda$ case, we conclude that
\begin{eqnarray}
& & C_{ \tilde{\chi}_1^0 \tilde{\chi}_1^0 h_s} \simeq  C_{\tilde{\chi}_1^0 \tilde{\chi}_1^0 A_s } \simeq - \sqrt{2}\kappa, \quad  C_{h_s h_s h_s} \simeq \sqrt{2} \kappa ( 3 m_N + A_\kappa ), \nonumber \\ & &  C_{h_s A_s A_s} \simeq \sqrt{2} \kappa ( m_N - A_\kappa), \label{Coupling-size}
\end{eqnarray}
and the other interactions, such as $C_{\tilde {\chi}^0_1 \tilde {\chi}^0_1 h}$, $C_{\tilde {\chi}^0_1 \tilde {\chi}^0_1 Z}$, $C_{h h_s h_s}$, and $C_{h A_s A_s}$, are suppressed by $\lambda$. These characteristics are crucial to understand the results of this study.

\subsubsection{DM relic density}\label{sec:DMRD}

 The thermally averaged cross section for WIMP DM pair annihilation under a non-relativistic approximation and in the absence of co-annihilation can be expanded as~\cite{Griest:2000kj,Griest:1990kh}
\begin{equation}
\label{thermalcs}
\left\langle \sigma_Av \right\rangle = a + b\left\langle v^2 \right\rangle + \mathcal{O}(\left\langle v^4 \right\rangle) \approx a + 6 \frac{b}{x},
\end{equation}
where $a$ corresponds to the $s$-wave contribution at a zero relative velocity, $b$ encompasses contributions from both $s$-wave and $p$-wave processes, and $x\equiv m_{\tilde{\chi}_1^0}/T$, with $T$ denoting the temperature of the early Universe.
After integrating the density function from the freeze-out temperature $x_F=m/T_F$ to infinity, the present thermal relic abundance is given by~\cite{Baum:2017enm}
\begin{eqnarray}
    \Omega h^2 = 0.12\left(\frac{80}{g_*}\right)^{1/2}\left(\frac{x_F}{25}\right) \left( \frac{2.3\times 10^{-26} \mathrm{cm^3/s}}{\langle \sigma v\rangle_{x_F}}\right)\;,
\end{eqnarray}
where ${g_*} \sim 80 $ is the total number of effectively relativistic degrees of freedom at the time of freeze-out, and $x_F\sim 25$ is obtained by solving the freeze-out equation in \cite{Griest:1990kh}.

As indicated by the following study, the GNMSSM predicts a Singlino-dominated DM in most cases and  achieves the observed relic abundance primarily through the following channels:
\begin{enumerate}
\item  $\tilde{\chi}_1^0 \tilde{\chi}_1^0 \to h_s h_s$

This channel, which can be the most important annihilation of $\tilde{\chi}_1^0$ when $m_{\tilde{\chi}_1^0} > m_{h_s}$ and  $m_{A_s} > 2 m_{\tilde{\chi}_1^0} - m_{h_s} $, occurs by the $s$-channel exchange of CP-even Higgs bosons and $t$-channel exchange of neutralinos. The full analytic expression of $\langle \sigma v \rangle_{x_F}^{h_s h_s} $ is very complex~\cite{Nihei:2002ij}. However, this can be significantly simplified in the case of minor $\lambda$ paired with significant $|\kappa|$, where only the contributions from the exchanges of $h_s$ and $\tilde{\chi}_1^0$ are crucial. Correspondingly, $\langle \sigma v \rangle_{x_F}^{h_s h_s} $ is approximated by~\cite{Arcadi:2017kky}
\begin{eqnarray} \label{eq:sigvhshs}
\left\langle \sigma v \right\rangle_{x_F}^{h_s h_s} & \simeq & \frac{v^2_F}{192 \pi m_{\tilde{\chi}_1^0}^2} \sqrt{1 - \frac{m_{h_s}^2}{m_{\tilde{\chi}_1^0}^2}} \times \left \{ \frac{8 C_{h_s h_s h_s} C_{\tilde{\chi}_1^0 \tilde{\chi}_1^0 h_s}^3 m_{\tilde{\chi}_1^0}^3 (2 m_{h_s}^2 - 5 m_{\tilde{\chi}_1^0}^2) }{(m_{h_s}^2 - 4 m_{\tilde{\chi}_1^0}^2)(m_{h_s}^2 - 2 m_{\tilde{\chi}_1^0}^2)}  \right . \nonumber \\
& & \left . + \frac{3 C_{h_s h_s h_s}^2 C_{\tilde{\chi}_1^0 \tilde{\chi}_1^0 h_s}^2 m_{\tilde{\chi}_1^0}^2}{(m_{h_s}^2 - 4 m_{\tilde{\chi}_1^0}^2)^2} + \frac{16 C_{\tilde{\chi}_1^0 \tilde{\chi}_1^0 h_s}^4 ( 9 m_{\tilde{\chi}_1^0}^8 - 8 m_{\tilde{\chi}_1^0}^6 m_{h_s}^2 + 2 m_{h_s}^8)}{(m_{h_s}^2 - 2 m_{\tilde{\chi}_1^0}^2)^4}  \right \}. \quad \quad \quad
\end{eqnarray}
This formula together with Eq.~(\ref{Coupling-size}) indicate that $\left\langle \sigma v \right\rangle_{x_F}^{h_s h_s}$ depends mainly on $\kappa$, $A_\kappa$, $m_{h_s}$, and $m_{\tilde{\chi}_1^0}$. In particular, $\kappa$ plays a critical role in determining the magnitude of $\left\langle \sigma v \right\rangle_{x_F}^{h_s h_s} $, since most of its dominant contributions are proportional to $\kappa^4$. If $C_{h_s h_s h_s}$ is sufficiently small so that the s-channel contribution can be neglected, we have
\begin{eqnarray}
\left\langle \sigma v \right\rangle_{x_F}^{h_s h_s} & \simeq & \frac{3 v_F^2 \kappa^4 }{16 \pi m_{\tilde{\chi}_1^0}^2},
\end{eqnarray}
if $m_{\tilde{\chi}_1^0}^2 \gg m_{h_s}^2$, which implies that
\begin{eqnarray}
|\kappa| \sim 0.23 \times \left ( \frac{m_{\tilde{\chi}_1^0}}{300~{\rm GeV}} \right )^{1/2}  \label{Kappa-approximation-1},
\end{eqnarray}
allowing the observed density to be acquired.

\item  $\tilde{\chi}_1^0 \tilde{\chi}_1^0 \to A_s A_s$

This channel can be the dominant annihilation of $\tilde{\chi}_1^0$ if $m_{\tilde{\chi}_1^0} > m_{A_s}$ and $m_{h_s} > 2 m_{\tilde{\chi}_1^0} - m_{A_s} $. It proceeds in a way similar to  $\tilde{\chi}_1^0 \tilde{\chi}_1^0 \to h_s h_s$, and $\left\langle \sigma v \right\rangle_{x_F}^{A_s A_s}$ is given by~\cite{Arcadi:2017kky}
\begin{eqnarray} \label{eq:sigvAsAs}
\left\langle \sigma v \right\rangle_{x_F}^{A_s A_s} & \simeq & \frac{v^2_F}{128 \pi m_{\tilde{\chi}_1^0}^2} \sqrt{1 - \frac{m_{A_s}^2}{m_{\tilde{\chi}_1^0}^2}} \times \nonumber \\
& & \left \{ \frac{32}{3} \frac{C_{\tilde{\chi}_1^0 \tilde{\chi}_1^0 A_s}^4 m_{\tilde{\chi}_1^0}^4 (m_{A_s}^2 - m_{\tilde{\chi}_1^0}^2)^2}{(m_{A_s}^2 - 2 m_{\tilde{\chi}_1^0}^2)^4} + \frac{4 C_{\tilde{\chi}_1^0 \tilde{\chi}_1^0 h_s}^2 C_{\tilde{\chi}_1^0 \tilde{\chi}_1^0 A_s}^2 m_{\tilde{\chi}_1^0}^2 m_{h_s}^2}{(m_{h_s}^2 - 4 m_{\tilde{\chi}_1^0}^2)^2 + m_{h_s}^2 \Gamma_{h_s}^2 }  \right \}. \quad \quad \quad
\end{eqnarray}
Compared with $\left \langle \sigma v \right\rangle_{x_F}^{h_s h_s}$, $\left \langle \sigma v \right\rangle_{x_F}^{A_s A_s}$ relies on one more parameter, $m_{A_s}$, and for $m_{h_s} \simeq 2 m_{\tilde{\chi}_1^0}$, it can be resonantly enhanced. As a result, it possesses more complex features than $\left \langle \sigma v \right\rangle_{x_F}^{h_s h_s}$. Similar to the previous discussion, if we neglect the s-channel contribution and assume $m_{\tilde{\chi}_1^0}^2 \gg m_{A_s}^2$, we conclude that
\begin{eqnarray}
\left\langle \sigma v \right\rangle_{x_F}^{A_s A_s} & \simeq & \frac{v_F^2 \kappa^4}{48 \pi m_{\tilde{\chi}_1^0}^2}
\end{eqnarray}
and
\begin{eqnarray}
|\kappa| \sim 0.40 \times \left ( \frac{m_{\tilde{\chi}_1^0}}{300~{\rm GeV}} \right )^{1/2}  \label{Kappa-approximation-2}
\end{eqnarray}
to predict the observed density.

\item  $\tilde{\chi}_1^0 \tilde{\chi}_1^0 \to h_s A_s$

This process occurs if $2 m_{\tilde{\chi}_1^0} > m_{h_s} + m_{A_s}$ and  proceeds through the $s$-channel exchange of $Z$ and CP-odd Higgs bosons and the $t$-channel exchange of neutralinos. Compared with the previous two annihilations, the full analytic expression of $\langle \sigma v \rangle_{x_F}^{h_s A_s} $ is much more complex~\cite{Nihei:2002ij}. However, given that the contributions from the exchanges of $A_s$ and $\tilde{\chi}_1^0$ are dominant for the small $\lambda$ and sizable $|\kappa|$ case, it can be simplified as follows~\cite{Baum:2017enm,Griest:1990kh}
\begin{eqnarray} \label{eq:sigvPhiPhi}
\left\langle \sigma v \right\rangle_{x_F}^{h_s A_s} \simeq && \frac{1}{64 \pi m_{\tilde{\chi}_1^0}^2} \left\{ \left[1-\frac{\left(m_{h_s} + m_{A_s}\right)^2}{4 m_{\tilde{\chi}_1^0}^2}\right] \left[1-\frac{\left(m_{h_s} - m_{A_s}\right)^2}{4 m_{\tilde{\chi}_1^0}^2}\right] \right\}^{1/2} | {\cal{A}}_s + {\cal{A}}_t |^2, \nonumber
\end{eqnarray}
where the $s$- and $t$-channel contributions are approximated by
\begin{eqnarray}
{\cal{A}}_s &\simeq & \frac{-2 m_{\tilde{\chi}_1^0} C_{\tilde {\chi}^0_1 \tilde {\chi}^0_1 A_s }  C_{h_s A_s A_s}}{m_{A_s}^2 - 4  m_{\tilde{\chi}_1^0}^2}, \nonumber \\
{\cal{A}}_t &\simeq & - 2 C_{\tilde{\chi}_1^0\tilde{\chi}_1^0 h_s} \, C_{\tilde{\chi}_1^0\tilde{\chi}_1^0 A_s} \left[ 1 + \frac{ 2 m_{A_s}^2}{ 4 m_{\tilde{\chi}_1^0}^2 - \left(m_{h_s}^2 + m_{A_s}^2\right) } \right],
\end{eqnarray}
respectively, assuming that $h_s$ is not exceptionally light to forbid $A_s$-mediated resonant annihilation. Furthermore, from the formula of $C_{h_s A_s A_s}$ in Eq.~(\ref{Coupling-size}), one can infer that
if $|m_{\tilde{\chi}_1^0} - A_\kappa | \ll (4 m_{\tilde{\chi}_1^0}^2 - m_{A_s}^2)/|m_{\tilde{\chi}_1^0}|$,  $|{\cal{A}}_t|$ is much larger than $|{\cal{A}}_s|$, reflecting that the $t$-channel contribution prevails. In this case, $\left\langle \sigma v \right\rangle_{x_F}^{h_s A_s}$ is simplified as
\begin{eqnarray}
\left\langle \sigma v \right\rangle_{x_F}^{h_s A_s} & \simeq & \frac{\kappa^4}{4 \pi m_{\tilde{\chi}_1^0}^2},
\end{eqnarray}
if $|m_{\tilde{\chi}_1^0}|$ is much larger than $m_{A_s}$ and $m_{h_s}$, and
\begin{eqnarray}
|\kappa| \sim 0.15 \times \left ( \frac{m_{\tilde{\chi}_1^0}}{300~{\rm GeV}} \right )^{1/2}  \label{Kappa-approximation-3}
\end{eqnarray}
to predict the observed density.
\end{enumerate}

Notably, once the process $\tilde{\chi}_1^0 \tilde{\chi}_1^0 \to h_s A_s$ is kinematically allowed, at least one of the annihilations $\tilde{\chi}_1^0 \tilde{\chi}_1^0 \to h_s h_s$  and $\tilde{\chi}_1^0 \tilde{\chi}_1^0 \to A_s A_s$ is open. Since the former channel is dominated by s-wave contributions while the latter ones are p-wave processes, $\tilde{\chi}_1^0 \tilde{\chi}_1^0 \to h_s A_s$ dominantly contributes to $\left\langle \sigma v \right\rangle_{x_F}$.  Additionally, the annihilation $\tilde{\chi}_1^0 \tilde{\chi}_1^0 \to h A_s$, which proceeds in a manner similar to $\tilde{\chi}_1^0 \tilde{\chi}_1^0 \to h_s A_s$, may provide substantial contributions to $\left\langle \sigma v \right\rangle_{x_F}$ when $\lambda$ and $V^S_h$ take exceptionally large values. Such a possibility, however, has been rigorously bounded by the results of the LZ experiment and the measurements of Higgs property at the LHC.

To achieve the observed relic abundance, the Singlino-dominated DM can also co-annihilate with electroweakinos. Corresponding processes include $\tilde{\chi}_i \tilde{\chi}_i$, $\tilde{\chi}_i \tilde{\chi}_j$, $\tilde{\chi}_j \tilde{\chi}_j \to X X^\prime$, where $\tilde{\chi}_i$ and $\tilde{\chi}_j$ denote either the LSP or next-to-lightest supersymmetric particle (NLSP) and $X X^\prime$ represents SM particles. Basically, these annihilations influence the abundance only when the mass difference between $\tilde{\chi}_1^0$ and its co-annihilation partner is less than approximately $10\%$~\cite{Griest:1990kh,Baker:2015qna}, which necessitates the tuning of the two theoretically independent masses and thus leads to the suppression of Bayesian evidence.

\subsubsection{DM-nucleon scattering}\label{sec:DMN SCS}

In the regime of heavy squarks, the SI scattering of DM with nucleons predominantly arises from the $t$-channel exchange of $CP$-even Higgs bosons. The cross section takes the following form~\cite{Badziak:2015nrb,Pierce:2013rda}
\begin{eqnarray}
   \sigma^{\rm SI}_{N} = \frac{4 \mu_r^2}{\pi} |f^{N}|^2, \quad
	f^{N} =  \sum_{i}^3 f^{N}_{h_i} = \sum_{i}^3 \frac{C_{ \tilde {\chi}^0_1 \tilde {\chi}^0_1 h_i} C_{N N h_i }}{2m^2_{h_i} }, \label{SI-cross}
\end{eqnarray}
where $N=p, n$ denotes a proton ($p$) or neutron ($n$), and $\mu_r \equiv m_N m_{\tilde {\chi}^0_1} /(m_N+m_{\tilde {\chi}^0_1})$ is the reduced mass of the DM-nucleon composite. $C_{NN h_i}$ represents the strength of Higgs coupling to the nucleon, given by
\begin{eqnarray}
C_{NN h_i} = -\frac{m_N}{v}
\left[ F^{N}_d \left( V_{h_i}^{\rm SM}- \tan\beta V_{h_i}^{\rm NSM}\right)+F^{N}_u \left(V_{h_i}^{\rm SM}+\frac{1}{\tan\beta} V_{h_i}^{\rm NSM} \right)
\right]\,. \quad \label{C-hNN_S}
\end{eqnarray}
Here, $F^N_d$ and $F^N_u$ are defined as $F^{N}_d \equiv f^{(N)}_d+f^{(N)}_s+\frac{2}{27}f^{(N)}_G$ and $F^{N}_u \equiv f^{(N)}_u+\frac{4}{27}f^{(N)}_G$, where
the nucleon form factors $f^{(N)}_q \equiv m_N^{-1}\left<N|m_qq\bar{q}|N\right>$ for $q=u,d,s$ denote the normalized light quark contribution to the nucleon mass and $f^{(N)}_G \equiv 1-\sum_{q=u,d,s}f^{(N)}_q$ represents other heavy quarks' mass fraction in the nucleon~\cite{Drees:1993bu,Drees:1992rr}. For the default settings of the package micrOMEGAs for $f_q^{(N)}$~\cite{Belanger:2008sj,Alarcon:2011zs,Alarcon:2012nr}, $F_u^{p} \simeq F_u^n \simeq 0.15$, $F_d^{p} \simeq F_d^n \simeq 0.13$, and therefore, $\sigma^{\rm SI}_{\tilde {\chi}^0_1-p} \simeq \sigma^{\rm SI}_{\tilde {\chi}^0_1-n}$.

Because the $H$-mediated contribution to $\sigma^{\rm SI}_{N}$  usually plays a minor role\footnote{This contribution is proportional to $\tan^2 \beta/m_H^4$ and is typically less than $10^{-49}~{\rm cm^2}$ for $\tan \beta \leq 5$, as indicated by Figure 6 of~\cite{Baum:2017enm}. With the further increase of $\tan \beta$, $m_H$ is stringently limited by the LHC search for extra Higgs bosons (see, e.g., Figure 22 of~\cite{ATLAS:2024lyh} for the MSSM results.). Consequently, the contribution has difficulty reaching $10^{-47}~{\rm cm^2}$. Therefore, barring the blind spots discussed in~\cite{Huang:2014xua}, it is less crucial.}, we focus on the contribution from the $t$-channel exchange of the SM-like Higgs boson $h$ and the singlet Higgs boson $h_s$. Using the formulas from Eqs.(\ref{Neutralino-Mixing})--(\ref{eq:hiasas}), we obtain the couplings $C_{\tilde {\chi}^0_1 \tilde {\chi}^0_1 h}$ and $C_{\tilde {\chi}^0_1 \tilde {\chi}^0_1 h_s}$ for the Singlino-dominated $\tilde{\chi}_1^0$ as follows:
\begin{eqnarray}
C_{\tilde {\chi}^0_1 \tilde {\chi}^0_1 h}
& \simeq & \frac{\mu_{tot}}{ v}\,\big( \frac{\lambda v}{\mu_{tot}} \big)^2\, \frac { Z_s V_{h}^{\rm SM}(m_{\tilde{\chi}_1^0}/\mu_{tot} -\sin 2 \beta)}{1-(m_{\tilde{\chi}_1^0}/\mu_{tot})^2}  + \frac{\lambda}{2 \sqrt{2}} \big( \frac{\lambda v}{\mu_{tot}} \big)^2 \frac{Z_s V_{h}^{\rm S} \sin2\beta}{\big[ 1-(m_{\tilde{\chi}_1^0}/\mu_{tot})^2 \big]}\nonumber \\
& & -\sqrt{2}\kappa Z_s V_h^{\rm S} \left[1+ \big( \frac{\lambda v}{\sqrt{2}\mu_{tot}} \big)^2\frac{1}{1-(m_{\tilde{\chi}_1^0}/\mu_{tot})^2} \frac{\mu_{eff}}{\mu_{tot}} \right],
\label{C_xsxshsm} \\
C_{\tilde {\chi}^0_1 \tilde {\chi}^0_1 h_{s}} & \simeq &
\frac{\mu_{tot}}{ v}\,\big( \frac{\lambda v}{\mu_{tot}} \big)^2\, \frac { Z_s V_{h_{s}}^{\rm SM}(m_{\tilde{\chi}_1^0}/\mu_{tot} -\sin 2 \beta)}{1-(m_{\tilde{\chi}_1^0}/\mu_{tot})^2}  + \frac{\lambda}{2 \sqrt{2}} \big( \frac{\lambda v}{\mu_{tot}} \big)^2 \frac{Z_s V_{h_{s}}^{\rm S} \sin2\beta}{\big[ 1-(m_{\tilde{\chi}_1^0}/\mu_{tot})^2 \big]}\nonumber \\
& & -\sqrt{2}\kappa Z_s V_{h_{s}}^{\rm S} \left[1+ \big( \frac{\lambda v}{\sqrt{2} \mu_{tot}} \big)^2\frac{1}{1-(m_{\tilde{\chi}_1^0}/\mu_{tot})^2} \frac{\mu_{eff}}{\mu_{tot}}  \right],
\label{C_xsxshs}
\end{eqnarray}
and the expression of the SI cross section is~\cite{Cao:2021ljw}
\begin{eqnarray}
\sigma^{\rm SI}_{\tilde{\chi}_1^0-N} & \simeq & 5 \times 10^{-45}\,{\rm cm^2}~\left(\frac{\cal{A}}{0.1}\right)^2,
\end{eqnarray}
where
\begin{eqnarray}
{\cal{A}} = \left( \frac{125 \rm GeV}{m_{h}}\right)^2 V_h^{\rm SM} C_{ \tilde {\chi}^0_1 \tilde {\chi}^0_1 h} +  \left( \frac{125 \rm GeV}{m_{h_s}} \right)^2  V_{h_s}^{\rm SM} C_{ \tilde {\chi}^0_1 \tilde {\chi}^0_1 h_{s}}~.
\end{eqnarray}

Given the approximations of $V_h^{\rm S}$ and $V_{h_s}^{\rm SM}$ in Eq.~(\ref{Approximations}), these formulas suggest that the primary contribution to $\cal{A}$ in the series expansion of $\lambda$ is proportional to $\lambda \kappa$ if the singlet--doublet mixing is substantial and the contribution mediated by $h$ significantly cancels out that mediated by $h_s$. This dynamic is critical only when $h_s$ is moderately light. In contrast, if $h_s$ is tremendously massive and consequently $h$ is purely SM like (i.e., $V_h^{\rm SM}\simeq 1$, $V_h^{\rm S} \simeq 0$, and $V_{h_s}^{\rm SM} \simeq 0$), ${\cal{A}}$'s expression
is significantly simplified:
\begin{eqnarray}
{\cal{A}} = \left( \frac{125 \rm GeV}{m_{h}}\right)^2 \frac{\mu_{tot}}{v}\,\big( \frac{\lambda v}{\mu_{tot}} \big)^2\, \frac { Z_s (m_{\tilde{\chi}_1^0}/\mu_{tot} -\sin 2 \beta)}{1-(m_{\tilde{\chi}_1^0}/\mu_{tot})^2},
\end{eqnarray}
indicating that ${\cal{A}}$ is proportional to $\lambda^4$. Either a small $\lambda$ or the correlation $m_{\tilde{\chi}_1^0}/\mu_{\rm tot} \simeq \sin2\beta$, known as the blind-spot condition of the NMSSM~\cite{Badziak:2015exr}, is required to suppress the SI cross section.

Regarding the spin-dependent (SD) scattering cross section, only the $t$-channel exchange of a $Z$ boson contributes in the limit of very massive squarks. It is approximated by~\cite{Badziak:2015nrb,Badziak:2017uto}
\begin{eqnarray}
    \sigma_{\tilde{\chi}_1^0-N}^{\rm SD} &\simeq & C_N \times 10^{-4}~{\rm pb} \times \left(\frac{N_{13}^2-N_{14}^2}{0.1}\right)^2, \\
	&\simeq & C_N \times 10^{-2}~{\rm pb} \times \left( \frac{\lambda v}{\sqrt{2} \mu_{\rm tot}} \right)^4 \left( \frac{N_{15}^2 \cos{2\beta}} {1-(m_{\tilde{\chi}_1^0}/\mu_{\rm tot})^2} \right)^2, \label{SDCS}
\end{eqnarray}
where $C_N \simeq 4.0$ for protons and $C_N \simeq 3.1$ for neutrons. This formula indicates that the SD cross section is proportional to $(\lambda v/\mu_{tot})^4$ and increases as $m_{\tilde{\chi}_1^0}$ approaches $\mu_{tot}$ from below. It is the same as the expression of the SD scattering cross section for Bino-dominated DM,
\begin{eqnarray}
    \sigma_{\tilde{B}_1^0-N}^{\rm SD} &\simeq & C_N \times 10^{-2}~{\rm pb} \times \left( \frac{v}{\sqrt{2} \mu_{\rm tot}} \right)^4 \left( \frac{\cos{2\beta}} {1-(m_{\tilde{\chi}_1^0}/\mu_{\rm tot})^2} \right)^2, \label{SDCS-Bino}
\end{eqnarray}
except for an additional factor $\lambda^4 N_{15}^2$.

In summary, the aforementioned formulas suggest the plausible existence of a secluded DM sector, composed of singlet states~\cite{Pospelov:2007mp}. Specifically, the Singlino-dominated DM achieves the observed relic abundance primarily through the processes $\tilde{\chi}_1^0 \tilde{\chi}_1^0 \to h_s A_s, h_s h_s, A_sA_s$, or $h_s/A_s$-funnel by tuning $\kappa$. Since the DM sector communicates with the SM sector mainly by the weak singlet--doublet Higgs mixing, the DM-nucleon scattering is suppressed by $\lambda$.

\section{Numerical results}\label{sec:NR}

Here the sampling strategy of this study is introduced and  the features of DM physics in the GNMSSM are demonstrated. The model file of the GNMSSM was generated by the package \textsf{SARAH\, 4.14.3}~\cite{Staub:2008uz, Staub:2012pb, Staub:2013tta, Staub:2015kfa}, the particle spectrum and low-energy flavor measurements were calculated using the \textsf{SPheno\, 4.0.4}~\cite{Porod:2003um,Porod:2011nf} and \textsf{FlavorKit}~\cite{Porod:2014xia} code, respectively, and DM observables were computed with the package  \textsf{micrOMEGAs\, 5.0.4}~\cite{Belanger:2001fz, Belanger:2005kh, Belanger:2006is, Belanger:2010pz, Belanger:2013oya, Barducci:2016pcb}. We analyzed the acquired samples using the marginal posterior probability density function (PDF) in Bayesian inference and the profile likelihood (PL) in frequentist statistics~\cite{Fowlie:2016hew}\footnote{In frequentist statistics, the one-dimensional PL refers to the maximum likelihood value in a specific parameter space~\cite{Fowlie:2016hew}. For a given set of input parameters $\Theta \equiv\left(\Theta_1, \Theta_2, \cdots\right)$, the one-dimensional PL, denoted as
\begin{equation}
\mathcal{L}\left(\Theta_A\right)=\max _{\Theta_1, \cdots, \Theta_{A-1}, \Theta_{A+1}, \cdots} \mathcal{L}(\Theta),
\end{equation}
is obtained by maximizing the likelihood function while varying the other parameters. At a given point $\Theta_A$, $\mathcal{L}\left(\Theta_A\right)$ reflects the capability of that point within the theory to explain the experimental data. The one-dimensional marginal posterior PDF is derived through the integration of the posterior PDF from the Bayesian theorem with respect to the remaining inputs of the model, i.e.,
\begin{equation}
P\left(\Theta_A\right)=\int P(\Theta) d \Theta_1 d \Theta_2 \cdots d \Theta_{A-1} d \Theta_{A+1} \cdots \cdots.
\end{equation}
The PL reflects the data's preference for the parameter space, while the posterior PDF represents the preference of the samples obtained during the scan.}.

\subsection{Research strategy}\label{sec:RS}

\begin{table}[tbp]
\caption{In this study, the parameter space was explored for the $h_1$ scenario, assuming all the inputs were flatly distributed in the prior since they have clear physical meanings in the small $\lambda$ case. We assumed that the soft trilinear coefficients for the third-generation squarks were equal, i.e., $A_t = A_b$, for simplicity and let them vary because they could significantly influence the Standard Model (SM)-like Higgs boson mass by radiative corrections. We also fixed $M_3 =3~{\rm TeV}$ and selected a shared value of $2~{\rm TeV}$ for the other unmentioned dimensional parameters to be consistent with the Large Hadron Collider (LHC) search for new physics, noting that they are not crucial to this study. We define all these parameters at the renormalization scale $Q_{inp} = 1~{\rm TeV}$.  {\bf The $h_2$ scenario corresponds to the same parameter space as that of the $h_1$ scenario except that $ 1~{\rm GeV} \leq m_B \leq 130~{\rm GeV}$}.
\label{tab:1}}
\centering

\vspace{0.3cm}

\resizebox{0.7\textwidth}{!}{
\begin{tabular}{c|c|c|c|c|c}
\hline
Parameter & Prior & Range & Parameter & Prior & Range   \\
\hline
$\lambda$ & Flat & $0$--$0.75$ & $\kappa$ & Flat & $-0.75$--$0.75$  \\
$\tan \beta$ & Flat & $1$--$60$ & $v_s/{\rm TeV}$ & Flat & $ 10^{-3}$--$1.0 $ \\
$\mu_{\rm tot}/{\rm TeV}$ & Flat & $0.1$--$1.0$ & $m_N/{\rm TeV}$ & Flat & $-1.0$--$1.0$   \\
$m_B/{\rm TeV}$ & Flat & $0.1$--$1.0$ &$m_C/{\rm TeV}$ & Flat & $ 10^{-3}$--$1.0 $ \\
$M_1/{\rm TeV}$ & Flat & $-1.0$--$1.0 $ & $M_2/{\rm TeV}$ & Flat & $0.1$--$1.0$ \\
$A_t/{\rm TeV}$ & Flat & $-5.0$--$5.0$ & $A_\lambda/{\rm TeV}$ & Flat & $ -2.5$--$2.5$ \\
\hline
\end{tabular}}
\end{table}

We started by constructing a likelihood function that incorporated the DM relic abundance, results from the LZ experiment~\cite{LZ:2022lsv}, and other relevant constraints on the GNMSSM. The likelihood function is given by
\begin{eqnarray}
\mathcal{L} & \equiv & \mathcal{L}_{\Omega h^2} \times \mathcal{L}_{\rm LZ} \times \mathcal{L}_{\rm Const},  \nonumber \\
\mathcal{L}_{\Omega h^2} &=& \exp\left[ -\frac{1}{2}\left( \frac{\Omega h^2-0.120}{0.012 } \right)^2 \right], \quad  \mathcal{L}_{\rm LZ}= \exp \left[-\frac{\sigma^{\rm SI}_{\tilde{\chi}_1^0-p}}{2\delta_{\sigma}^2} \right ], \nonumber \\
\mathcal{L}_{\rm Const} &=& \left \{ \begin{aligned} & 1 & &{\rm if\ satisfying\ all\ experimental\ constraints} \\ &\exp\left[-100\right] & &{\rm otherwise} \end{aligned} \right .. \label{Likelihood}
\end{eqnarray}
In this formulation, we assumed the relic abundance $\Omega h^2$ followed a Gaussian distribution. We used its central value from the Planck experiment~\cite{Planck:2018vyg} with a theoretical uncertainty of $10\%$ in its calculation. Regarding the likelihood function of the LZ experiment, $\mathcal{L}_{\rm LZ}$, we modeled it with a Gaussian distribution centered at zero and defined $\delta_{\sigma}^2 = {\rm UL}_\sigma^2/1.64^2 + (0.2\sigma)^2$, where ${\rm UL}_\sigma$ refers to the upper limit of the LZ results on the DM-nucleon SI scattering rate at a $90\%$ C. L., and $0.2 \sigma$ accounts for theoretical uncertainties~\cite{Matsumoto:2016hbs}.
$\mathcal{L}_{\text{Const}}$ includes the following experimental constraints:
\begin{itemize}
    \item \textbf{Higgs data fit:} The properties of the SM-like Higgs boson $h$ must align with the LHC Higgs data at a $95\%$ confidence level. This condition was examined using the \textsf{HiggsSignal 2.6.2}code, requiring the fit's $p$-value to exceed 0.05~\cite{HS2013xfa,HSConstraining2013hwa,HS2014ewa,HS2020uwn}.
    \item \textbf{Extra Higgs searches:} This aspect was tested using the \textsf{HiggsBounds 5.10.2} code to conduct a comprehensive search for additional Higgs bosons beyond the SM at LEP, Tevatron, and the LHC~\cite{HB2008jh,HB2011sb,HBHS2012lvg,HB2013wla,HB2020pkv}.
   \item \textbf{Indirect DM searches:} The Fermi-LAT collaboration has made years of observations of dwarf galaxies, limiting the annihilation cross section as a function of the DM mass\footnote{see website: www-glast.stanford.edu/pub data/1048}. We employed the likelihood function proposed in Ref.~\cite{Carpenter:2016thc,Huang:2016tfo} to implement this constraint.
    \item \textbf{$B$-physics observables:} The branching ratios of $B_s \to \mu^+ \mu^-$ and $B \to X_s \gamma$ should remain consistent with their experimental measurements at the $2\sigma$ level~\cite{pdg2018}.
    \item \textbf{Vacuum stability:} The vacuum state of the scalar potential should be either stable or long lived~\cite{Hollik:2018wrr}. This criterion was rigorously examined using \textsf{VevaciousPlusPlus}~\cite{VPP2014}, the C++ version of \textsf{Vevacious} ~\cite{Camargo-Molina:2013qva}. Note that its implications on the GNMSSM were comprehensively addressed in a recent study~\cite{Hollik:2018wrr}.
\end{itemize}

\begin{table}[]
	\caption{Experimental analyses included in the package \texttt{SModelS-2.1.1}.}
	\label{tab:SModelS}
	\vspace{0.1cm}
	\resizebox{1.0 \textwidth}{!}{
		\begin{tabular}{cccc}
			\hline\hline
			\texttt{Name} & \texttt{Scenario} &\texttt{Final State} &$\texttt{Luminosity} (\texttt{fb}^{\texttt{-1}})$ \\\hline
			\begin{tabular}[l]{@{}l@{}} CMS-SUS-17-010~\cite{CMS:2018xqw}\end{tabular}   &\begin{tabular}[c]{@{}c@{}}$\tilde{\chi}_1^{\pm}\tilde{\chi}_1^{\mp}\rightarrow W^{\pm}\tilde{\chi}_1^0 W^{\mp}\tilde{\chi}_1^0$\\$\tilde{\chi}_1^{\pm}\tilde{\chi}_1^{\mp}\rightarrow \nu\tilde{\ell} \ell\tilde{\nu}$ \\ \end{tabular}&2$ \ell$  + $E_{\rm T}^{\rm miss}$    & 35.9  \\ \\
			\begin{tabular}[l]{@{}l@{}} CMS-SUS-17-009~\cite{CMS:2018eqb}\end{tabular}   &$\tilde{\ell}\tilde{\ell}\rightarrow \ell\tilde{\chi}_1^0\ell\tilde{\chi}_1^0$ &2$\ell$ + $E_{\rm T}^{\rm miss}$    &  35.9               \\ \\
			\begin{tabular}[l]{@{}l@{}} CMS-SUS-17-004~\cite{CMS:2018szt}\end{tabular} &$\tilde{\chi}_{2}^0\tilde{\chi}_1^{\pm}\rightarrow Wh(Z)\tilde{\chi}_1^0\tilde{\chi}_1^0$ & n$ \ell$(n\textgreater{}=0) + nj(n\textgreater{}=0) + $ E_{\rm T}^{\rm miss}$   & 35.9               \\ \\
			\begin{tabular}[l]{@{}l@{}}CMS-SUS-16-045~\cite{CMS:2017bki}\end{tabular}          &$\tilde{\chi}_2^0\tilde{\chi}_1^{\pm}\rightarrow W^{\pm}\tilde{\chi}_1^0h\tilde{\chi}_1^0$& 1$ \ell$ 2b + $ E_{\rm T}^{\rm miss}$                           & 35.9               \\ \\
			\begin{tabular}[l]{@{}l@{}} CMS-SUSY-16-039~\cite{CMS:2017moi} \end{tabular}          &\begin{tabular}[c]{@{}c@{}c@{}c@{}c@{}} $\tilde{\chi}_2^0\tilde{\chi}_1^{\pm}\rightarrow \ell\tilde{\nu}\ell\tilde{\ell}$\\$\tilde{\chi}_2^0\tilde{\chi}_1^{\pm}\rightarrow\tilde{\tau}\nu\tilde{\ell}\ell$\\$\tilde{\chi}_2^0\tilde{\chi}_1^{\pm}\rightarrow\tilde{\tau}\nu\tilde{\tau}\tau$\\ $\tilde{\chi}_2^0\tilde{\chi}_1^{\pm}\rightarrow WZ\tilde{\chi}_1^0\tilde{\chi}_1^0$\\$\tilde{\chi}_2^0\tilde{\chi}_1^{\pm}\rightarrow WH\tilde{\chi}_1^0\tilde{\chi}_1^0$\end{tabular} & n$\ell(n\textgreater{}0)$($\tau$) + $E_{\rm T}^{\rm miss}$& 35.9\\ \\
			\begin{tabular}[l]{@{}l@{}}CMS-SUS-16-034~\cite{CMS:2017kxn}\end{tabular}&$\tilde{\chi}_2^0\tilde{\chi}_1^{\pm}\rightarrow W\tilde{\chi}_1^0Z(h)\tilde{\chi}_1^0$ & n$\ell$(n\textgreater{}=2) + nj(n\textgreater{}=1) $E_{\rm T}^{\rm miss}$       &               35.9               \\ \\
			\begin{tabular}[l]{@{}l@{}}ATLAS-1803-02762~\cite{ATLAS:2018ojr}\end{tabular} &\begin{tabular}[c]{@{}c@{}c@{}c@{}}$\tilde{\chi}_2^0\tilde{\chi}_1^{\pm}\rightarrow WZ\tilde{\chi}_1^0\tilde{\chi}_1^0$\\$\tilde{\chi}_2^0\tilde{\chi}_1^{\pm}\rightarrow \nu\tilde{\ell}l\tilde{\ell}$\\$\tilde{\chi}_1^{\pm}\tilde{\chi}_1^{\mp}\rightarrow \nu\tilde{\ell}\nu\tilde{\ell}$\\ $ \tilde{\ell}\tilde{\ell}\rightarrow \ell\tilde{\chi}_1^0\ell\tilde{\chi}_1^0$\end{tabular} & n$ \ell$ (n\textgreater{}=2) + $ E_{\rm T}^{\rm miss}$ & 36.1               \\ \\
			\begin{tabular}[l]{@{}l@{}}ATLAS-1812-09432~\cite{ATLAS:2018qmw}\end{tabular} &$\tilde{\chi}_2^0\tilde{\chi}_1^{\pm}\rightarrow Wh\tilde{\chi}_1^0\tilde{\chi}_1^0$ & n$ \ell$ (n\textgreater{}=0) + nj(n\textgreater{}=0) + nb(n\textgreater{}=0) + n$\gamma$(n\textgreater{}=0) + $E_{\rm T}^{\rm miss}$ & 36.1               \\ \\
			\begin{tabular}[l]{@{}l@{}}ATLAS-1806-02293~\cite{ATLAS:2018eui}\end{tabular} &$\tilde{\chi}_2^0\tilde{\chi}_1^{\pm}\rightarrow WZ\tilde{\chi}_1^0\tilde{\chi}_1^0$ &n$\ell$(n\textgreater{}=2) + nj(n\textgreater{}=0) + $ E_T^{miss}$ & 36.1               \\ \\
			\begin{tabular}[l]{@{}l@{}}ATLAS-1912-08479~\cite{ATLAS:2019wgx}\end{tabular}          &$\tilde{\chi}_2^0\tilde{\chi}_1^{\pm}\rightarrow W(\rightarrow l\nu)\tilde{\chi}_1^0Z(\rightarrow\ell\ell)\tilde{\chi}_1^0$& 3$\ell $ + $ E_{\rm T}^{\rm miss}$                           & 139               \\ \\
			\begin{tabular}[l]{@{}l@{}}ATLAS-1908-08215~\cite{ATLAS:2019lff}\end{tabular}   &\begin{tabular}[c]{@{}c@{}}$\tilde{\ell}\tilde{\ell}\rightarrow \ell\tilde{\chi}_1^0\ell\tilde{\chi}_1^0$\\$\tilde{\chi}_1^{\pm}\tilde{\chi}_1^{\mp}$ \\ \end{tabular} & 2$\ell$ + $ E_{\rm T}^{\rm miss}$ & 139               \\ \\
			\begin{tabular}[l]{@{}l@{}}ATLAS-1909-09226~\cite{Aad:2019vvf}\end{tabular}          & $\tilde{\chi}_{2}^0\tilde{\chi}_1^{\pm}\rightarrow Wh\tilde{\chi}_1^0\tilde{\chi}_1^0$                            & 1$ \ell$ + h($\bm \rightarrow$ bb) + $ E_{\rm T}^{\rm miss}$    & 139               \\ \hline\\
			
	\end{tabular}} 
\end{table}

\begin{table}[]
	\caption{Experimental analyses of the electroweakino production processes considered in this study, categorized by the topologies of the electroweakino's signal.}
	\label{tab:LHC1}
	\vspace{0.2cm}
	\resizebox{0.97\textwidth}{!}{
		\begin{tabular}{llll}
			\hline\hline
			\texttt{Scenario} & \texttt{Final State} &\multicolumn{1}{c}{\texttt{Name}}\\\hline
			\multirow{6}{*}{$\tilde{\chi}_{2}^0\tilde{\chi}_1^{\pm}\rightarrow WZ\tilde{\chi}_1^0\tilde{\chi}_1^0$}&\multirow{6}{*}{$n\ell (n\geq2) + nj(n\geq0) + \text{E}_\text{T}^{\text{miss}}$}&\texttt{CMS-SUS-20-001($137fb^{-1}$)}~\cite{CMS:2020bfa}\\&&\texttt{ATLAS-2106-01676($139fb^{-1}$)}~\cite{ATLAS:2021moa}\\&&\texttt{CMS-SUS-17-004($35.9fb^{-1}$)}~\cite{CMS:2018szt}\\&&\texttt{CMS-SUS-16-039($35.9fb^{-1}$)}~\cite{CMS:2017moi}\\&&\texttt{ATLAS-1803-02762($36.1fb^{-1}$)}~\cite{ATLAS:2018ojr}\\&&\texttt{ATLAS-1806-02293($36.1fb^{-1}$)}~\cite{ATLAS:2018eui}\\\\
			\multirow{2}{*}{$\tilde{\chi}_2^0\tilde{\chi}_1^{\pm}\rightarrow \ell\tilde{\nu}\ell\tilde{\ell}$}&\multirow{2}{*}{$n\ell (n=3) + \text{E}_\text{T}^{\text{miss}}$}&\texttt{CMS-SUS-16-039($35.9fb^{-1}$)}~\cite{CMS:2017moi}\\&&\texttt{ATLAS-1803-02762($36.1fb^{-1}$)}~\cite{ATLAS:2018ojr}\\\\
			$\tilde{\chi}_2^0\tilde{\chi}_1^{\pm}\rightarrow \tilde{\tau}\nu\ell\tilde{\ell}$&$2\ell + 1\tau + \text{E}_\text{T}^{\text{miss}}$&\texttt{CMS-SUS-16-039($35.9fb^{-1}$)}~\cite{CMS:2017moi}\\\\
			$\tilde{\chi}_2^0\tilde{\chi}_1^{\pm}\rightarrow \tilde{\tau}\nu\tilde{\tau}\tau$&$3\tau + \text{E}_\text{T}^{\text{miss}}$&\texttt{CMS-SUS-16-039($35.9fb^{-1}$)}~\cite{CMS:2017moi}\\\\
			\multirow{6}{*}{$\tilde{\chi}_{2}^0\tilde{\chi}_1^{\pm}\rightarrow Wh\tilde{\chi}_1^0\tilde{\chi}_1^0$}&\multirow{6}{*}{$n\ell(n\geq1) + nb(n\geq0) + nj(n\geq0) + \text{E}_\text{T}^{\text{miss}}$}&\texttt{ATLAS-1909-09226($139fb^{-1}$)}~\cite{ATLAS:2020pgy}\\&&\texttt{CMS-SUS-17-004($35.9fb^{-1}$)}~\cite{CMS:2018szt}\\&&\texttt{CMS-SUS-16-039($35.9fb^{-1}$)}~\cite{CMS:2017moi}\\
			&&\texttt{ATLAS-1812-09432($36.1fb^{-1}$)}\cite{ATLAS:2018qmw}\\&&\texttt{CMS-SUS-16-034($35.9fb^{-1}$)}\cite{CMS:2017kxn}\\&&\texttt{CMS-SUS-16-045($35.9fb^{-1}$)}~\cite{CMS:2017bki}\\\\
			\multirow{2}{*}{$\tilde{\chi}_1^{\mp}\tilde{\chi}_1^{\pm}\rightarrow WW\tilde{\chi}_1^0 \tilde{\chi}_1^0$}&\multirow{2}{*}{$2\ell + \text{E}_\text{T}^{\text{miss}}$}&\texttt{ATLAS-1908-08215($139fb^{-1}$)}~\cite{ATLAS:2019lff}\\&&\texttt{CMS-SUS-17-010($35.9fb^{-1}$)}~\cite{CMS:2018xqw}\\\\
			\multirow{2}{*}{$\tilde{\chi}_1^{\mp}\tilde{\chi}_1^{\pm}\rightarrow 2\tilde{\ell}\nu(\tilde{\nu}\ell)$}&\multirow{2}{*}{$2\ell + \text{E}_\text{T}^{\text{miss}}$}&\texttt{ATLAS-1908-08215($139fb^{-1}$)}~\cite{ATLAS:2019lff}\\&&\texttt{CMS-SUS-17-010($35.9fb^{-1}$)}~\cite{CMS:2018xqw}\\\\
			\multirow{1}{*}{$\tilde{\chi}_2^{0}\tilde{\chi}_1^{\pm}\rightarrow ZW\tilde{\chi}_1^0\tilde{\chi}_1^0$}&\multirow{2}{*}{$2j(\text{large}) + \text{E}_\text{T}^{\text{miss}}$}&\multirow{2}{*}{\texttt{ATLAS-2108-07586($139fb^{-1}$)}~\cite{ATLAS:2021yqv}}\\{$\tilde{\chi}_1^{\pm}\tilde{\chi}_1^{\mp}\rightarrow WW\tilde{\chi}_1^0\tilde{\chi}_1^0$}&&\\\\
			\multirow{1}{*}{$\tilde{\chi}_2^{0}\tilde{\chi}_1^{\pm}\rightarrow (h/Z)W\tilde{\chi}_1^0\tilde{\chi}_1^0$}&\multirow{2}{*}{$j(\text{large}) + b(\text{large}) + \text{E}_\text{T}^{\text{miss}}$}&\multirow{2}{*}{\texttt{ATLAS-2108-07586($139fb^{-1}$)}~\cite{ATLAS:2021yqv}}\\{$\tilde{\chi}_2^{0}\tilde{\chi}_3^{0}\rightarrow (h/Z)Z\tilde{\chi}_1^0\tilde{\chi}_1^0$}&&\\\\
			$\tilde{\chi}_2^{0}\tilde{\chi}_1^{\mp}\rightarrow h/ZW\tilde{\chi}_1^0\tilde{\chi}_1^0,\tilde{\chi}_1^0\rightarrow \gamma/Z\tilde{G}$&\multirow{2}{*}{$2\gamma + n\ell(n\geq0) + nb(n\geq0) + nj(n\geq0) + \text{E}_\text{T}^{\text{miss}}$}&\multirow{2}{*}{\texttt{ATLAS-1802-03158($36.1fb^{-1}$)}~\cite{ATLAS:2018nud}}\\$\tilde{\chi}_1^{\pm}\tilde{\chi}_1^{\mp}\rightarrow WW\tilde{\chi}_1^0\tilde{\chi}_1^0,\tilde{\chi}_1^0\rightarrow \gamma/Z\tilde{G}$&&\\\\
			$\tilde{\chi}_2^{0}\tilde{\chi}_1^{\pm}\rightarrow ZW\tilde{\chi}_1^0\tilde{\chi}_1^0,\tilde{\chi}_1^0\rightarrow h/Z\tilde{G}$&\multirow{4}{*}{$n\ell(n\geq4) + \text{E}_\text{T}^{\text{miss}}$}&\multirow{4}{*}{\texttt{ATLAS-2103-11684($139fb^{-1}$)}~\cite{ATLAS:2021yyr}}\\$\tilde{\chi}_1^{\pm}\tilde{\chi}_1^{\mp}\rightarrow WW\tilde{\chi}_1^0\tilde{\chi}_1^0,\tilde{\chi}_1^0\rightarrow h/Z\tilde{G}$&&\\$\tilde{\chi}_2^{0}\tilde{\chi}_1^{0}\rightarrow Z\tilde{\chi}_1^0\tilde{\chi}_1^0,\tilde{\chi}_1^0\rightarrow h/Z\tilde{G}$&&\\$\tilde{\chi}_1^{\mp}\tilde{\chi}_1^{0}\rightarrow W\tilde{\chi}_1^0\tilde{\chi}_1^0,\tilde{\chi}_1^0\rightarrow h/Z\tilde{G}$&&\\\\
			\multirow{3}{*}{$\tilde{\chi}_{i}^{0,\pm}\tilde{\chi}_{j}^{0,\mp}\rightarrow \tilde{\chi}_1^0\tilde{\chi}_1^0+\chi_{soft}\rightarrow ZZ/H\tilde{G}\tilde{G}$}&\multirow{3}{*}{$n\ell(n\geq2) + nb(n\geq0) + nj(n\geq0) + \text{E}_\text{T}^{\text{miss}}$}&\texttt{CMS-SUS-16-039($35.9fb^{-1}$)}~\cite{CMS:2017moi}\\&&\texttt{CMS-SUS-17-004($35.9fb^{-1}$)}~\cite{CMS:2018szt}\\&&\texttt{CMS-SUS-20-001($137fb^{-1}$)}~\cite{CMS:2020bfa}\\\\
			\multirow{2}{*}{$\tilde{\chi}_{i}^{0,\pm}\tilde{\chi}_{j}^{0,\mp}\rightarrow \tilde{\chi}_1^0\tilde{\chi}_1^0+\chi_{soft}\rightarrow HH\tilde{G}\tilde{G}$}&\multirow{2}{*}{$n\ell(n\geq2) + nb(n\geq0) + nj(n\geq0) + \text{E}_\text{T}^{\text{miss}}$}&\texttt{CMS-SUS-16-039($35.9fb^{-1}$)}~\cite{CMS:2017moi}\\&&\texttt{CMS-SUS-17-004($35.9fb^{-1}$)}~\cite{CMS:2018szt}\\\\
			$\tilde{\chi}_{2}^{0}\tilde{\chi}_{1}^{\pm}\rightarrow W^{*}Z^{*}\tilde{\chi}_1^0\tilde{\chi}_1^0$&$3\ell + \text{E}_\text{T}^{\text{miss}}$&\texttt{ATLAS-2106-01676($139fb^{-1}$)}~\cite{ATLAS:2021moa}\\\\
			\multirow{3}{*}{$\tilde{\chi}_{2}^{0}\tilde{\chi}_{1}^{\pm}\rightarrow Z^{*}W^{*}\tilde{\chi}_1^0\tilde{\chi}_1^0$}&\multirow{2}{*}{$2\ell + nj(n\geq0) + \text{E}_\text{T}^{\text{miss}}$}&\texttt{ATLAS-1911-12606($139fb^{-1}$)}~\cite{ATLAS:2019lng}\\&&\texttt{ATLAS-1712-08119($36.1fb^{-1}$)}~\cite{ATLAS:2017vat}\\&&\texttt{CMS-SUS-16-048($35.9fb^{-1}$)}~\cite{CMS:2018kag}\\\\
			\multirow{3}{*}{$\tilde{\chi}_{2}^{0}\tilde{\chi}_{1}^{\pm}+\tilde{\chi}_{1}^{\pm}\tilde{\chi}_{1}^{\mp}+\tilde{\chi}_{1}^{\pm}\tilde{\chi}_{1}^{0}$}&\multirow{3}{*}{$2\ell + nj(n\geq0) + \text{E}_\text{T}^{\text{miss}}$}&\texttt{ATLAS-1911-12606($139fb^{-1}$)}~\cite{ATLAS:2019lng}\\&&\texttt{ATLAS-1712-08119($36.1fb^{-1}$)}~\cite{ATLAS:2017vat}\\&&\texttt{CMS-SUS-16-048($35.9fb^{-1}$)}~\cite{CMS:2018kag}\\\hline
	\end{tabular}}
\end{table}

Next, we performed sophisticated scans over the parameter space in Table~\ref{tab:1} for the $h_1$ and $h_2$ scenario\footnote{Given that the DM physics of the GNMSSM involves numerous free parameters and $\xi^\prime$ only appears in the tadpole equation to determine the soft squared mass $m_S^2$, we simplify our study by assuming that the tadpole term in Eq.~(\ref{Soft-terms}) vanishes. Under this assumption, $A_\kappa$ is related to $v_s$ by the equation $\kappa A_\kappa = \sqrt{2} m^2_B/v_s + \lambda \mu v^2/v_s^2 - \lambda(A_\lambda + m_N - \sqrt{2} \sin{2\beta} \kappa v_s) v^2/(2 v_s^2) + \sqrt{2} \kappa^2 v_s - 3 \kappa m_N$, according to the expression of $\xi^\prime$ in Eq.~(\ref{Simplify-1}). As we will demonstrate below, this specific parameter configuration suggests that the GNMSSM predominantly predicts a Singlino-dominated DM. We emphasize that this conclusion remains valid even when $A_\kappa$ is considered a free parameter, leading to a nonzero $\xi^\prime$. The primary reason for this is that $A_\kappa$ only influences the triple Higgs couplings in Eq.~(\ref{Coupling-size}), and allowing it to vary freely enables the theory to more easily achieve the observed relic abundance when any of the processes $\tilde{\chi}_1^0 \tilde{\chi}_1^0 \to h_s A_s, h_s h_s, A_s A_s$ is the dominant annihilation channel. We verified that $\kappa A_\kappa$ in this study is within a reasonable range, satisfying $-800~{\rm GeV} \lesssim \kappa A_\kappa \lesssim 1200~{\rm GeV}$ for the $2\sigma$ credible region. }, using the MultiNest algorithm with ${\it{nlive}} = 8000$~\cite{MultiNest2009}\footnote{{\it{Nlive}} in the MultiNest method represents the number of active or live points to determine the iso-likelihood contour in each iteration~\cite{MultiNest2009,Importance2019}.
 The larger it is, the more detailed the scan results will be in surveying the parameter space.} and the likelihood function in Eq.~(\ref{Likelihood}).
Regarding the acquired samples, we were particularly interested in those that can explain the relic abundance at the $2 \sigma$ level, satisfy the LZ bounds, and be consistent with all the experimental constraints. We will analyze their features using statistical quantities and illustrate the differences of the two scenarios.

Finally, we assessed the alignment of the samples with the results of the LHC search for electroweakinos. To reduce the calculation time, we initially utilized the program \textsf{SModelS-2.1.1}~\cite{Khosa:2020zar}, which encoded various event-selection efficiencies by the topologies of the electroweakino signals listed in Table~\ref{tab:SModelS} to implement the task. Given that the exclusion capability of \textsf{SModelS-2.1.1} on the samples is limited by its database and strict working prerequisites, we further surveyed the surviving samples by simulating the analyses listed in Table~\ref{tab:LHC1}. We finished this task by the following procedures. We concentrated on the electroweakino production processes
\begin{equation}\begin{split}
pp \to \tilde{\chi}_i^0\tilde{\chi}_j^{\pm} &, \quad i = 2, 3, 4, 5; \quad j = 1, 2 \\
pp \to \tilde{\chi}_i^{\pm}\tilde{\chi}_j^{\mp} &, \quad i,j = 1, 2 \\
pp \to \tilde{\chi}_i^{0}\tilde{\chi}_j^{0} &, \quad i,j = 2, 3, 4, 5
\end{split}\end{equation}
and calculated their cross sections to next-to-leading order using the program \texttt{Prospino2}~\cite{Beenakker:1996ed}.
Subsequently, we generated $10^5$ events for these processes by \texttt{MadGraph\_aMC@NLO}~\cite{Alwall:2011uj, Conte:2012fm} and furnished their parton shower and hadronization by the program \texttt{PYTHIA8}~\cite{Sjostrand:2014zea}. Detector simulations were implemented with the program \texttt{Delphes}~\cite{deFavereau:2013fsa}. Finally, we fed the event files into the package \texttt{CheckMATE-2.0.29}~\cite{Drees:2013wra,Dercks:2016npn, Kim:2015wza} to calculate the $R$ value defined by $R \equiv max\{S_i/S_{i,obs}^{95}\}$ for all the involved analyses, where $S_i$ represents the simulated event number of the $i$-th signal region (SR), and $S_{i,obs}^{95}$ is
the corresponding $95\%$ confidence level upper limit. $R > 1 $ means that the sample is experimentally excluded if the involved uncertainties are neglected~\cite{Cao:2021tuh}, while $R < 1$ indicates that it is consistent with the experimental analyses.

\subsection{Global features}\label{sec:nr}

\begin{table}[tbp]
    \caption{Primary annihilation channels for Singlino-dominated dark matter (DM) in the $h_1$ scenario and their marginal posterior probability density functions (PDFs) normalized to the total Bayesian evidence of the scenario. The second line takes into account all 21097 samples acquired from the scan, while the third line only includes those further satisfying the LHC constraints, which resulted in smaller PDFs than their corresponding ones in the second line. Regarding the co-annihilation in this table, the annihilation partners were Higgsino-like electroweakinos in most cases and Wino-like electroweakinos in a few cases, as indicated by the samples.  \label{tab:tab4}}
    \centering
    \vspace{0.3cm}
    \resizebox{0.7\textwidth}{!}{
\begin{tabular}{cccc}
\hline
\multicolumn{1}{c|}{$\tilde{\chi}_1^0 \tilde{\chi}_1^0  \to  h_s  A_s$} & \multicolumn{1}{c|}{$\tilde{\chi}_1^0 \tilde{\chi}_1^0  \to  A_s  A_s$} & \multicolumn{1}{c|}{Co-annihilation} & \multicolumn{1}{c}{$\tilde{\chi}_1^0 \tilde{\chi}_1^0  \to  h_s  h_s$} \\ \hline
\multicolumn{1}{c|}{$58.1\%$}                                   & \multicolumn{1}{c|}{$34.8\%$} & \multicolumn{1}{c|}{$2.5\%$} & \multicolumn{1}{c}{$2.3\%$}  \\ \hline
\multicolumn{1}{c|}{$58.0\%$}                                   & \multicolumn{1}{c|}{$32.7\%$} & \multicolumn{1}{c|}{$2.4\%$} & \multicolumn{1}{c}{$2.3\%$}  \\ \hline
\end{tabular}}
\end{table}

\begin{table}[tbp]
    \caption{Similar to Table \ref{tab:tab4}, but presenting the $h_2$ scenario results. The channel $\tilde{\chi}_1^0 \tilde{\chi}_1^0  \to  q  \bar{q}$ in this table proceeded primarily through exchanging a resonant $h_s$, $A_s$, or $h$. \label{tab:tab5}}
    \centering
    \vspace{0.3cm}
    \resizebox{0.7\textwidth}{!}{
\begin{tabular}{cccc}
\hline
\multicolumn{1}{c|}{$\tilde{\chi}_1^0 \tilde{\chi}_1^0  \to  h_s  h_s$} & \multicolumn{1}{c|}{$\tilde{\chi}_1^0 \tilde{\chi}_1^0  \to  h_s  A_s$} & \multicolumn{1}{c|}{Co-annihilation} & \multicolumn{1}{c}{$\tilde{\chi}_1^0 \tilde{\chi}_1^0  \to  q  \bar{q}$} \\ \hline
\multicolumn{1}{c|}{$42.3\%$}                                   & \multicolumn{1}{c|}{$12.6\%$} & \multicolumn{1}{c|}{$5.9\%$} & \multicolumn{1}{c}{$2.1\%$}  \\ \hline
\multicolumn{1}{c|}{$18.3\%$}                                   & \multicolumn{1}{c|}{$9.7\%$} & \multicolumn{1}{c|}{$4.1\%$} & \multicolumn{1}{c}{$1.4\%$}  \\ \hline
\end{tabular}}
\end{table}

\begin{figure*}[t]
		\centering
\includegraphics[width=0.50\textwidth]{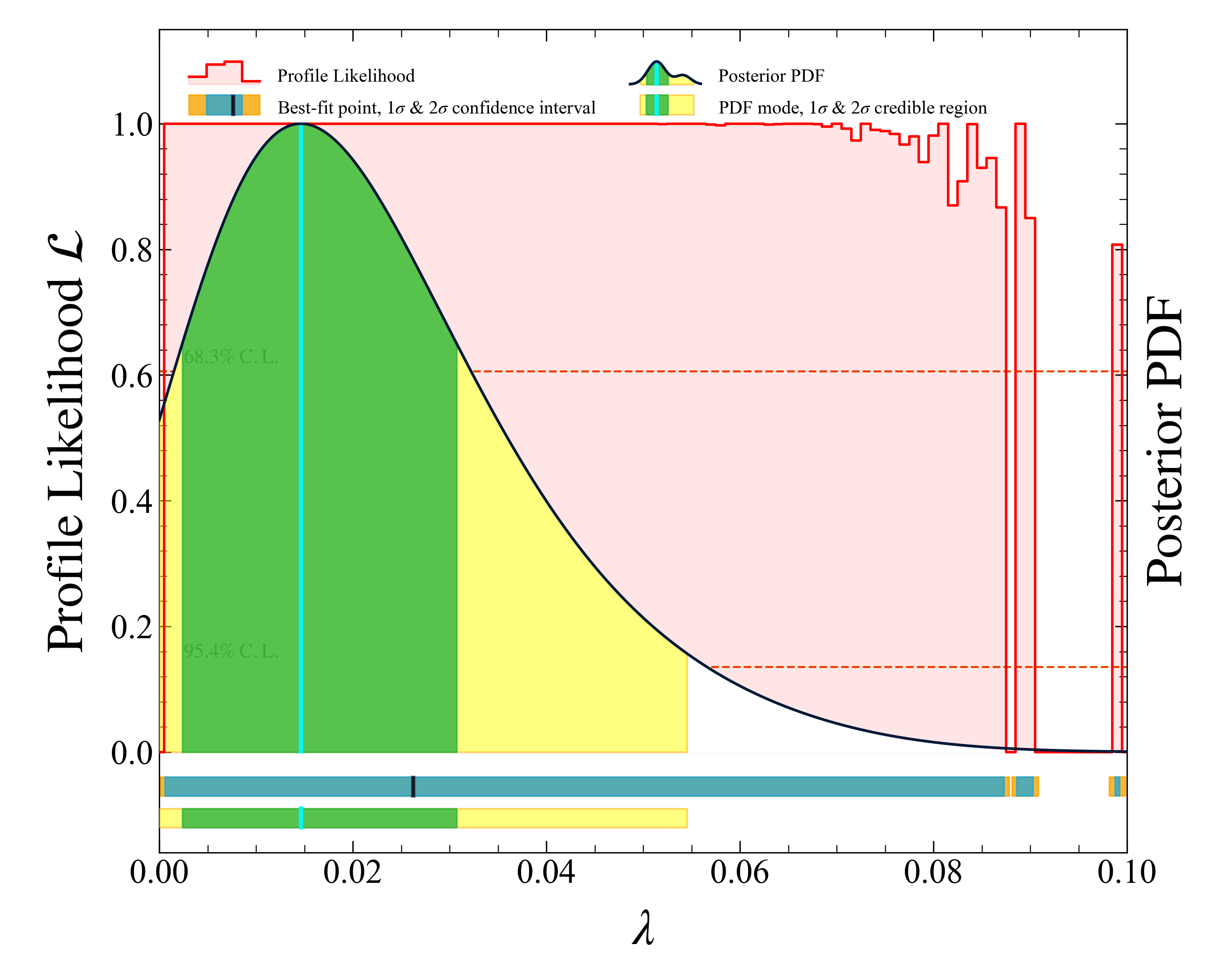}\hspace{-0.2cm}	\includegraphics[width=0.50\textwidth]{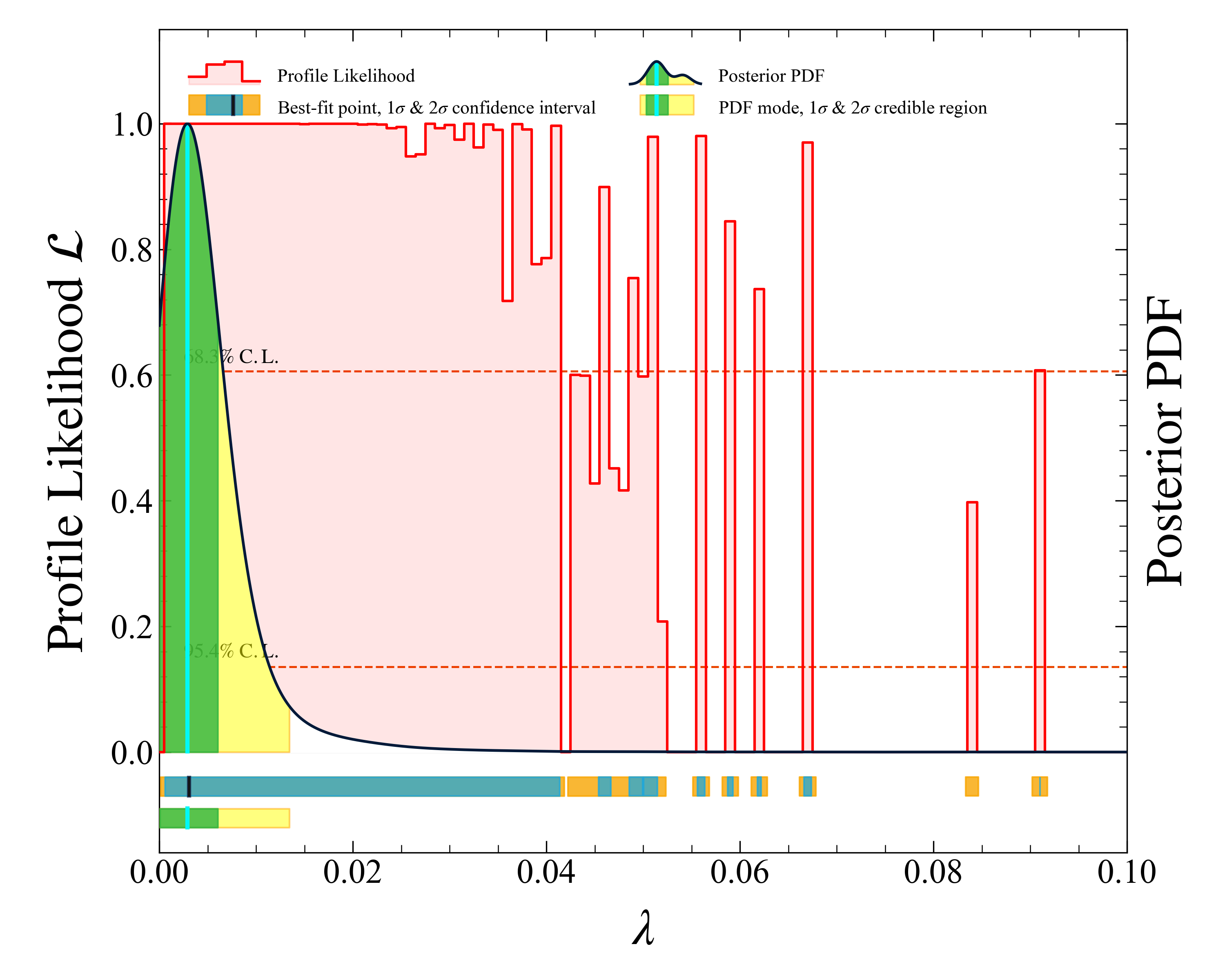} 
    \\
\includegraphics[width=0.50\textwidth]{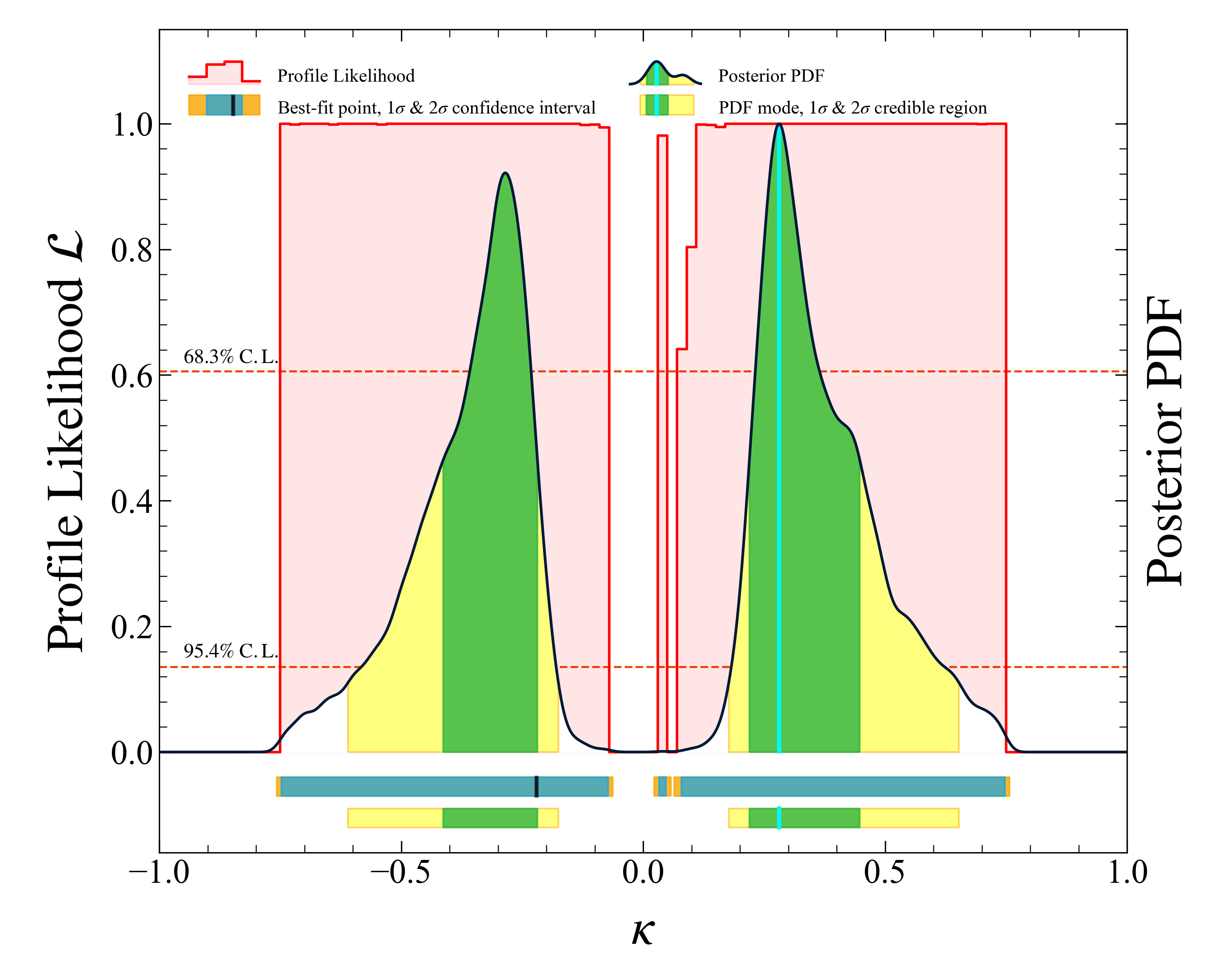}\hspace{-0.2cm}	\includegraphics[width=0.50\textwidth]{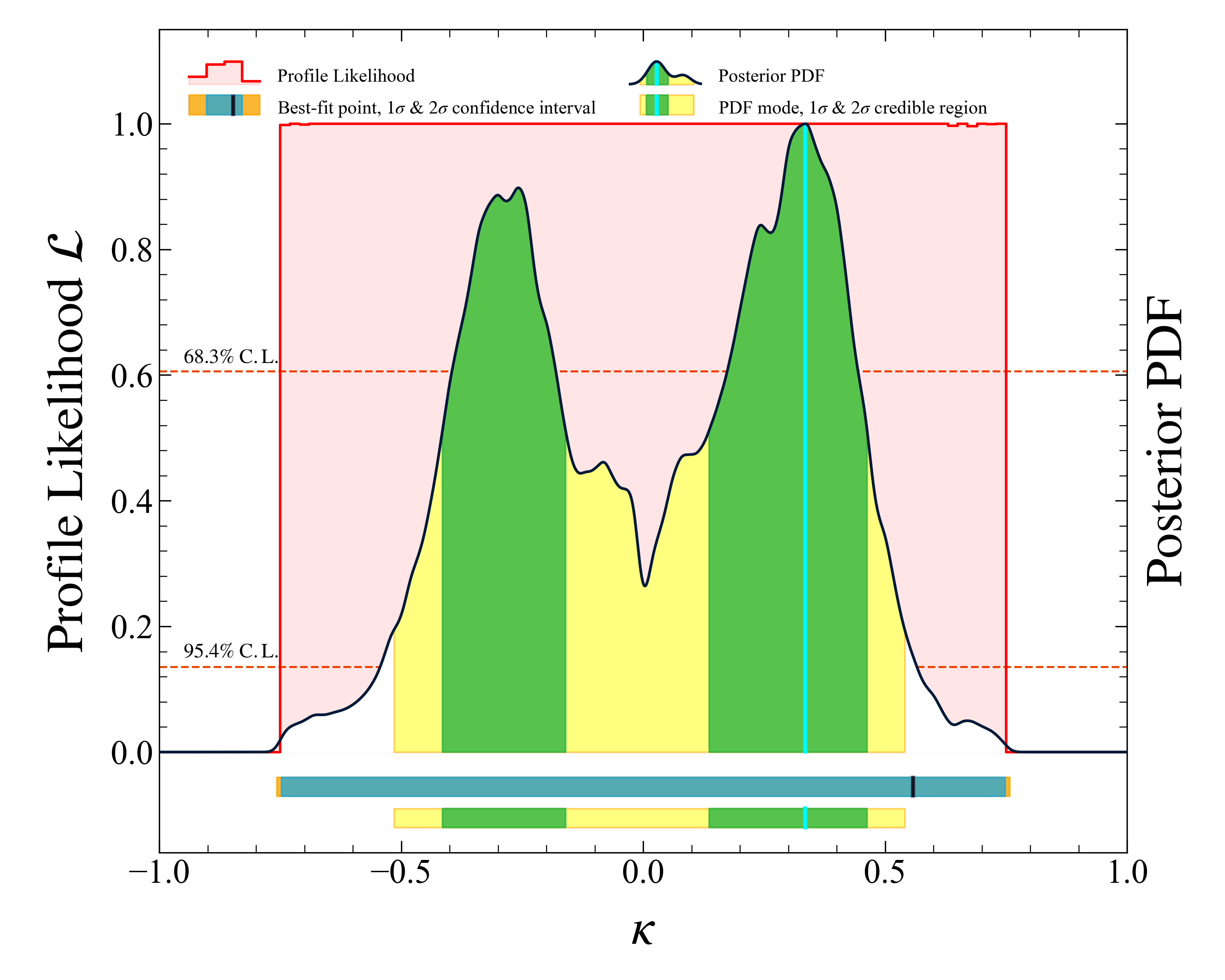}
\\
\includegraphics[width=0.50\textwidth]{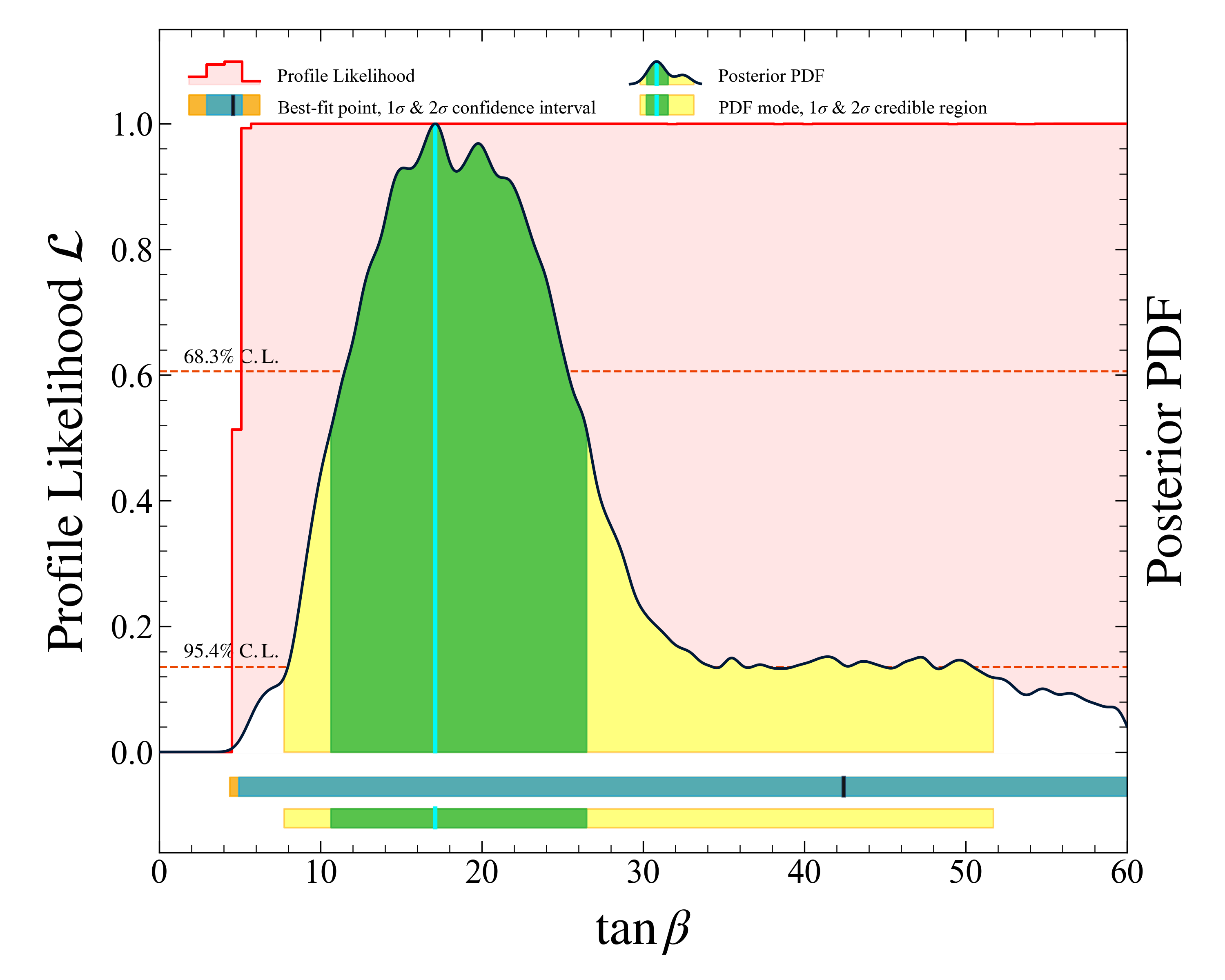}\hspace{-0.2cm}	\includegraphics[width=0.50\textwidth]{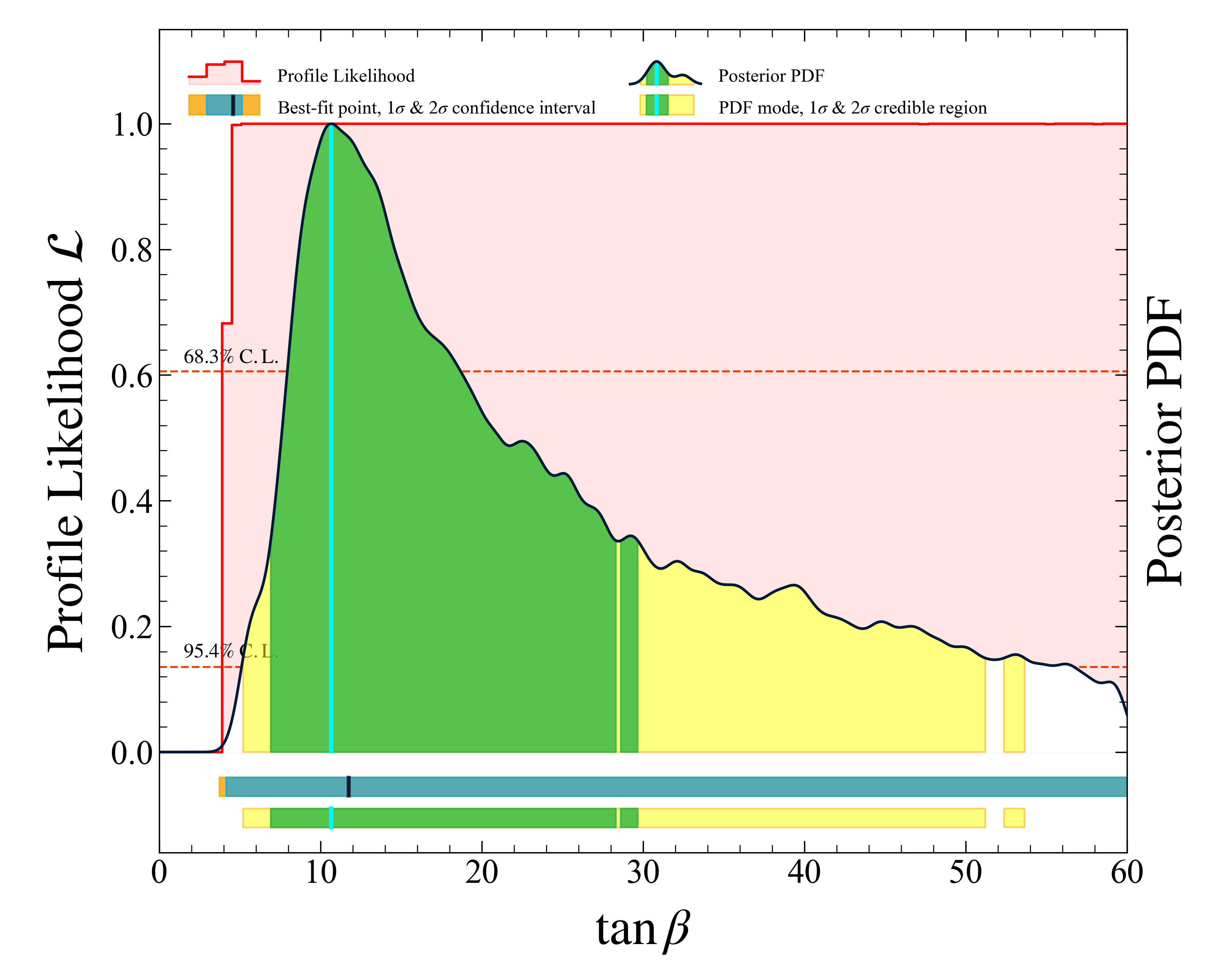}
\\

\vspace{-0.5cm}

\caption{ One-dimensional PLs (red line) and marginal posterior PDFs (black line) of $\lambda$, $\kappa$, and $\tan \beta$ before implementing the LHC constraints. The left panels show the results of the $h_1$ scenario, while the right panels depict those of the $h_2$ scenario. Statistical measures such as the confidence interval and credible region were introduced in Ref.~\cite{Fowlie:2016hew}. \label{Fig1}}
\end{figure*}

\begin{figure*}[t]
		\centering

\includegraphics[width=0.50\textwidth]{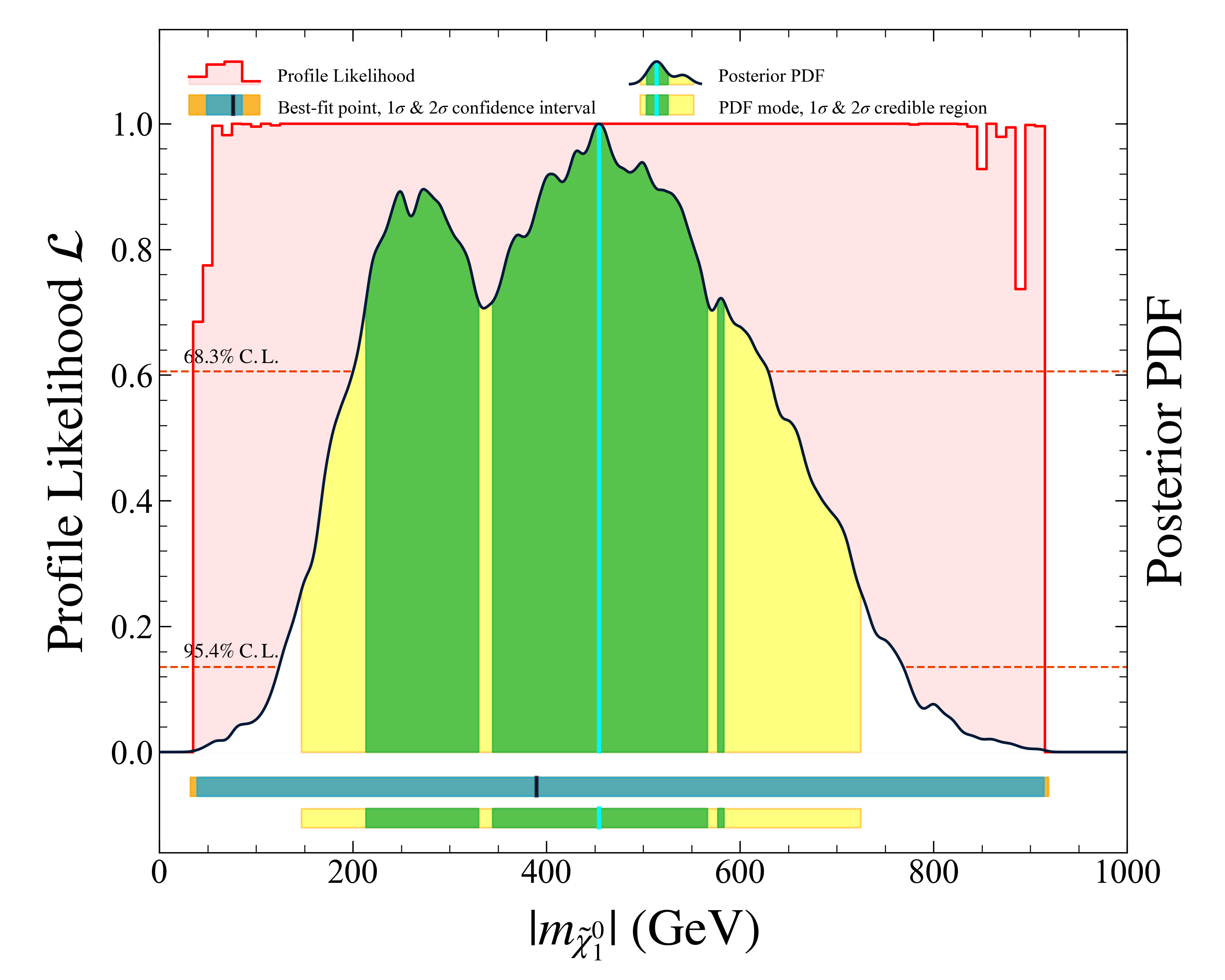}\hspace{-0.2cm}	\includegraphics[width=0.50\textwidth]{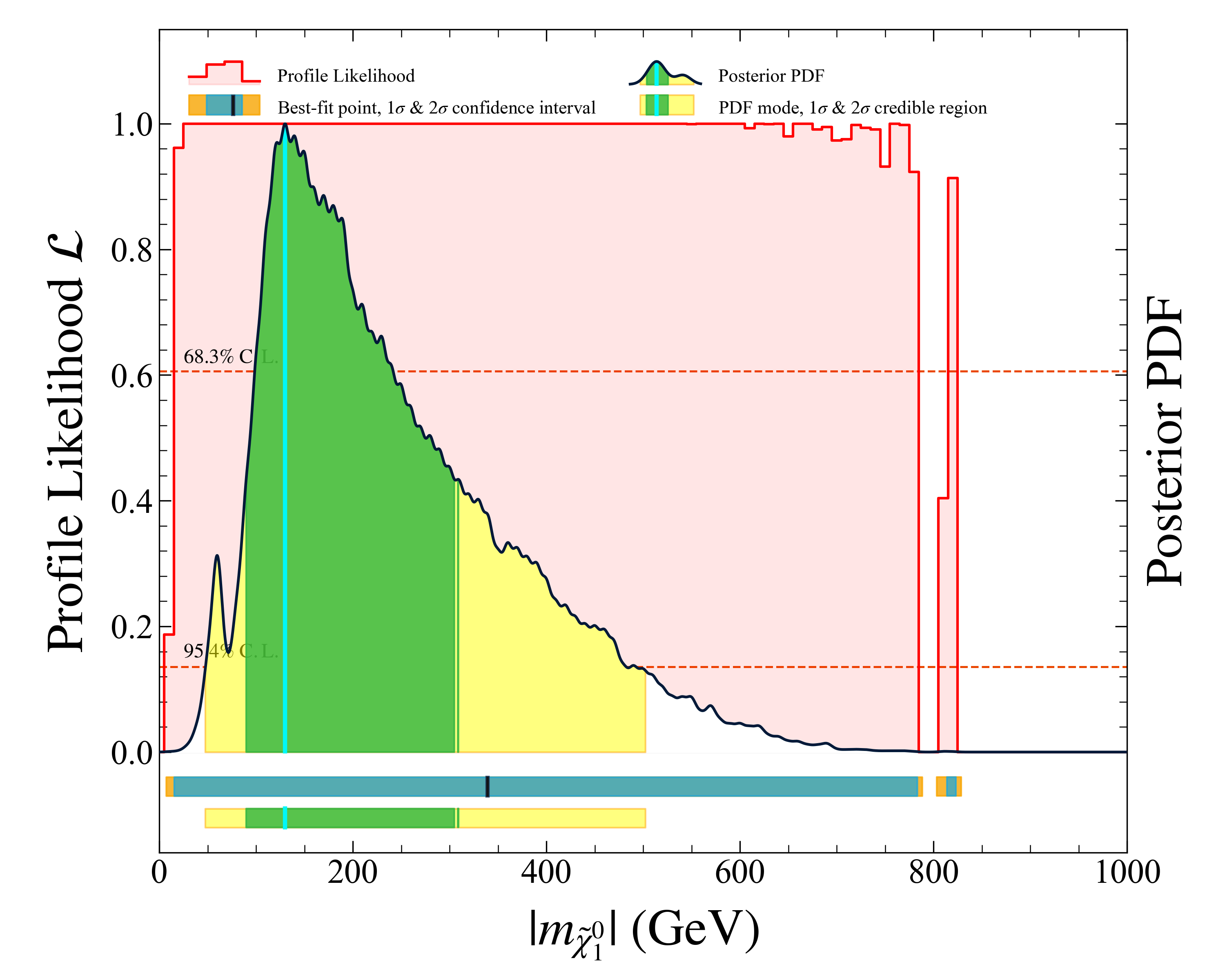} 
    \\
\includegraphics[width=0.50\textwidth]{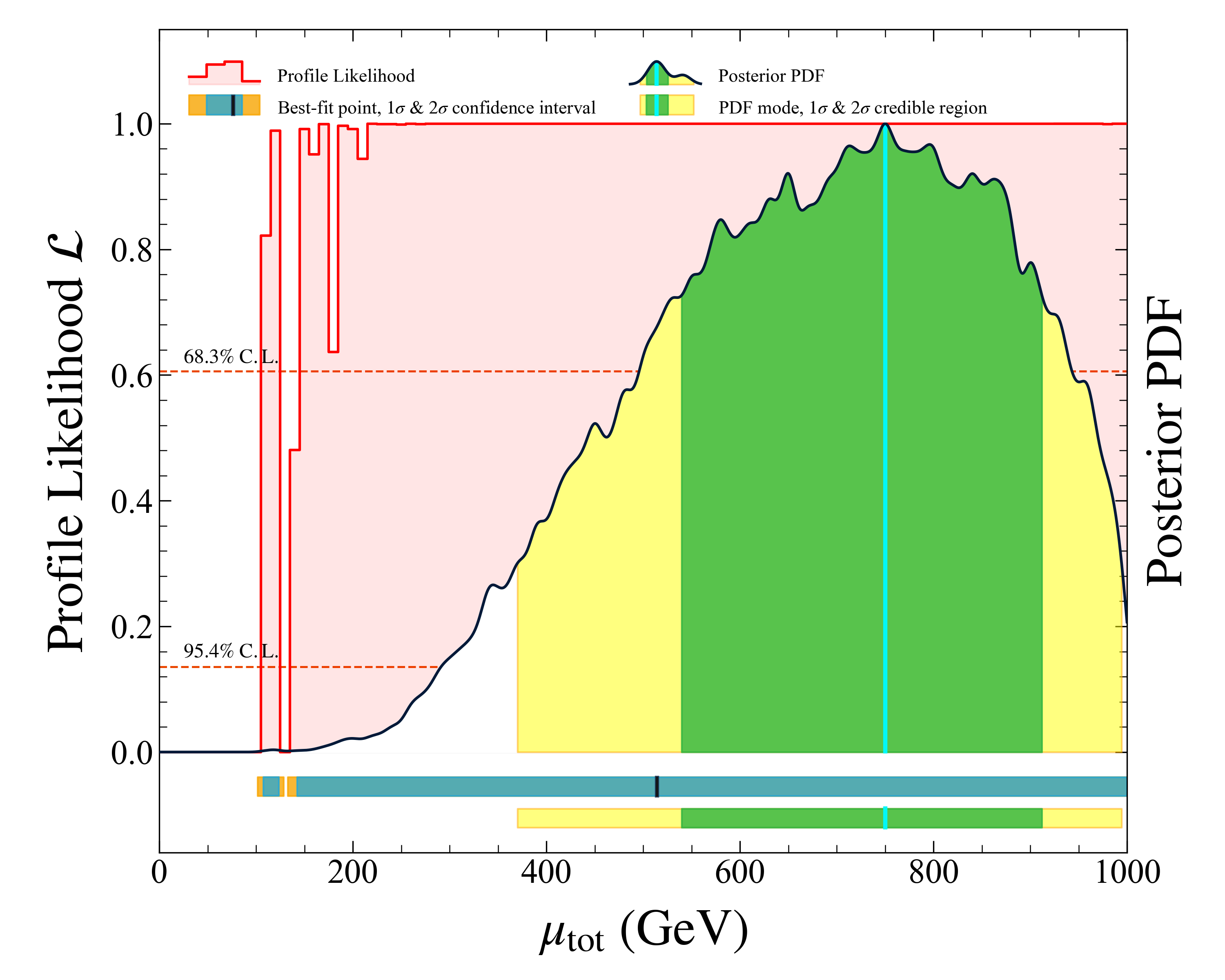}\hspace{-0.2cm}	\includegraphics[width=0.50\textwidth]{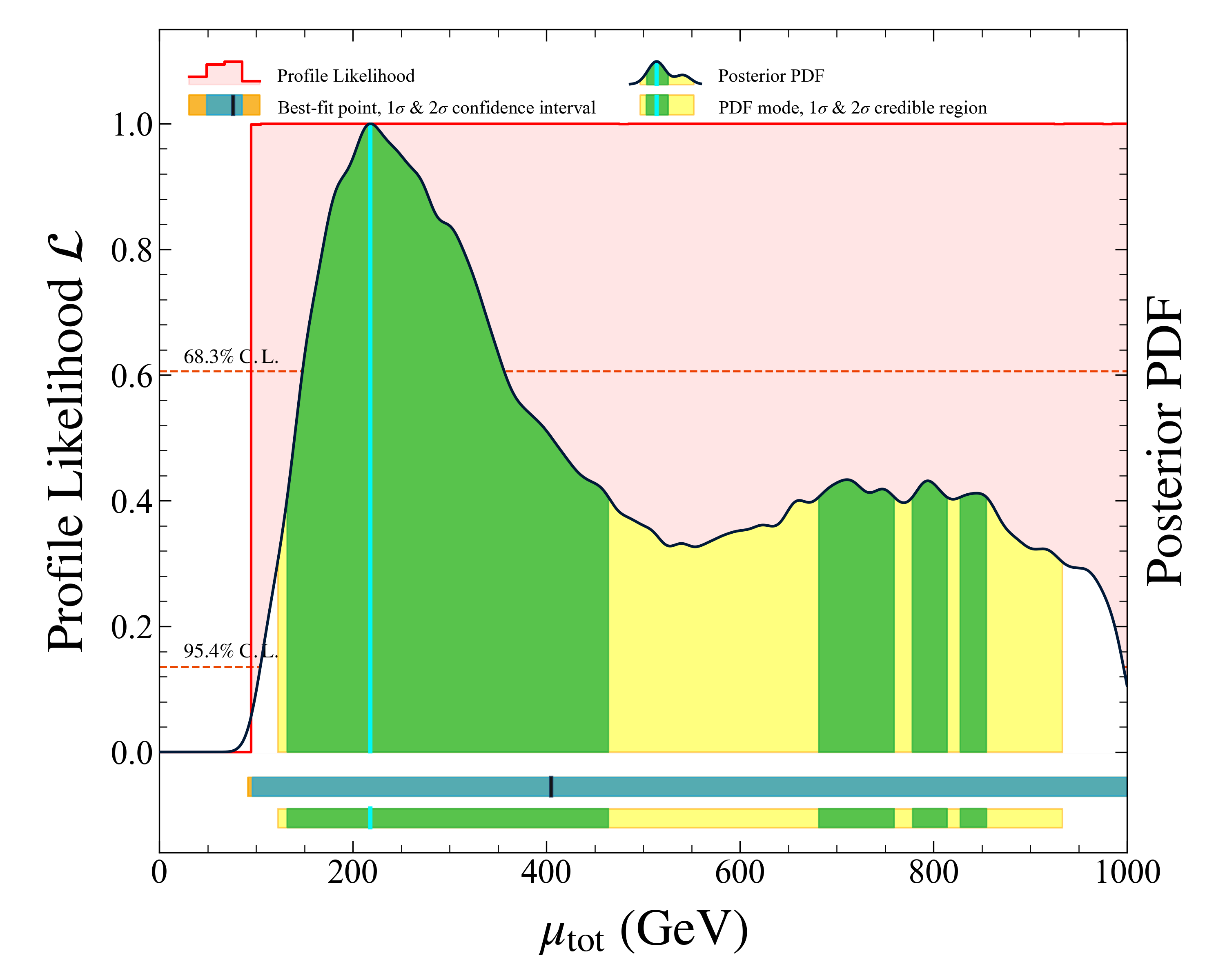}

\vspace{-0.5cm}

\caption{Same as Fig.~\ref{Fig1}, but for the distributions of $m_{\tilde{\chi}_1^0}$ and $\mu_{\rm tot}$.   \label{Fig2} }
\setlength{\abovecaptionskip}{-2.0cm}
\end{figure*}

By analyzing the samples acquired by the scans, we made the following observations:
\begin{itemize}
\item We obtained 21229 samples for the $h_1$ scenario and 20511 samples for the $h_2$ scenario. These sample sets accounted for the relic abundance at the $2 \sigma$ level, remained consistent with the LZ bounds, and satisfied all the experimental constraints encoded in $\mathcal{L}_{\rm Const}$. The Bayesian evidences were $\ln Z_{h_1} = -11.28 \pm 0.037$ and $\ln Z_{h_2} = -12.12 \pm 0.038$. The Jeffrey's scale, defined as the ratio of $\ln Z_{h_1}$ to $\ln Z_{h_2}$, was equal to 0.84; thus, the $h_1$ scenario was better suited to account for experimental results than the $h_2$ scenario~\cite{MR1647885}. The primary reason was that the $h_2$ scenario introduced additional non-trivial theoretical constraints, specifically $m_B \lesssim 130~{\rm GeV}$ and $\lambda \mu_{\rm tot} \lesssim 30~{\rm GeV}$, based on the Higgs data fit of this analysis, leading to a more stringent limitation on its parameter space.

\item The DM candidate may have been dominated by the Singlino or Bino field in its components. In the $h_1$ scenario, 21097 out of 21229 samples predicted a Singlino-dominated $\tilde{\chi}_1^0$, contributing to $99.3\%$ of the Bayesian evidence. In contrast, 13113 out of 20511 samples in the $h_2$ scenario predicted a Singlino-dominated $\tilde{\chi}_1^0$, accounting for approximately $64.8\%$ of the Bayesian evidence.

\item After including constraints from the LHC search for supersymmetry, the sample numbers were reduced to 20709 and 12604 for the scenarios $h_1$ and $h_2$, respectively. Correspondingly, the Bayesian evidences decreased by $2.4\%$ and $38.8\%$, respectively. These results indicated that the LHC constraint had significant impacts on the DM physics of the $h_2$ scenario, but hardly affected that of the $h_1$ scenario.

\item As discussed in Section~\ref{sec:DMRD}, the relative importance between the annihilation channels $\tilde{\chi}_1^0 \tilde{\chi}_1^0 \to h_s h_s, A_s A_s, h_s A_s$ for the Singlino-dominated DM depended on the mass spectrum and couplings of the samples. We have categorized the samples based on their dominant annihilation processes and evaluated their impact on the Bayesian evidence. Tables~\ref{tab:tab4} and~\ref{tab:tab5} detail  their contributions to the total Bayesian evidence for the $h_1$ and $h_2$ scenario, respectively, both before and after incorporating the LHC constraints. These tables reveal that most samples in the $h_1$ scenario predominantly annihilated through $\tilde{\chi}_1^0 \tilde{\chi}_1^0 \to h_s A_s$, aligning with the observed relic abundance. In contrast, the $h_2$ scenario showed that a substantial fraction of the samples primarily annihilated via $\tilde{\chi}_1^0 \tilde{\chi}_1^0 \to h_s h_s$, facing stringent restrictions from the LHC search for the supersymmetry.

\item In the $h_2$ scenario, a significant fraction of samples predicted a Bino-dominated $\tilde{\chi}_1^0$, accounting  for $35.2\%$ and $28.0\%$ of the total Bayesian evidence before and after taking the LHC constraints into consideration, respectively. Such a DM achieved the observed relic abundance primarily through co-annihilating with Wino-like electroweakinos. It preferred a negative value of $M_1$ to suppress the SI DM-nucleon scattering by canceling different contributions~\cite{Cao:2019qng}. The LZ experiment alone set a lower bound of approximately $380~{\rm GeV}$ on $\mu_{\rm tot}$~\cite{He:2023lgi}, and the LHC search for the compressed spectrum of supersymmetry by leptonic signals required $|m_{\tilde{\chi}_1^0}| \gtrsim 220~{\rm GeV}$~\cite{ATLAS:2021moa}. Our results confirmed these characteristics.

\end{itemize}

These aforementioned conclusions indicated that the GNMSSM preferred a Singlino-dominated DM, instead of a Bino-dominated DM, across a wide range of parameters. This feature distinguished it from the MSSM and $\mathbb{Z}_3$-NMSSM.

\begin{figure*}[t]
		\centering

\includegraphics[width=0.50\textwidth]{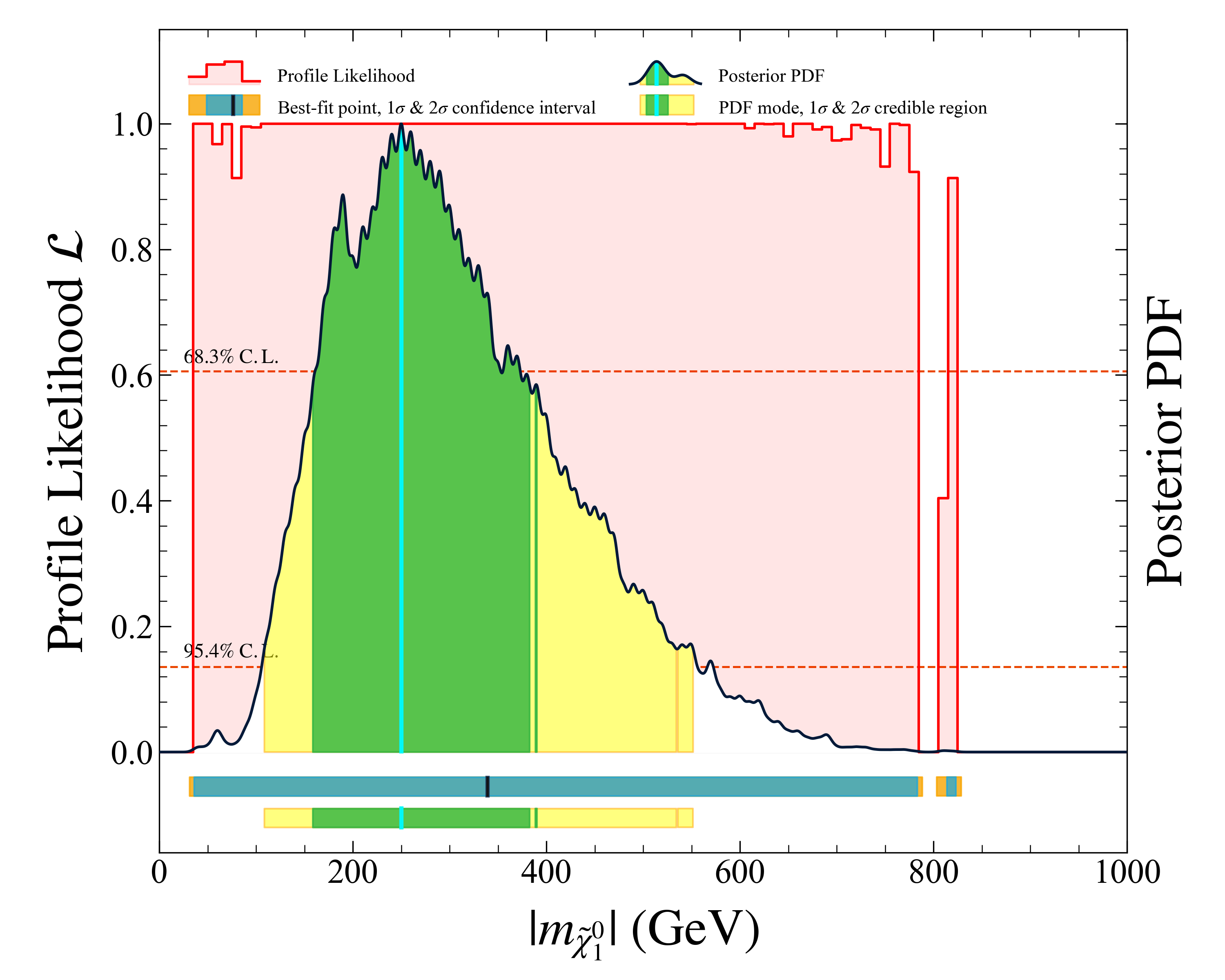}\hspace{-0.2cm}	\includegraphics[width=0.50 \textwidth]{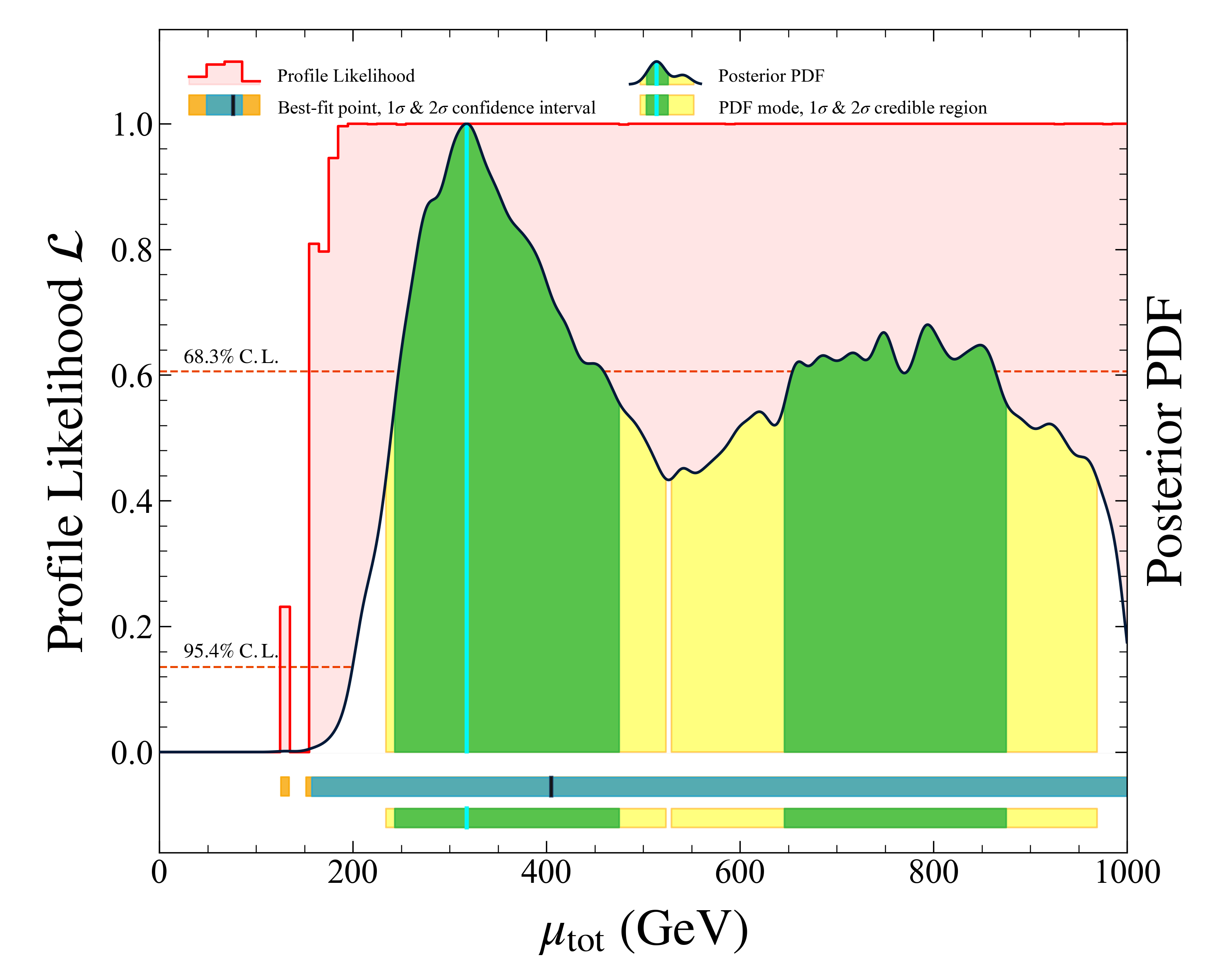} 

\vspace{-0.5cm}

\caption{Same as Fig.~\ref{Fig2}, but for the results of the $h_2$ scenario after considering the LHC constraints.  \label{Fig3} }
\setlength{\abovecaptionskip}{-2.0cm}
\end{figure*}

In the analysis discussed below, we studied the statistical distributions of the input parameters that determined the property of $\tilde{\chi}_1^0$. We plotted one-dimensional PLs and marginal posterior PDFs of $\lambda$, $\kappa$, and $\tan \beta$ in Fig.~\ref{Fig1} and those of $m_{\tilde{\chi}_1^0}$ and $\mu_{\rm tot}$ in Fig.~\ref{Fig2}. The left panels in both figures show the results of the $h_1$ scenario without considering the LHC constraints, while the right panels depict those of the $h_2$ scenario. From these plots, several insights can be inferred:
\begin{itemize}
\item $\lambda$'s PL and posterior PDF distributions indicated that $\lambda \lesssim 0.09$ for almost all samples in the $h_1$ scenario and $\lambda \lesssim 0.05$ for almost all samples in the $h_2$ scenario. These ranges are notably narrower than those listed in Table~\ref{tab:1}, and the small values of $\lambda$ are a distinctive feature of the GNMSSM. This trend primarily stemmed from the LZ constraint discussed in Section~\ref{sec:theory}: in scenarios where $\tilde{\chi}_1^0$ is Singlino dominated, the main contribution to $\sigma^{\rm SI}_{\tilde{\chi}_1^0-N}$ is proportional to $\lambda^2 \kappa^2$ for substantial singlet--doublet Higgs mixing and $\lambda^4$ in the heavy $h_s$ limit. In contrast, when $\tilde{\chi}_1^0$ is Bino dominated, as significant in the $h_2$ scenario, the LZ constraint required $\mu_{\rm tot}$ to exceed $380~{\rm GeV}$, and the Higgs data fit limited $\lambda \mu_{\rm tot}$ in ${\cal{M}}_{\rm S,23}^2$ to approximately $30~{\rm GeV}$ or less. In any scenario, a small $\lambda$ is preferred.

\item The value of $\kappa$ varied widely from $-0.75$ to $0.75$. Notably, in the $h_1$ scenario, there were gaps around $\kappa = 0$ in the PL and posterior PDF distributions, while they were continuous in the $h_2$ scenario. This distinction arose because the Singlino component predominantly contributed to the DM candidate for nearly all samples in the $h_1$ scenario, necessitating a non-zero $\kappa$ to account for the observed relic abundance. In contrast, in the $h_2$ scenario, a significant portion of samples featured Bino-dominated DM, whose properties were less influenced by $\kappa$.

\item For both the $h_1$ and $h_2$ scenarios, $\tan \beta$ spanned a broad range from 5 to 60. The distributions did not show a significant preference for specific regions, suggesting that the DM physics are not sensitive to $\tan \beta$.

\item The PL distributions suggest that $\mu_{\rm tot}$ can range from $100~{\rm GeV}$ to $1~{\rm TeV}$ in both scenarios. However, $\mu_{\rm tot}$'s posterior PDF distributions indicated a preference for smaller $\mu_{\rm tot}$ in the $h_2$ scenario. This preference was driven by the constraints from the Higgs data fit, which limited $\lambda \mu_{\rm tot}$ in ${\cal{M}}_{\rm S,23}^2$ to be no more than approximately $30~{\rm GeV}$, a restriction not applicable to the $h_1$ scenario.

\item Similarly, a relatively light DM was favored in the $h_2$ scenario, stemming from the condition $|m_{\tilde{\chi}_1^0}| < \mu_{\rm tot}$ and the observed preference for a smaller $\mu_{\rm tot}$ in this scenario. Additionally, a light DM is self-consistent with the annihilation $\tilde{\chi}_1^0 \tilde{\chi}_1^0 \to h_s h_s$, where $h_s$ could be as light as several GeV in the $h_2$ scenario.

\end{itemize}

We also investigated the impact of the LHC search for supersymmetry on the distributions. In the $h_1$ scenario, we observed that it only shifted the lower bounds of $m_{\tilde{\chi}_1^0}$ and $\mu_{\rm tot}$ in the PL distribution from $34$ and $105~{\rm GeV}$ to approximately $48$ and $165~{\rm GeV}$, respectively. In contrast, in the $h_2$ scenario, apart from altering the lower bounds on $m_{\tilde{\chi}_1^0}$ and $\mu_{\rm tot}$ from $14$ and $100~{\rm GeV}$ to approximately $40$ and $155~{\rm GeV}$, respectively, the LHC constraint significantly modified their posterior PDFs by shifting the peaks of their distribution toward higher mass values. The updated results are shown in Fig.~\ref{Fig3} compared to their corresponding ones in Fig.~\ref{Fig2}. This phenomenon arose due to the crucial nature of the LHC constraints when the DM and Higgsinos had moderately low masses. This effect was less pronounced on other distributions of the $h_2$ scenario.

\begin{figure}[t]
	\centering	\includegraphics[width=0.45\textwidth]{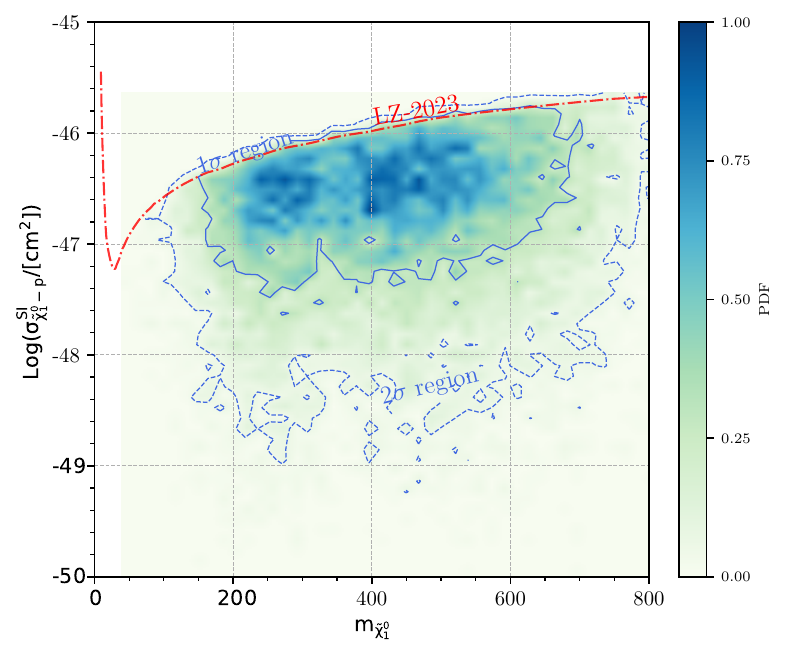}\hspace{-0.3cm}	\includegraphics[width=0.45\textwidth]{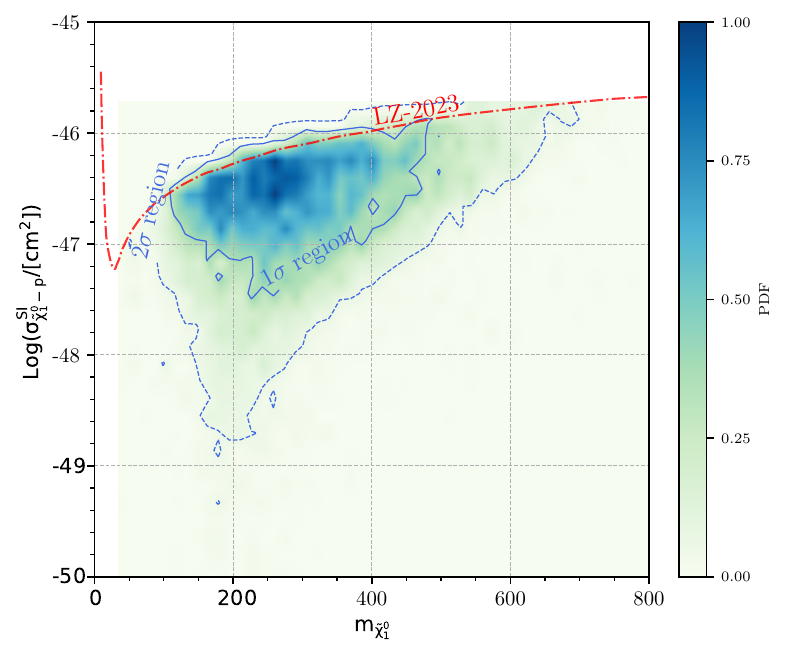} 
    \\
\includegraphics[width=0.45\textwidth]{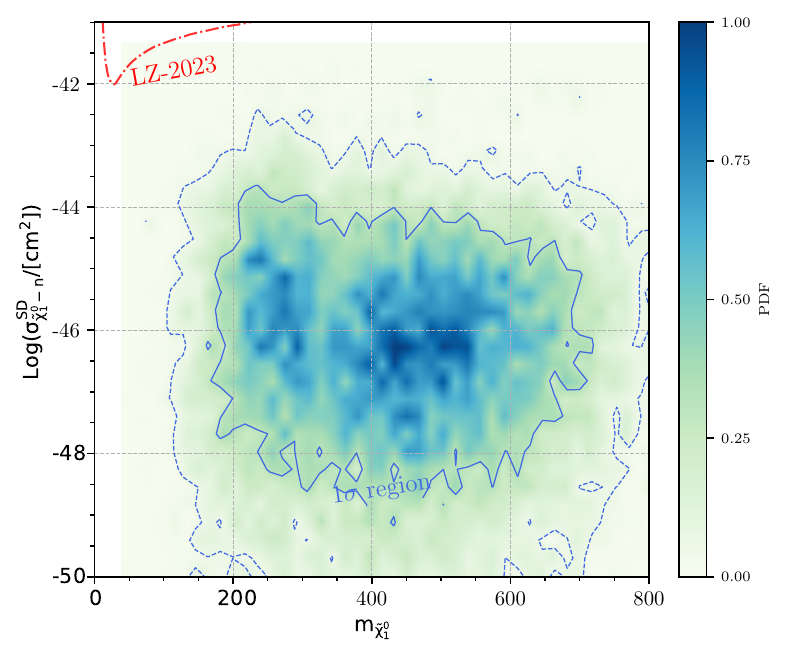}\hspace{-0.3cm}	\includegraphics[width=0.45\textwidth]{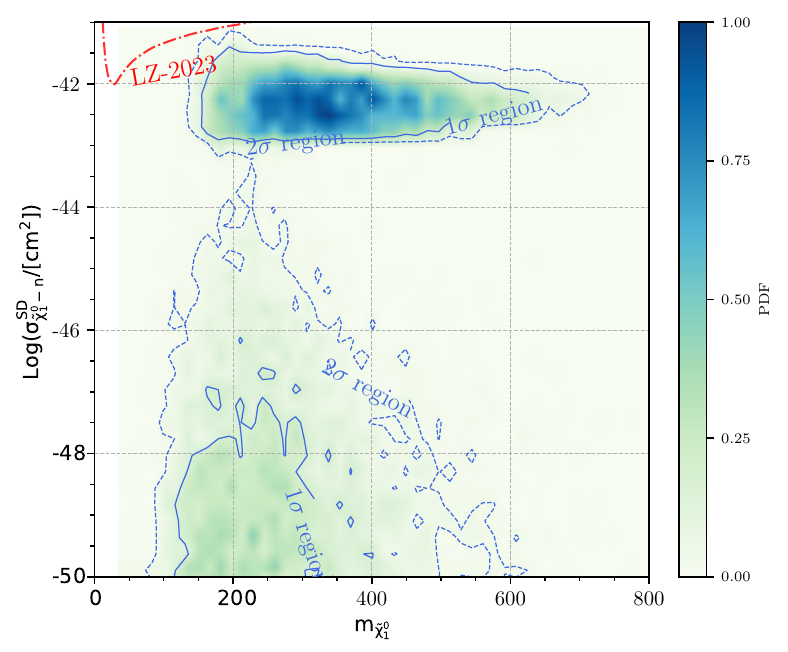}
\\
\vspace{-0.5cm}
	\caption{\label{fig4}
		Two-dimensional distributions of the marginal posterior PDF for the likelihood $\mathcal{L}$ in Eq.~\ref{Likelihood}, projected onto the $|m_{\tilde{\chi}_1^0} | -\sigma^{\rm SI}_{\tilde{\chi}_1^0-p}$ and  $|m_{\tilde{\chi}_1^0}| -\sigma^{\rm SD}_{\tilde{\chi}_1^0-n}$ planes. The left panels show the results of the $h_1$ scenario, while the right panels show those of the $h_2$ scenario.
The solid and dashed lines surround the $1 \sigma$ and $2 \sigma$ credible regions, respectively, and the samples falling within these regions contributed $65.3\%$ and $95.4\%$ of the total Bayesian evidence~\cite{Fowlie:2016hew}. All the considered samples were consistent with the LHC constraint. }
\end{figure}

\begin{figure*}[t]
		\centering
\includegraphics[width=0.50\textwidth]{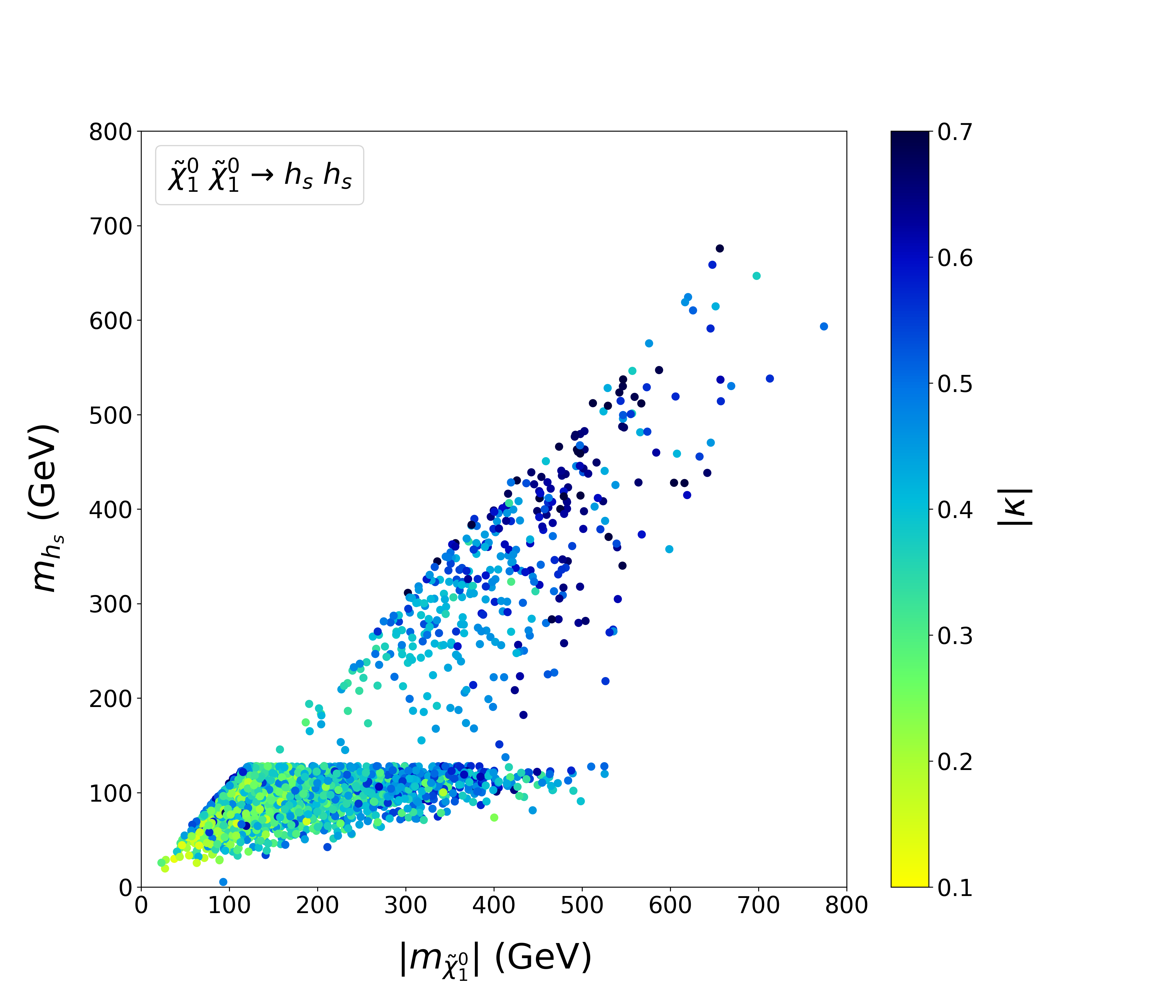}\hspace{-0.5cm}	\includegraphics[width=0.50\textwidth]{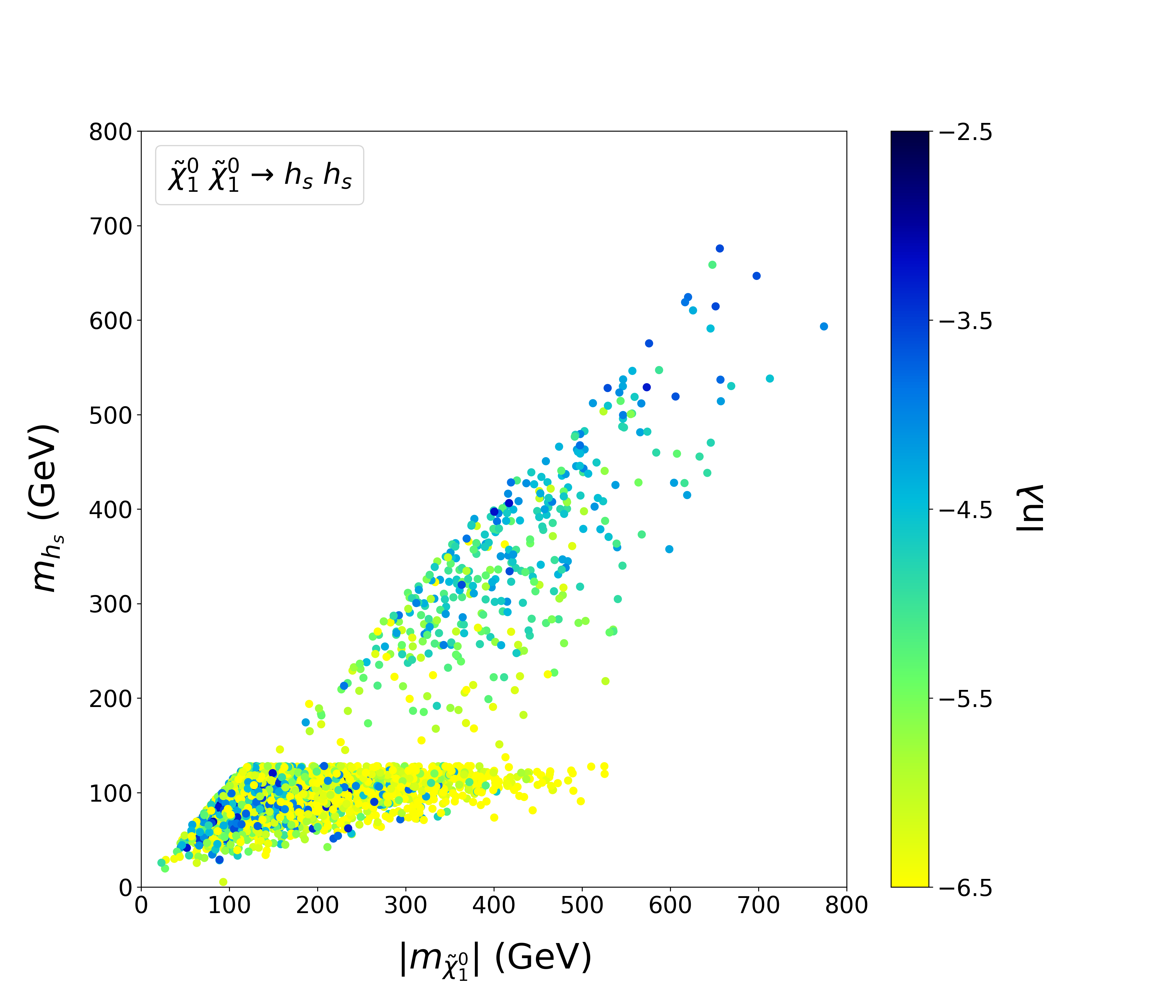}

\vspace{-0.9cm}

\includegraphics[width=0.50\textwidth]{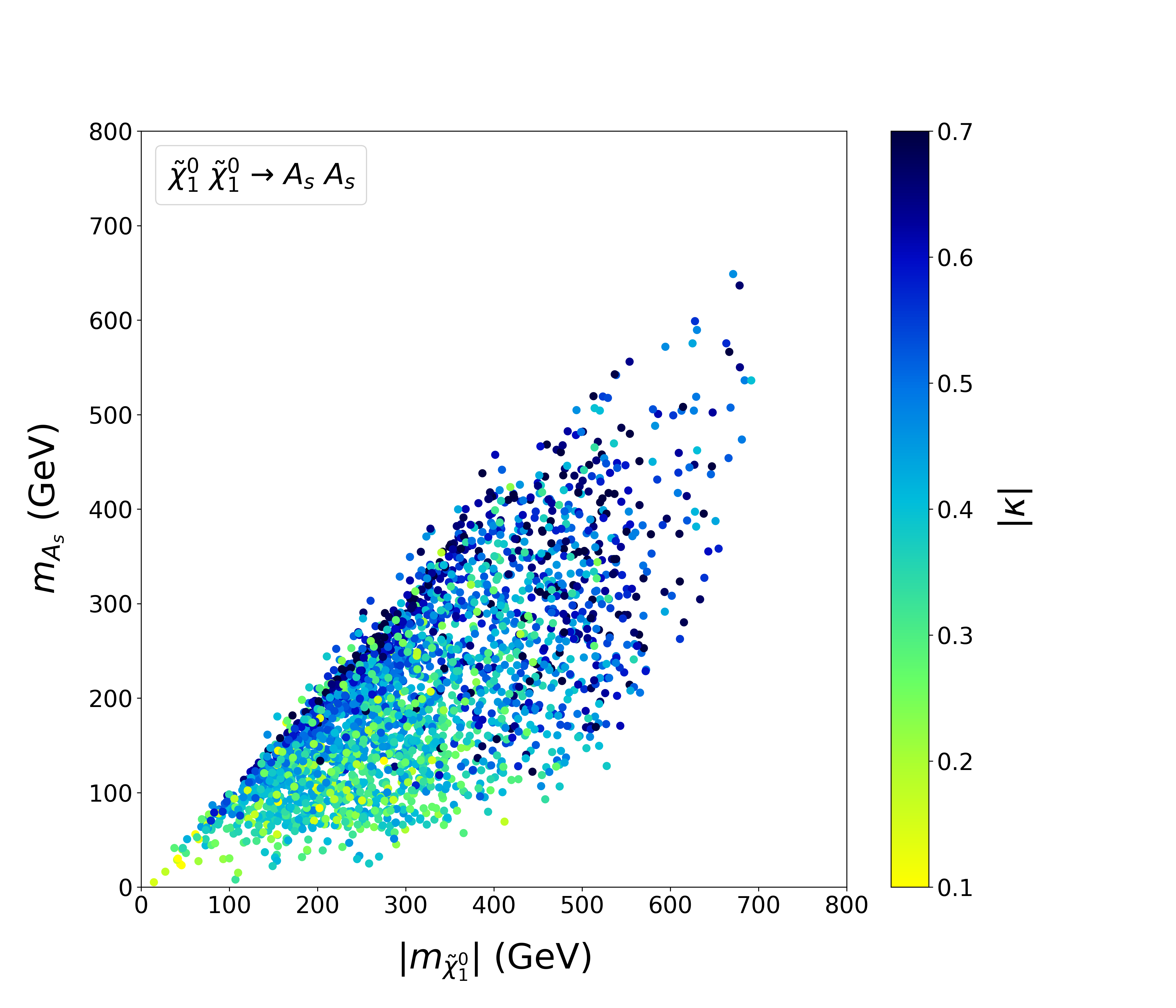}\hspace{-0.5cm}	\includegraphics[width=0.50\textwidth]{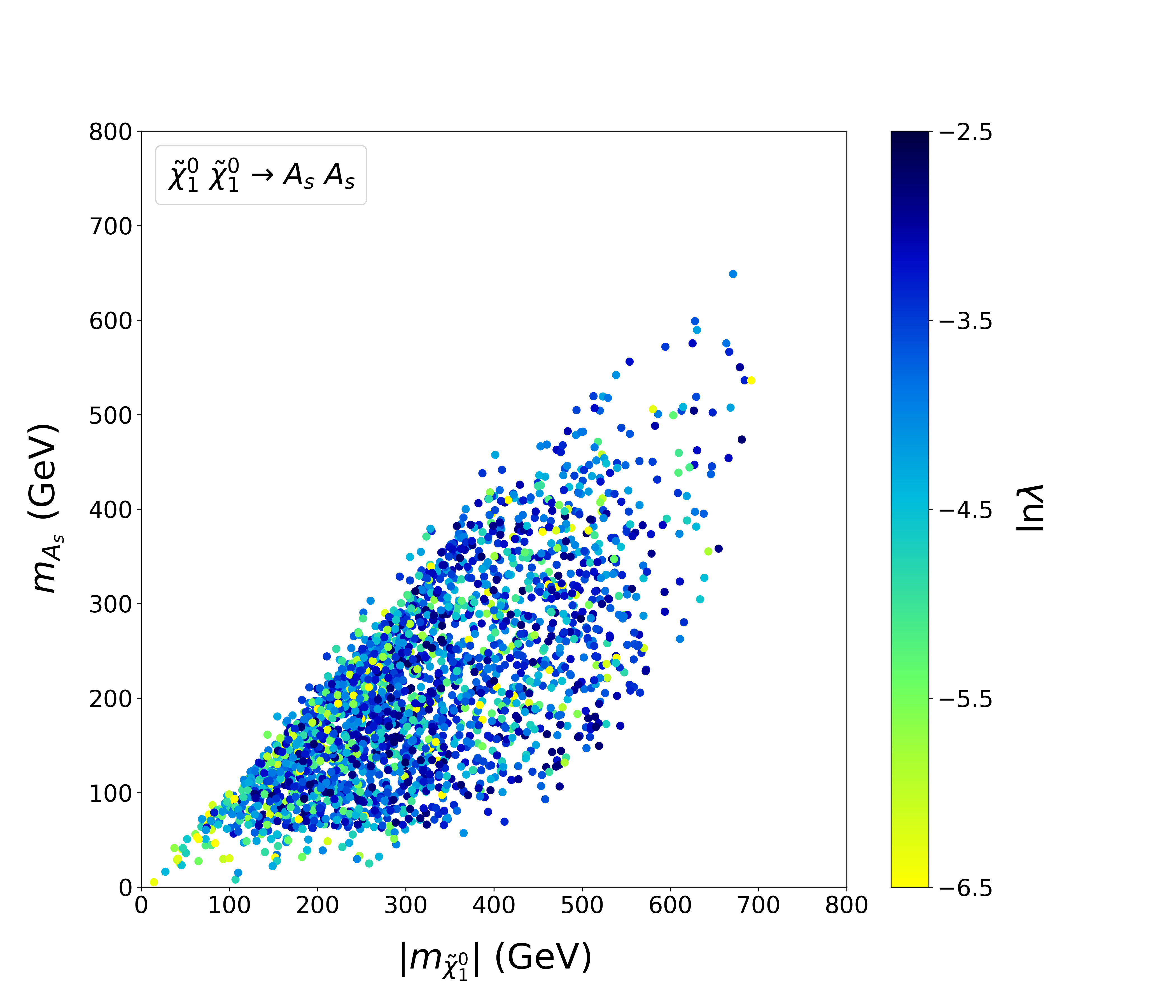}

\vspace{-0.9cm}

\includegraphics[width=0.50\textwidth]{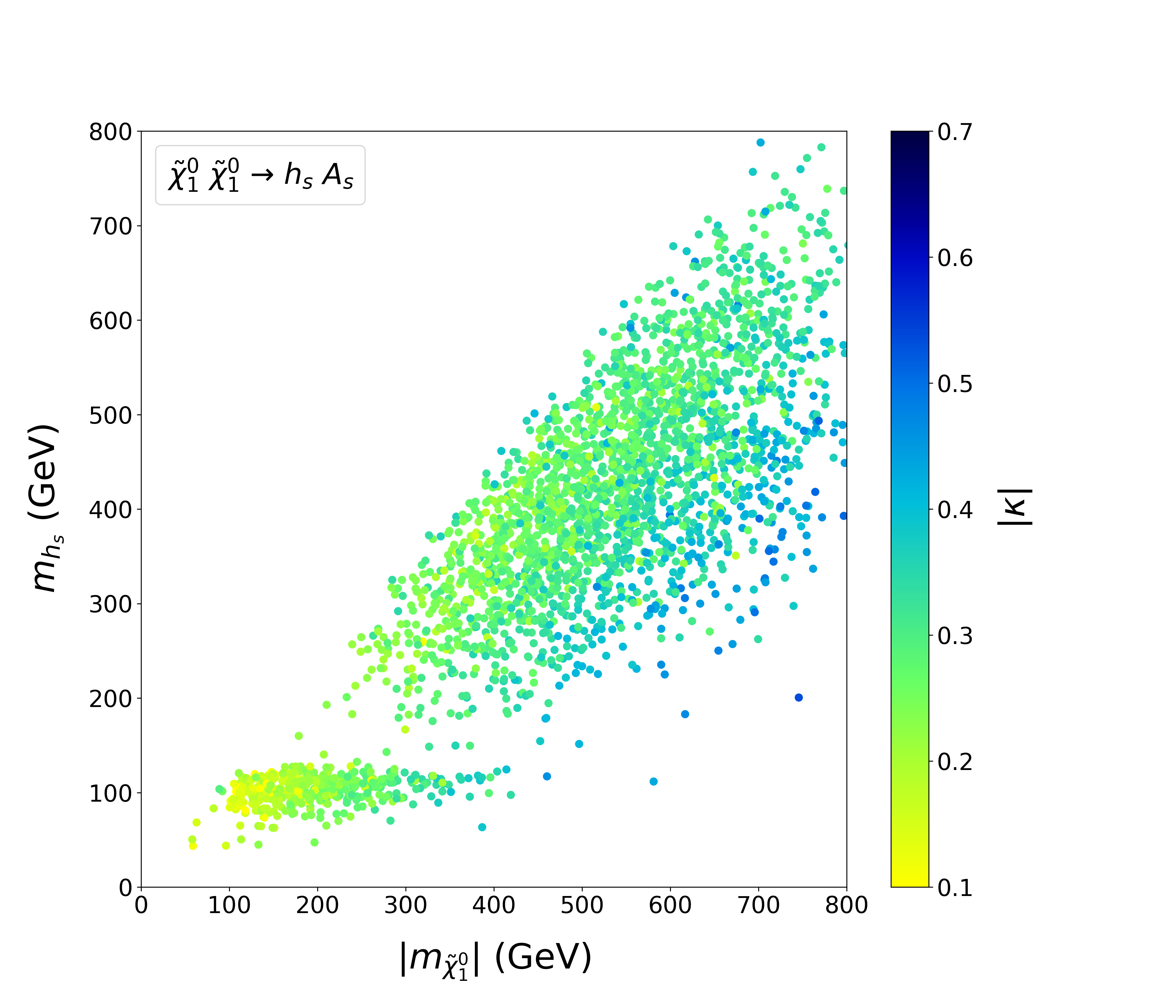}\hspace{-0.5cm}	\includegraphics[width=0.50\textwidth]{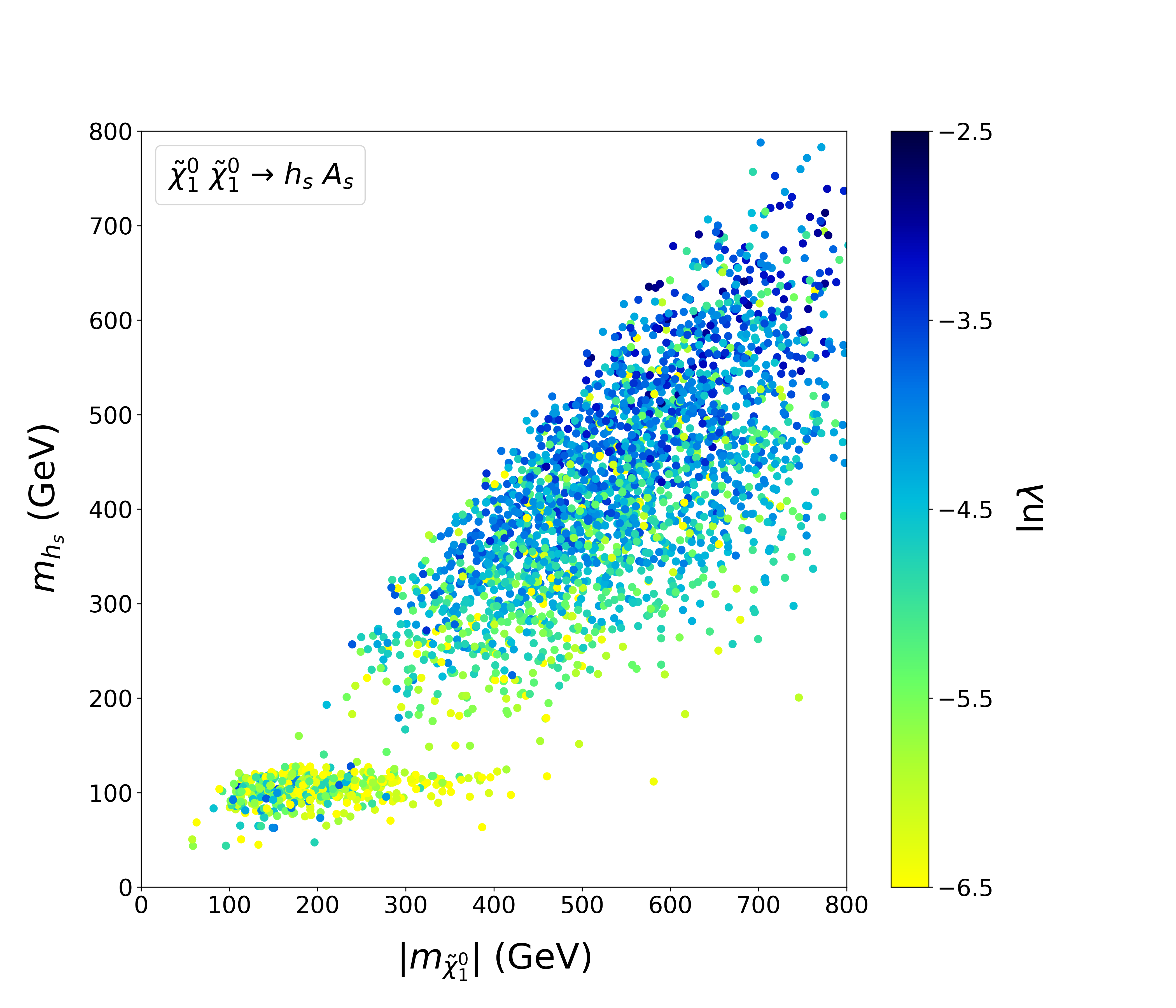}
\\

\vspace{-0.5cm}

\caption{Scattering plots of the samples that primarily annihilated through $\tilde{\chi}_1^0 \tilde{\chi}_1^0 \to h_s h_s, A_s A_s, h_s A_s$ channels to achieve the observed relic abundance, projected onto the $|m_{\tilde{\chi}_1^0}|-m_{h_s}$, $|m_{\tilde{\chi}_1^0}|-m_{A_s}$, and $|m_{\tilde{\chi}_1^0}|-m_{h_s}$ planes, respectively. The color bars on the left panels denote the values of $|\kappa|$, while those on the right panels represent the values of $\ln \lambda$. \label{Fig5}}
\end{figure*}

\begin{figure*}[t]
		\centering
\includegraphics[width=0.48\textwidth]{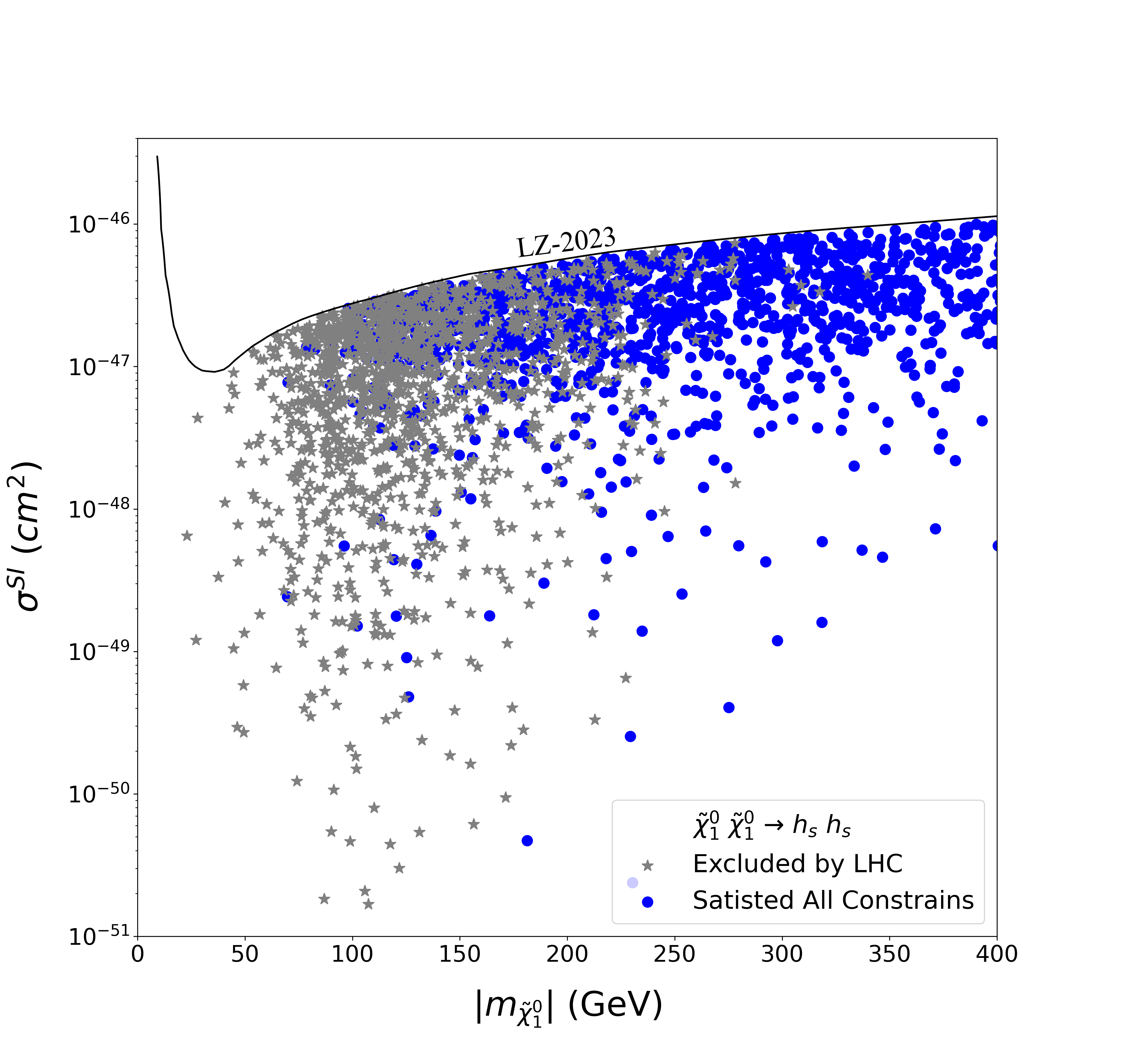}\hspace{-0.5cm}	\includegraphics[width=0.48\textwidth]{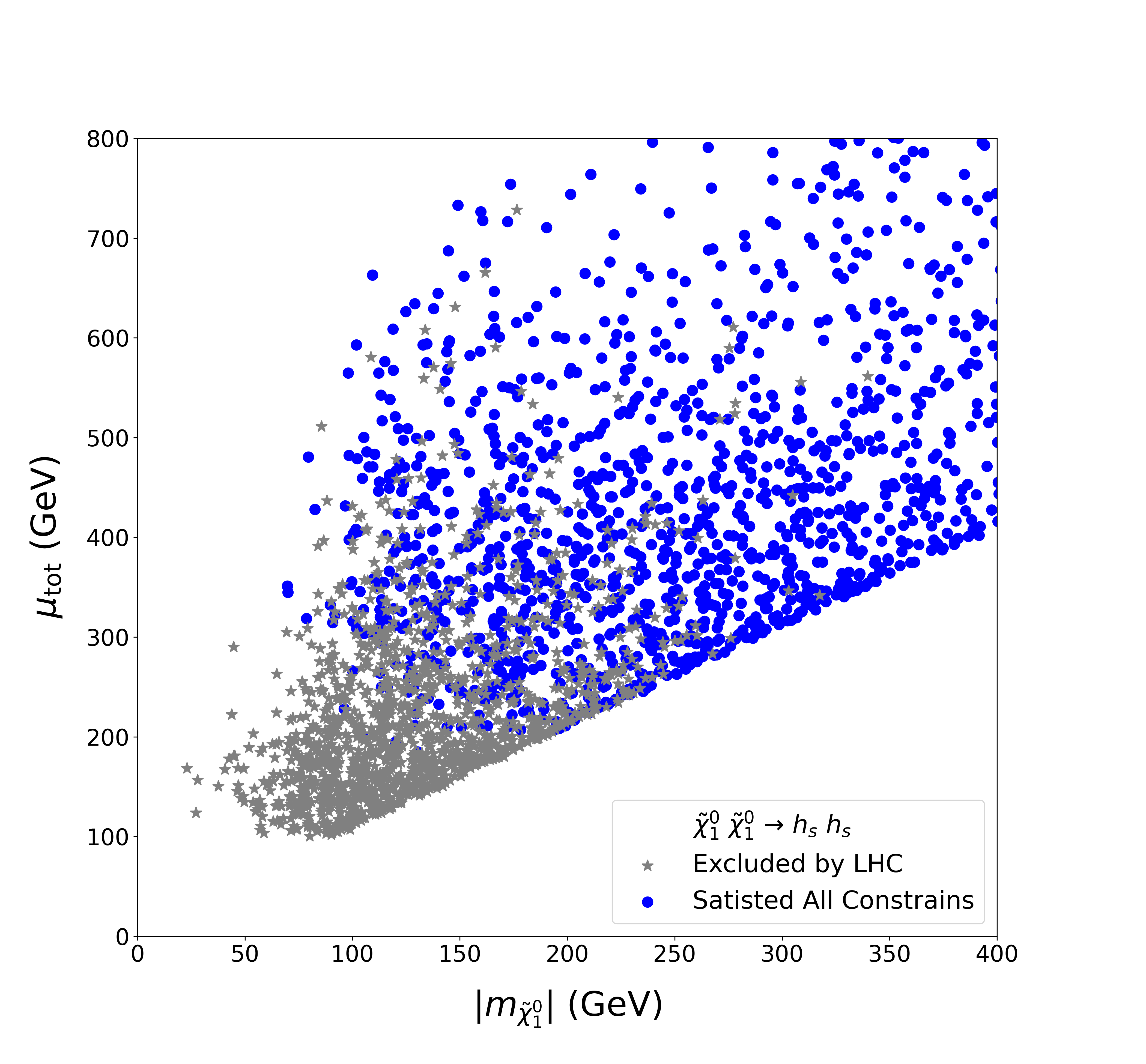} 

\vspace{-0.9cm}

\includegraphics[width=0.48\textwidth]{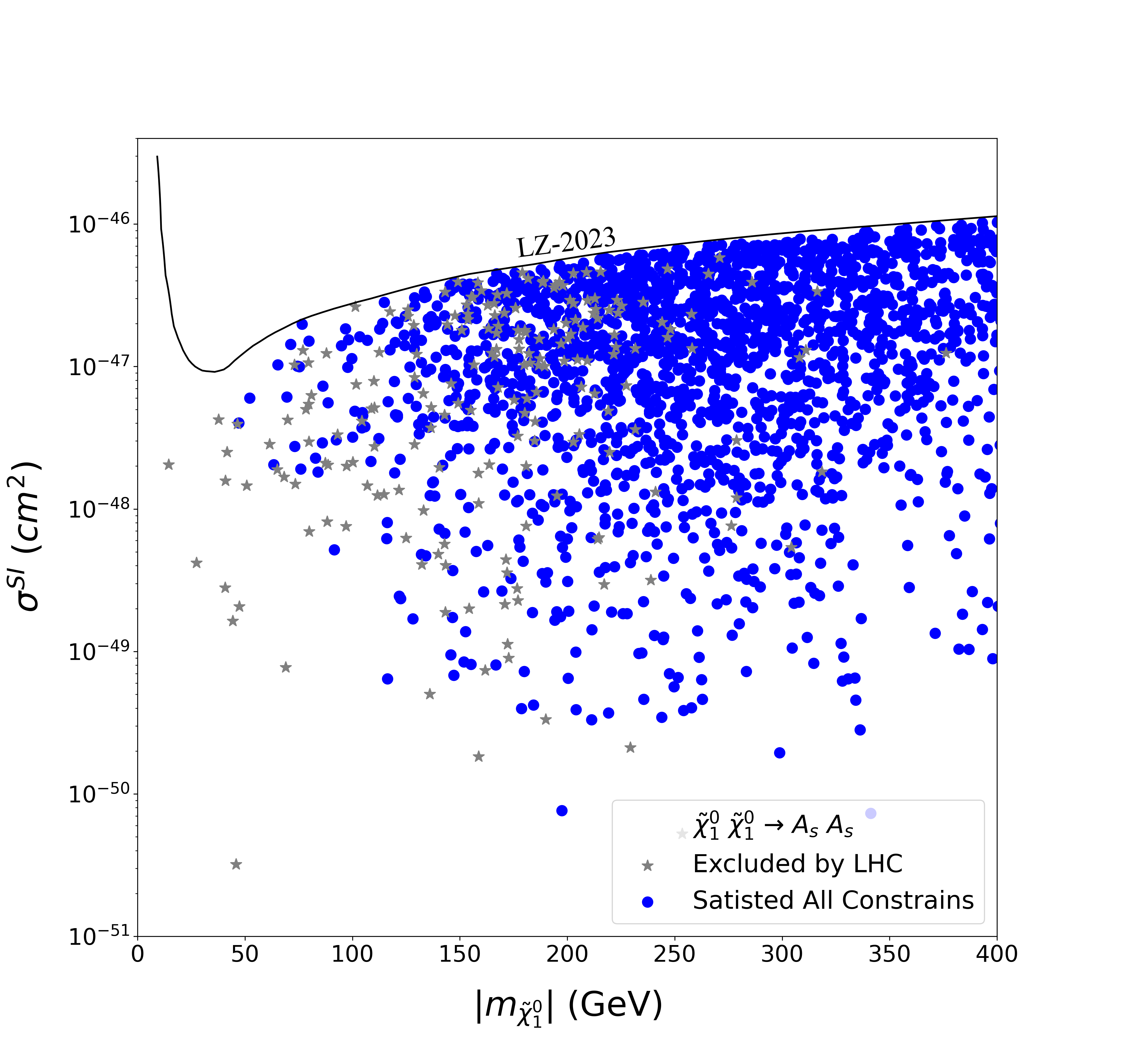}\hspace{-0.5cm}	\includegraphics[width=0.48\textwidth]{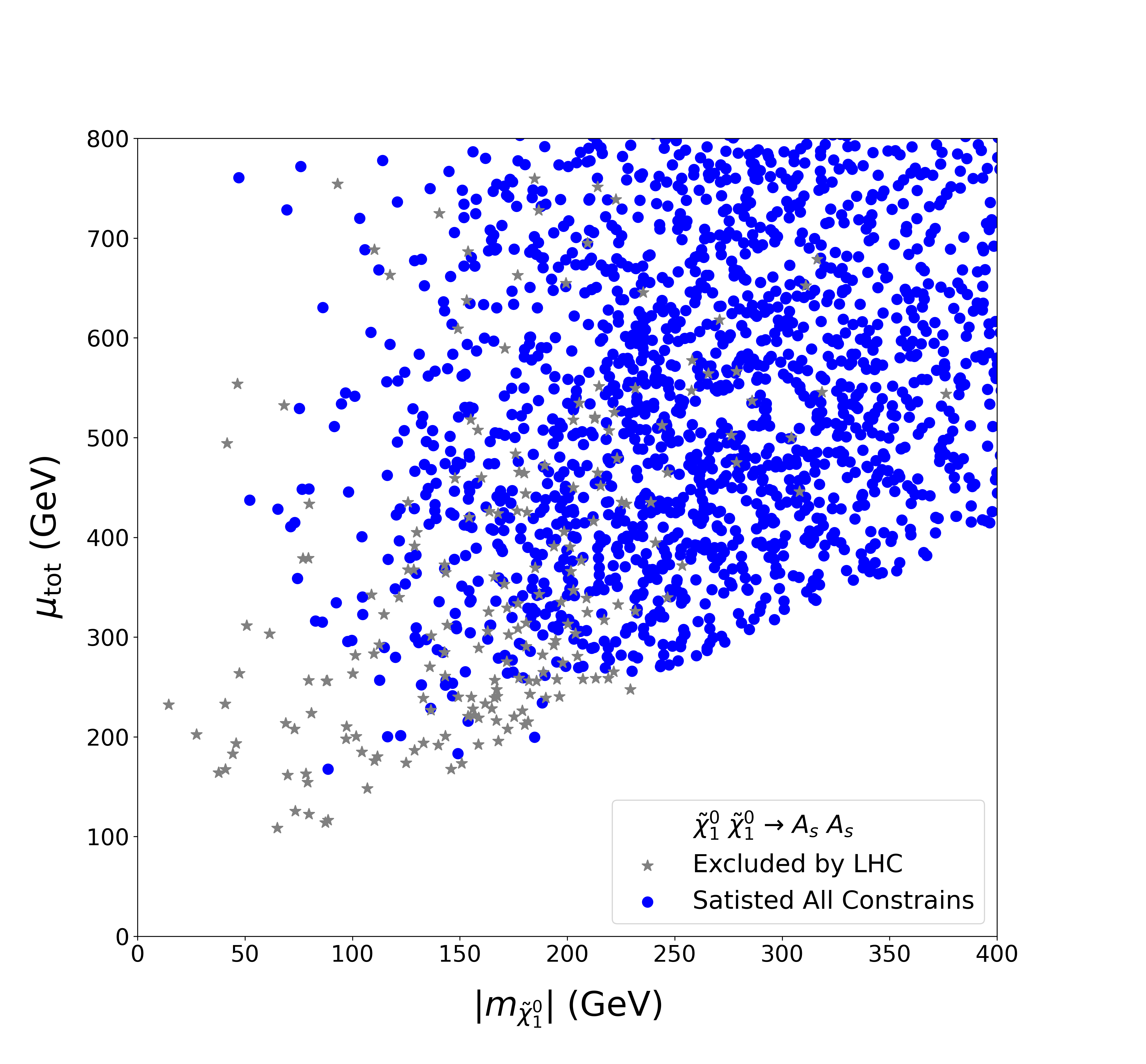}

\vspace{-0.9cm}

\includegraphics[width=0.48\textwidth]{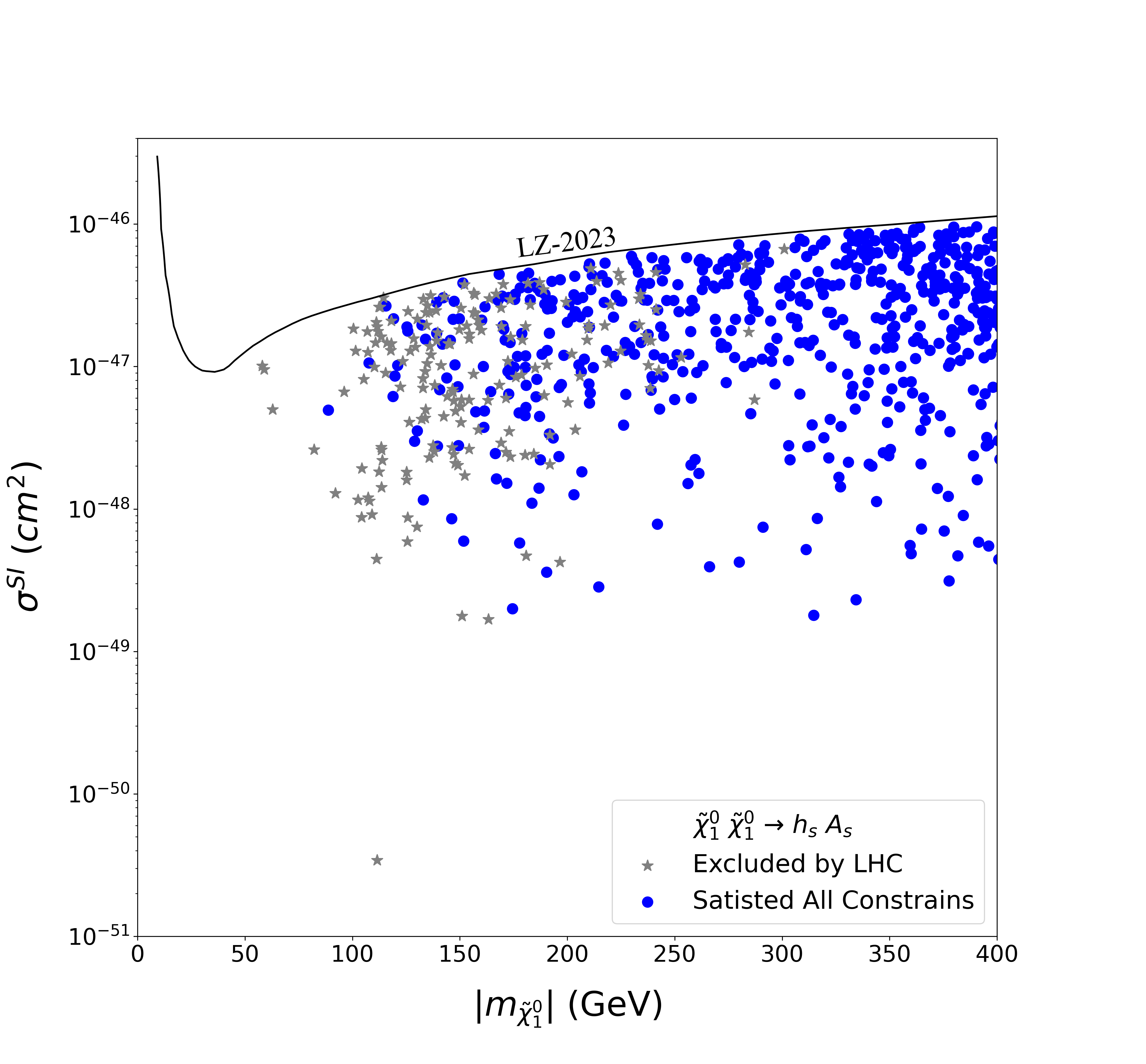}\hspace{-0.5cm}	\includegraphics[width=0.48\textwidth]{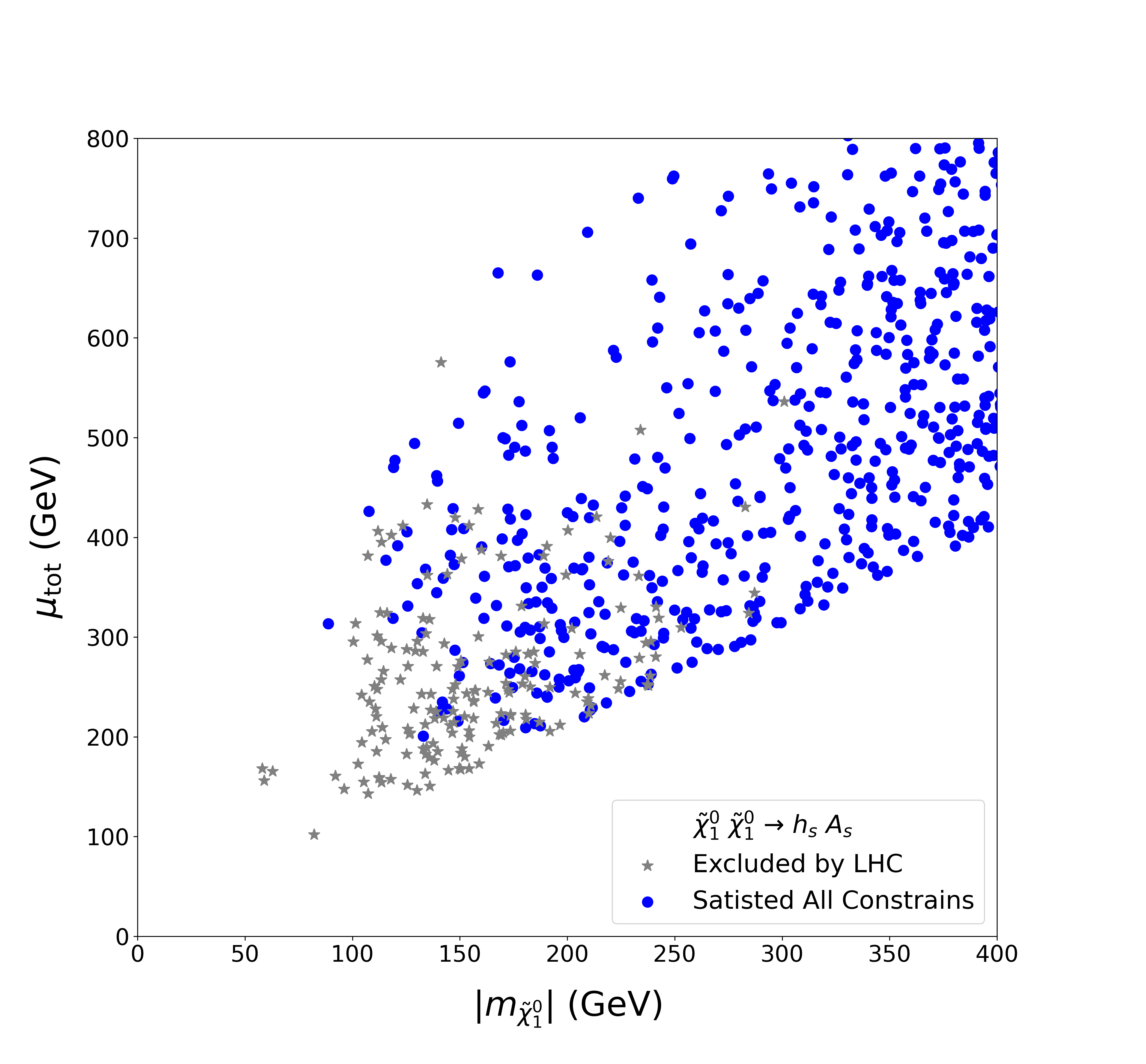}

\vspace{-0.5cm}

\caption{Similar to Fig.~\ref{Fig5}, except that the results are shown on the $|m_{\tilde{\chi}_1^0}|-\sigma^{\rm SI}_{\tilde{\chi}_1^0-p}$ (left panels) and $|m_{\tilde{\chi}_1^0}|-\mu_{\rm tot}$ (right panels) planes. Samples marked by gray were excluded by the LHC constraint, while those in blue were consistent with all the experimental constraints. \label{Fig6}}
\end{figure*}

Finally, we projected the samples passing the LHC constraint onto the $|m_{\tilde{\chi}_1^0}|-\sigma^{\rm SI}_{\tilde{\chi}_1^0-p}$ and $|m_{\tilde{\chi}_1^0}|-\sigma^{\rm SD}_{\tilde{\chi}_1^0-n}$ planes to obtain two-dimensional posterior PDF maps, which are shown in Fig.~\ref{fig4}. This figure reveals that the samples in the $2 \sigma$ credible region predicted  $1.0 \times 10^{-49} \lesssim \sigma^{\rm SI}_{\tilde{\chi}_1^0-p}/{\rm cm^2} \lesssim 2.5 \times 10^{-46} $ and  $1.0 \times 10^{-50} \lesssim \sigma^{\rm SD}_{\tilde{\chi}_1^0-n}/{\rm cm^2} \lesssim 3.2 \times 10^{-43} $ in the $h_1$ scenario.
In contrast, analyzing the $h_2$ scenario yielded  $ 5.0 \times 10^{-49} \lesssim \sigma^{\rm SI}_{\tilde{\chi}_1^0-p}/{\rm cm^2} \lesssim 2.0 \times 10^{-46}$ and $ 1.0 \times 10^{-52} \lesssim \sigma^{\rm SD}_{\tilde{\chi}_1^0-n}/{\rm cm^2} \lesssim 1.0 \times 10^{-43}$ for Singlino-like DM and  $ 3.0 \times 10^{-49} \lesssim \sigma^{\rm SI}_{\tilde{\chi}_1^0-p}/{\rm cm^2} \lesssim 2.0 \times 10^{-46}$, $ 1.0 \times 10^{-43} \lesssim \sigma^{\rm SD}_{\tilde{\chi}_1^0-n}/{\rm cm^2} \lesssim 5.0 \times 10^{-42}$ for Bino-like DM. These results showed significant difference of $\sigma^{\rm SD}_{\tilde{\chi}_1^0-n}$ for Singlino- and Bino-like DM, which was explained in Eqs.~(\ref{SDCS}) and (\ref{SDCS-Bino}). Additionally, they revealed that with the future increase in experimental sensitivities to $\sigma^{\rm SI}_{\tilde{\chi}_1^0-p}$, a significant portion of the samples would become disfavored. Numerically speaking, we observed that the Bayesian evidences decreased by $46\%$ and $62\%$ for the $h_1$ and $h_2$ scenarios, respectively, if the sensitivities were improved by a factor of 5. Consequently, the upper bounds of $\lambda$ were reduced to approximately $0.05$ and $0.02$, respectively.

\subsection{Three annihilation channels}

As introduced in Section~\ref{DM-physics}, the relic abundance and nucleon scattering cross sections of the Singlino-dominated DM depend not only on parameters within the neutralino sector, including $\lambda$, $\kappa$, $\tan \beta$, $\mu_{\rm tot}$, and $m_{\tilde{\chi}_1^0}$, but also on those within the Higgs sector, such as $v_s$, $A_\lambda$, $m_B$ (or equivalently $m_{h_s}$), and $m_C$ ($m_{A_s}$). Consequently, the characteristics of the DM are complex. Nevertheless, valuable insights can still be obtained from scattering plots derived from our sample sets. We categorized the samples based on their predominant annihilation processes for this study's purposes. We particularly focus on samples where the primary annihilation mechanisms were $\tilde{\chi}_1^0 \tilde{\chi}_1^0 \to h_s h_s$, $\tilde{\chi}_1^0 \tilde{\chi}_1^0 \to A_s A_s$, and $\tilde{\chi}_1^0 \tilde{\chi}_1^0 \to h_s A_s$, and label them as Type-I, Type-II, and Type-III samples, respectively. Their respective counts were approximately 8000, 8000, and 16000 samples obtained by the scans, which created a challenge in plot representations and complicated our analyses due to their sheer volume. To address this issue, we applied a method introduced in Ref.~\cite{He:2023lgi} to reduce the sample density for the first two types of samples, while for the latter type, we controlled its sample size by implementing a mass constraint of $0.8 m_{h_s} \lesssim m_{A_s} \lesssim 1.2 m_{h_s}$ to simplify our analysis. By comparing plots derived from the full sample sets with those from the reduced sample sets, we observed that the latter efficiently captured essential features of the annihilations without losing the generality of our findings.

\begin{table}[t]
\centering
\resizebox{1\textwidth}{!}
{
\begin{tabular}{lrlr|lrlr}
\hline \hline
\multicolumn{4}{c|}{\bf Point P1}     & \multicolumn{4}{c}{\bf Point P2}   \\
\hline
$\lambda$          & 0.012        & $m_{h_s}$                 & 269.8~GeV                           &            $\lambda$          & 0.003        & $m_{h_s}$                 & 119.3~GeV     \\
$\kappa$           & -0.40         & $m_{h}$                   & 125.6~GeV                             &            $\kappa$           & 0.56         & $m_{h}$                   & 124.9~GeV    \\
$\tan{\beta}$      & 16.58       & $m_{H}$                   & 1826 ~GeV                        &              $\tan{\beta}$      & 11.75       & $m_{H}$                   & 1871 ~GeV      \\
$v_s$       	    & 692.3 ~GeV       & $m_{A_s}$                 & 737.5~GeV                            &            $v_s$       	      & 609.2 ~GeV       & $m_{A_s}$                 & 735.6~GeV    \\
$\mu_{\rm tot}$    & 390.4 ~GeV    & $m_{A_H}$	              & 1826~GeV                           &             $\mu_{\rm tot}$    & 405.2 ~GeV    & $m_{A_H}$	                & 1871~GeV     \\
$M_1$	             & -747.5 ~GeV       & $m_{H^\pm}$	              & 1827~GeV                          &             $M_1$	            & -718.8 ~GeV       & $m_{H^\pm}$	             & 1872~GeV   \\
$M_2$ 	          & 822.5 ~GeV       & $m_{\tilde{\chi}_1^0}$    & 324.2~GeV                        &              $M_2$ 	            & 614.6 ~GeV       & $m_{\tilde{\chi}_1^0}$    & 339~GeV   \\
$A_t$              & -3983 ~GeV       & $m_{\tilde{\chi}_2^0}$    & 393.2~GeV                        &              $A_t$              & -4233 ~GeV       & $m_{\tilde{\chi}_2^0}$    & 400.5~GeV    \\
$A_\lambda$        & 1654 ~GeV  & $m_{\tilde{\chi}_3^0}$	  & -400~GeV                         &             $A_\lambda$        & 1562 ~GeV  & $m_{\tilde{\chi}_3^0}$	 & -414.6~GeV   \\
$m_A$              & 2000 ~GeV       & $m_{\tilde{\chi}_4^0}$ 	  & -755.3~GeV                      &                $m_A$              & 2000 ~GeV       & $m_{\tilde{\chi}_4^0}$ 	 & 660.3~GeV   \\
$m_B$              & 280.8 ~GeV       & $m_{\tilde{\chi}_5^0}$ 	  & 865.1~GeV                      &                $m_B$              & 109.5 ~GeV       & $m_{\tilde{\chi}_5^0}$ 	 & -726.7~GeV   \\
$m_N$              & 326.3 ~GeV       & $m_{\tilde{\chi}_1^\pm}$  & 395~GeV                        &              $m_N$              & 345.1 ~GeV       & $m_{\tilde{\chi}_1^\pm}$  & 402.5~GeV    \\
$\Omega h^2$       & 0.1199         & $m_{\tilde{\chi}_2^\pm}$  & 865.3~GeV                        &              $\Omega h^2$       & 0.12         & $m_{\tilde{\chi}_2^\pm}$  & 660.6~GeV       \\
$\sigma^{\rm SD}_{\tilde{\chi}_1^0-n}$  & 1.73$\times 10^{-46}{\rm ~cm^2}$  &$\sigma^{\rm SI}_{\tilde{\chi}_1^0-p}$  & 2.24$\times 10^{-47}{\rm ~cm^2}$ &
$\sigma^{\rm SD}_{\tilde{\chi}_1^0-n}$  & 7.42$\times 10^{-49}{\rm ~cm^2}$  &$\sigma^{\rm SI}_{\tilde{\chi}_1^0-p}$  & 7.19$\times 10^{-47}{\rm ~cm^2}$ \\
\hline
\multicolumn{2}{l}{$V_{h_s}^{\rm S}, ~V_{h_s}^{\rm SM}, ~V_{h}^{\rm S}, ~V_{h}^{\rm SM}$}     & \multicolumn{2}{l|}{~-0.999, ~-0.019, ~0.019,~-0.999}&
\multicolumn{2}{l}{$V_{h_s}^{\rm S}, ~V_{h_s}^{\rm SM}, ~V_{h}^{\rm S}, ~V_{h}^{\rm SM}$}     & \multicolumn{2}{l} {~~0.982, ~-0.190, ~~0.190,~~0.982} \\
\multicolumn{2}{l}{$N_{11}, ~N_{12}, ~N_{13}, ~N_{14}, ~N_{15}$}      &\multicolumn{2}{l|}{ ~0.001, ~0.002,  -0.012, ~0.015, -0.999}&
 \multicolumn{2}{l}{$N_{11}, ~N_{12}, ~N_{13}, ~N_{14}, ~N_{15}$}      &\multicolumn{2}{l}{ ~0.001, ~0.001,  -0.003, ~0.004, -0.999}    \\
\multicolumn{2}{l}{$N_{21}, ~N_{22}, ~N_{23}, ~N_{24}, ~N_{25}$}      &\multicolumn{2}{l|}{ -0.028, -0.126,  ~0.706, -0.696, -0.020}&
 \multicolumn{2}{l}{$N_{21}, ~N_{22}, ~N_{23}, ~N_{24}, ~N_{25}$}      &\multicolumn{2}{l}{ -0.028, -0.239,  ~0.696, -0.676, -0.005}    \\
\multicolumn{2}{l}{$N_{31}, ~N_{32}, ~N_{33}, ~N_{34}, ~N_{35}$}      &\multicolumn{2}{l|}{ -0.081, -0.042,  -0.704, -0.704, -0.002}&
 \multicolumn{2}{l}{$N_{31}, ~N_{32}, ~N_{33}, ~N_{34}, ~N_{35}$}      &\multicolumn{2}{l}{ -0.090, -0.048,  -0.703, -0.704, -0.001}    \\
\multicolumn{2}{l}{$N_{41}, ~N_{42}, ~N_{43}, ~N_{44}, ~N_{45}$}      &\multicolumn{2}{l|}{ ~0.996, -0.004,  -0.037, -0.078, -0.001}&
 \multicolumn{2}{l}{$N_{41}, ~N_{42}, ~N_{43}, ~N_{44}, ~N_{45}$}      &\multicolumn{2}{l}{ ~0.007, -0.970,  -0.136, ~0.202, ~0.001}    \\
\multicolumn{2}{l}{$N_{51}, ~N_{52}, ~N_{53}, ~N_{54}, ~N_{55}$}      &\multicolumn{2}{l|}{ ~0.003, -0.991,  -0.060, ~0.118, ~0.001}&
 \multicolumn{2}{l}{$N_{51}, ~N_{52}, ~N_{53}, ~N_{54}, ~N_{55}$}      &\multicolumn{2}{l}{ ~0.996, -0.005,  -0.043, -0.084, -0.001}    \\
\hline
\multicolumn{2}{l}{annihilations}                            & \multicolumn{2}{l|}{Fractions [\%]} 			&
\multicolumn{2}{l}{annihilations}                            & \multicolumn{2}{l}{Fractions [\%]} \\
\multicolumn{2}{l}{$\tilde{\chi}_1^0\tilde{\chi}_1^0 \to h_s h_s $} 					   & \multicolumn{2}{l|}{99.9}&
\multicolumn{2}{l}{$\tilde{\chi}_1^0\tilde{\chi}_1^0 \to h_s h_s /h_s h  $} 	   & \multicolumn{2}{l}{92.7/7.02} \\
\hline
\multicolumn{2}{l}{Decays}   & \multicolumn{2}{l|}{Branching ratios [\%]} 	 &
\multicolumn{2}{l}{Decays}   & \multicolumn{2}{l} {Branching ratios [\%]}  \\
\multicolumn{2}{l}{$\tilde{\chi}^0_2 \to \tilde{\chi}^0_1 Z^\ast$}       & \multicolumn{2}{l|}{~100}  &
\multicolumn{2}{l}{$\tilde{\chi}^0_2 \to \tilde{\chi}^0_1 Z^\ast$}       & \multicolumn{2}{l} {~100}    \\
\multicolumn{2}{l}{$\tilde{\chi}^0_3 \to \tilde{\chi}^0_1 Z^\ast$} & \multicolumn{2}{l|}{~96.5} &
\multicolumn{2}{l}{$\tilde{\chi}^0_3 \to \tilde{\chi}^\mp_1(W^\pm)^\ast/ \tilde{\chi}^0_2 Z^\ast/\tilde{\chi}^0_1 Z^\ast   $}   & \multicolumn{2}{l}{~51.8/40.7/8.5}    \\
\multicolumn{2}{l}{$\tilde{\chi}^0_4 \to \tilde{\chi}^\pm_1 W^\mp /\tilde{\chi}^0_3 h /\tilde{\chi}^0_2 Z/\tilde{\chi}^0_3 Z$}         & \multicolumn{2}{l|}{~50/22.5/21.5/3}  &
\multicolumn{2}{l}{$\tilde{\chi}^0_4 \to \tilde{\chi}^\pm_1 W^\mp / \tilde{\chi}^0_3 Z / \tilde{\chi}^0_2 h$}         & \multicolumn{2}{l}{~55.8/22.5/19.2}    \\
\multicolumn{2}{l}{$\tilde{\chi}^0_5 \to \tilde{\chi}^\pm_1 W^\mp/\tilde{\chi}^0_3 Z/\tilde{\chi}^0_2 h$}         & \multicolumn{2}{l|}{~51.7/22.2/21}  &
\multicolumn{2}{l}{$\tilde{\chi}^0_5 \to \tilde{\chi}^\pm_1 W^\mp/\tilde{\chi}^0_5 h/\tilde{\chi}^0_2 Z$}         & \multicolumn{2}{l}{~50.4/21.8/21.6}    \\
\multicolumn{2}{l}{$\tilde{\chi}^+_1 \to \tilde{\chi}^0_1 (W^+)^\ast$}         & \multicolumn{2}{l|}{~100}&
\multicolumn{2}{l}{$\tilde{\chi}^+_1 \to \tilde{\chi}^0_1 (W^+)^\ast$}         & \multicolumn{2}{l}{~100}     \\
\multicolumn{2}{l}{$\tilde{\chi}^+_2 \to \tilde{\chi}^0_2 W^+/\tilde{\chi}^0_3 W^+/\tilde{\chi}^+_1 Z/\tilde{\chi}^+_1 h/$}         & \multicolumn{2}{l|}{~25.4/25.4/25.1/23.7}  &
\multicolumn{2}{l}{$\tilde{\chi}^+_2 \to \tilde{\chi}^0_2 W^+ / \tilde{\chi}^+_1 Z / \tilde{\chi}^0_3 W^+ / \tilde{\chi}^+_1 h$}         & \multicolumn{2}{l}{~27.3/25.8/25.6/20.5}
\\
\multicolumn{2}{l}{$h_s \to W^+ W^-/ZZ/hh{b}$}       & \multicolumn{2}{l|}{~56.4/23/20.4}  &
\multicolumn{2}{l}{$h_s \to b\bar{b}/WW^\ast/gg/\tau^+ \tau^-$}       & \multicolumn{2}{l} {~62.9/16.2/9.1/7.2}    \\
\multicolumn{2}{l}{$h \to b\bar{b}/WW^\ast/gg/\tau^+ \tau^-$}       & \multicolumn{2}{l|}{~53.7/26.6/8.2/6.2}  &
\multicolumn{2}{l}{$h \to b\bar{b}/WW^\ast/gg/\tau^+ \tau^-/c\bar{c}$}       & \multicolumn{2}{l} {~54.9/25.2/8.3/6.4/2.5}    \\
\multicolumn{2}{l}{$H \to b\bar{b}/\tilde{\chi}^\pm_1\tilde{\chi}^\mp_2 /\tilde{\chi}^0_3\tilde{\chi}^0_5/\tilde{\chi}^0_2\tilde{\chi}^0_5/\tau^+ \tau^-$}       & \multicolumn{2}{l|}{~49.3/26.2/13.7/6/4.9}  &
\multicolumn{2}{l}{$H \to b\bar{b}/\tilde{\chi}^\pm_1\tilde{\chi}^\mp_2/\tilde{\chi}^0_3\tilde{\chi}^0_4/\tilde{\chi}^0_2\tilde{\chi}^0_4/\tilde{\chi}^+_1\tilde{\chi}^-_1$}       & \multicolumn{2}{l} {~46.6/15.9/12.4/7/3.7}    \\
\multicolumn{2}{l}{$A_H \to b\bar{b}/\tilde{\chi}^\pm_1\tilde{\chi}^\mp_2 /\tilde{\chi}^0_2\tilde{\chi}^0_5$}       & \multicolumn{2}{l|}{~38.5/26.3/13.5}  &
\multicolumn{2}{l}{$A_H \to b\bar{b}/\tilde{\chi}^\pm_1\tilde{\chi}^\mp_2/\tilde{\chi}^0_2\tilde{\chi}^0_4/\tilde{\chi}^+_1\tilde{\chi}^-_1/\tilde{\chi}^0_2\tilde{\chi}^0_5$}       & \multicolumn{2}{l} {~42.9/14.1/12.4/5.8/2.9}    \\
\multicolumn{2}{l}{$H^+ \to t\bar{b}/\tilde{\chi}^0_2\tilde{\chi}^+_2/\tilde{\chi}^0_5 \tilde{\chi}^+_5/\tilde{\chi}^0_3\tilde{\chi}^+_2/\tilde{\chi}^0_4\tilde{\chi}^+_1$}       & \multicolumn{2}{l|}{~27.2/20.3/19.7/19.3/6.8}  &
\multicolumn{2}{l}{$H^+ \to \tilde{\chi}^0_2 \tilde{\chi}^+_2/\tilde{\chi}^0_4 \tilde{\chi}^+_1/\tilde{\chi}^0_3 \tilde{\chi}^+_2/t\bar{b}/\tilde{\chi}^0_5\tilde{\chi}^+_1$}       & \multicolumn{2}{l} {~25.9/24.4/22.9/14.8/6.3}    \\
\hline
\multicolumn{2}{l}{$R$ value: 0.16}     & \multicolumn{2}{l|}{Signal Region: SR-incWh-SFOS-09 in Ref.~\cite{ATLAS:2021moa}}&
\multicolumn{2}{l}{$R$ value: 0.33}     & \multicolumn{2}{l} {Signal Region: SR-G05 in Ref.~\cite{ATLAS:2021yqv}}    \\
\hline \hline
\end{tabular}
}
\caption{Benchmark points where the Singlino-dominated DM achieved the observed relic abundance primarily through the process $\tilde{\chi}_1^0 \tilde{\chi}_1^0 \to h_s h_s$. Both points satisfied all experimental constraints. In the left and right parts of this table, $h_1$ and $h_2$ were predicted to be the SM-like Higgs boson, respectively. \label{tab:6}}
\end{table}

\begin{table}[t]
\centering
\resizebox{1\textwidth}{!}
{
\begin{tabular}{lrlr|lrlr}
\hline \hline
\multicolumn{4}{c|}{\bf Point P3}     & \multicolumn{4}{c}{\bf Point P4}   \\
\hline
$\lambda$          & 0.006        & $m_{h_s}$                 & 818.5~GeV                           &            $\lambda$          & 0.002        & $m_{h_s}$                 & 98.97~GeV     \\
$\kappa$           & 0.45         & $m_{h}$                   & 124.7~GeV                             &            $\kappa$           & 0.13         & $m_{h}$                   & 125.6~GeV    \\
$\tan{\beta}$      & 26.64       & $m_{H}$                   & 1670 ~GeV                        &              $\tan{\beta}$      & 25.25       & $m_{H}$                   & 1730 ~GeV      \\
$v_s$       	    & 886.7 ~GeV       & $m_{A_s}$                 & 119.5~GeV                            &            $v_s$       	      & 285.8 ~GeV       & $m_{A_s}$                 & 50.57~GeV    \\
$\mu_{\rm tot}$    & 569 ~GeV    & $m_{A_H}$	              & 1670~GeV                           &             $\mu_{\rm tot}$    & 428.2 ~GeV    & $m_{A_H}$	                & 1730~GeV     \\
$M_1$	             & 692.6 ~GeV       & $m_{H^\pm}$	              & 1665~GeV                          &             $M_1$	            & 732.2 ~GeV       & $m_{H^\pm}$	             & 1728~GeV   \\
$M_2$ 	          & 303.9 ~GeV       & $m_{\tilde{\chi}_1^0}$    & 217.5~GeV                        &              $M_2$ 	            & 656.5 ~GeV       & $m_{\tilde{\chi}_1^0}$    & -65.31~GeV   \\
$A_t$              & -2748 ~GeV       & $m_{\tilde{\chi}_2^0}$    & 314.8~GeV                        &              $A_t$              & -3168 ~GeV       & $m_{\tilde{\chi}_2^0}$    & 422.4~GeV    \\
$A_\lambda$        & -1307 ~GeV  & $m_{\tilde{\chi}_3^0}$	  & 583.5~GeV                         &             $A_\lambda$        & -731.2 ~GeV  & $m_{\tilde{\chi}_3^0}$	 & -441.9~GeV   \\
$m_A$              & 2000 ~GeV       & $m_{\tilde{\chi}_4^0}$ 	  & -584.1~GeV                      &                $m_A$              & 2000 ~GeV       & $m_{\tilde{\chi}_4^0}$ 	 & 699.8~GeV   \\
$m_B$              & 834.7 ~GeV       & $m_{\tilde{\chi}_5^0}$ 	  & 706.1~GeV                      &                $m_B$              & 101 ~GeV       & $m_{\tilde{\chi}_5^0}$ 	 & 742~GeV   \\
$m_N$              & 217.5 ~GeV       & $m_{\tilde{\chi}_1^\pm}$  & 315.2~GeV                        &              $m_N$              & -65.37 ~GeV       & $m_{\tilde{\chi}_1^\pm}$  & 428.5~GeV    \\
$\Omega h^2$       & 0.12         & $m_{\tilde{\chi}_2^\pm}$  & 596.2~GeV                        &              $\Omega h^2$       & 0.1119         & $m_{\tilde{\chi}_2^\pm}$  & 701.9~GeV       \\
$\sigma^{\rm SD}_{\tilde{\chi}_1^0-n}$  & 5.37$\times 10^{-49}{\rm ~cm^2}$  &$\sigma^{\rm SI}_{\tilde{\chi}_1^0-n}$  & 7.86$\times 10^{-49}{\rm ~cm^2}$ &
$\sigma^{\rm SD}_{\tilde{\chi}_1^0-n}$  & 6.11$\times 10^{-51}{\rm ~cm^2}$  &$\sigma^{\rm SI}_{\tilde{\chi}_1^0-n}$  & 1.03$\times 10^{-47}{\rm ~cm^2}$ \\
\hline
\multicolumn{2}{l}{$V_{h_s}^{\rm S}, ~V_{h_s}^{\rm SM}, ~V_{h}^{\rm S}, ~V_{h}^{\rm SM}$}     & \multicolumn{2}{l|}{~~0.999, ~~0.002, ~-0.002,~~0.999}&
\multicolumn{2}{l}{$V_{h_s}^{\rm S}, ~V_{h_s}^{\rm SM}, ~V_{h}^{\rm S}, ~V_{h}^{\rm SM}$}     & \multicolumn{2}{l} {~-0.999, ~~0.043, ~-0.043,~-0.999} \\
\multicolumn{2}{l}{$N_{11}, ~N_{12}, ~N_{13}, ~N_{14}, ~N_{15}$}      &\multicolumn{2}{l|}{ ~0.001, -0.002,  ~0.001, -0.002, ~0.999}&
 \multicolumn{2}{l}{$N_{11}, ~N_{12}, ~N_{13}, ~N_{14}, ~N_{15}$}      &\multicolumn{2}{l}{ -0.001, ~0.001,  ~0.001, ~0.001, -0.999}    \\
\multicolumn{2}{l}{$N_{21}, ~N_{22}, ~N_{23}, ~N_{24}, ~N_{25}$}      &\multicolumn{2}{l|}{ ~0.013, -0.975,  ~0.192, -0.108, -0.002}&
 \multicolumn{2}{l}{$N_{21}, ~N_{22}, ~N_{23}, ~N_{24}, ~N_{25}$}      &\multicolumn{2}{l}{ ~0.098, -0.212,  ~0.699, -0.676, -0.001}    \\
\multicolumn{2}{l}{$N_{31}, ~N_{32}, ~N_{33}, ~N_{34}, ~N_{35}$}      &\multicolumn{2}{l|}{ ~0.264, ~0.208,  ~0.664, -0.668, -0.002}&
 \multicolumn{2}{l}{$N_{31}, ~N_{32}, ~N_{33}, ~N_{34}, ~N_{35}$}      &\multicolumn{2}{l}{ -0.025, ~0.048,  ~0.703, ~0.709, ~0.001}    \\
\multicolumn{2}{l}{$N_{41}, ~N_{42}, ~N_{43}, ~N_{44}, ~N_{45}$}      &\multicolumn{2}{l|}{ -0.023, ~0.059,  ~0.703, ~0.708, ~0.001}&
 \multicolumn{2}{l}{$N_{41}, ~N_{42}, ~N_{43}, ~N_{44}, ~N_{45}$}      &\multicolumn{2}{l}{ ~0.173, ~0.965,  ~0.108, -0.166, -0.001}    \\
\multicolumn{2}{l}{$N_{51}, ~N_{52}, ~N_{53}, ~N_{54}, ~N_{55}$}      &\multicolumn{2}{l|}{ ~0.964, -0.042,  -0.167, ~0.201, ~0.001}&
 \multicolumn{2}{l}{$N_{51}, ~N_{52}, ~N_{53}, ~N_{54}, ~N_{55}$}      &\multicolumn{2}{l}{ ~0.980, -0.148,  -0.071, ~0.115, ~0.001}    \\
\hline
\multicolumn{2}{l}{annihilations}                            & \multicolumn{2}{l|}{Fractions [\%]} 								   & \multicolumn{2}{l}{annihilations}                                          & \multicolumn{2}{l}{Fractions [\%]}                                         \\
\multicolumn{2}{l}{$\tilde{\chi}_1^0\tilde{\chi}_1^0 \to A_s A_s   $} 						   & \multicolumn{2}{l|}{99.8}
   & \multicolumn{2}{l}{$\tilde{\chi}_1^0\tilde{\chi}_1^0 \to A_s A_s   $} 	   & \multicolumn{2}{l}{99.9}             \\
\hline
\multicolumn{2}{l}{Decays}   & \multicolumn{2}{l|}{Branching ratios [\%]} 	 &
\multicolumn{2}{l}{Decays}   & \multicolumn{2}{l} {Branching ratios [\%]}  \\
\multicolumn{2}{l}{$\tilde{\chi}^0_2 \to \tilde{\chi}^0_1 Z$}       & \multicolumn{2}{l|}{~100}  &
\multicolumn{2}{l}{$\tilde{\chi}^0_2 \to \tilde{\chi}^0_1 Z/\tilde{\chi}^0_1 h$}  & \multicolumn{2}{l} {~62.2/37.2}    \\
\multicolumn{2}{l}{$\tilde{\chi}^0_3 \to \tilde{\chi}^\mp_1 W^\pm/ \tilde{\chi}^0_2 h/ \tilde{\chi}^0_2 Z$} & \multicolumn{2}{l|}{~69.5/27.7/2.8} &
\multicolumn{2}{l}{$\tilde{\chi}^0_3 \to \tilde{\chi}^0_1 h/ \tilde{\chi}^0_1 Z/\tilde{\chi}^0_2 Z^\ast/\tilde{\chi}^\mp_1(W^\pm)^\ast$}   & \multicolumn{2}{l}{~45.3/29.5/14.6/5.8}    \\
\multicolumn{2}{l}{$\tilde{\chi}^0_4 \to \tilde{\chi}^\pm_1 W^\mp /\tilde{\chi}^0_2 Z /\tilde{\chi}^0_2 h$}         & \multicolumn{2}{l|}{~67.8/30.1/1.7}  &
\multicolumn{2}{l}{$\tilde{\chi}^0_4 \to \tilde{\chi}^\pm_1 W^\mp / \tilde{\chi}^0_3 Z / \tilde{\chi}^0_2 h$}         & \multicolumn{2}{l}{~65.1/17.7/15.5}    \\
\multicolumn{2}{l}{$\tilde{\chi}^0_5 \to \tilde{\chi}^\pm_2 W^\mp/\tilde{\chi}^\pm_1 W^\mp/\tilde{\chi}^0_4 Z/\tilde{\chi}^0_2 Z$}         & \multicolumn{2}{l|}{~37.1/29.7/19.4/1.8}  &
\multicolumn{2}{l}{$\tilde{\chi}^0_5 \to \tilde{\chi}^0_3 Z/ \tilde{\chi}^0_2 h/ \tilde{\chi}^\pm_2 W^\mp$}         & \multicolumn{2}{l}{~38.4/34.8/22.2}    \\
\multicolumn{2}{l}{$\tilde{\chi}^+_1 \to \tilde{\chi}^0_1 W^+$}         & \multicolumn{2}{l|}{~100}&
\multicolumn{2}{l}{$\tilde{\chi}^+_1 \to \tilde{\chi}^0_1 W^+$}         & \multicolumn{2}{l}{~100}     \\
\multicolumn{2}{l}{$\tilde{\chi}^+_2 \to \tilde{\chi}^0_2 W^+ / \tilde{\chi}^+_1 Z/ \tilde{\chi}^+_1 h$}         & \multicolumn{2}{l|}{~35.6/34.5/29.9}  &
\multicolumn{2}{l}{$\tilde{\chi}^+_2 \to \tilde{\chi}^0_2 W^+ / \tilde{\chi}^+_1 Z / \tilde{\chi}^0_3 W^+ / \tilde{\chi}^+_1 h$}         & \multicolumn{2}{l}{~26.4/26/25.4/22.1}
\\
\multicolumn{2}{l}{$h_s \to A_s A_s/\tilde{\chi}^0_1 \tilde{\chi}^0_1$}       & \multicolumn{2}{l|}{~84.4/15.6}  &
\multicolumn{2}{l}{$h_s \to b\bar{b}/\tau^+ \tau^-/gg/c\bar{c}$}       & \multicolumn{2}{l} {~79.6/8.9/7.2/3.4}    \\
\multicolumn{2}{l}{$h \to b\bar{b}/WW^\ast/gg/\tau^+ \tau^-$}       & \multicolumn{2}{l|}{~55.1/24.9/8.4/6.4}  &
\multicolumn{2}{l}{$h \to b\bar{b}/ A_s A_s /WW^\ast/gg/\tau^+ \tau^-$}       & \multicolumn{2}{l} {~39.3/26.6/19.6/6.1/4.6}    \\
\multicolumn{2}{l}{$H \to b\bar{b}/\tilde{\chi}^\pm_1\tilde{\chi}^\mp_2 /\tilde{\chi}^0_2\tilde{\chi}^0_4/\tau^+ \tau^-/\tilde{\chi}^0_5\tilde{\chi}^0_5$}       & \multicolumn{2}{l|}{~39.8/30/9.3/7.7/5.7}  &
\multicolumn{2}{l}{$H \to b\bar{b}/\tilde{\chi}^\pm_1\tilde{\chi}^\mp_2/\tilde{\chi}^0_3\tilde{\chi}^0_4\tau^+ \tau^-/\tilde{\chi}^0_3\tilde{\chi}^0_5$}       & \multicolumn{2}{l} {~41.2/28.6/7.8/7.8/4.5}    \\
\multicolumn{2}{l}{$A_H \to b\bar{b}/\tilde{\chi}^\pm_1\tilde{\chi}^\mp_2 /\tilde{\chi}^0_2\tilde{\chi}^0_3/\tau^+ \tau^-/\tilde{\chi}^0_2\tilde{\chi}^0_4$}       & \multicolumn{2}{l|}{~39.8/28.9/8.3/7.7/5.8}  &
\multicolumn{2}{l}{$A_H \to b\bar{b}/\tilde{\chi}^\pm_1\tilde{\chi}^\mp_2/\tau^+ \tau^-/\tilde{\chi}^0_2\tilde{\chi}^0_4$}       & \multicolumn{2}{l} {~41.3/26.8/7.8/6.9}    \\
\multicolumn{2}{l}{$H^+ \to t\bar{b}/\tilde{\chi}^0_3\tilde{\chi}^+_1/\tilde{\chi}^0_2 \tilde{\chi}^+_2/\tilde{\chi}^0_4\tilde{\chi}^+_1$}       & \multicolumn{2}{l|}{~39.5/16.4/16.3/15.4}  &
\multicolumn{2}{l}{$H^+ \to t\bar{b}/\tilde{\chi}^0_4 \tilde{\chi}^+_1/\tilde{\chi}^0_3 \tilde{\chi}^+_2/\tilde{\chi}^0_2 \tilde{\chi}^+_2/\tau^+ \nu_{\tau}$}       & \multicolumn{2}{l} {~41/17.6/14.2/15.3/8.6}    \\
\hline
\multicolumn{2}{l}{$R$ value: 0.79}     & \multicolumn{2}{l|}{Signal Region: SR-incWZ-03 in Ref.~\cite{ATLAS:2021moa}}&
\multicolumn{2}{l}{$R$ value: 0.85}     & \multicolumn{2}{l} {Signal Region: SRG07-0j-mll in Ref.~\cite{ATLAS:2021yqv}}    \\
\hline \hline
\end{tabular}
}
\caption{Same as Table \ref{tab:6} except that the Singlino-dominated DM achieved the observed relic abundance primarily through the process $\tilde{\chi}_1^0 \tilde{\chi}_1^0 \to A_s A_s$. \label{tab:7}}
\end{table}

\begin{table}[t]
\centering
\resizebox{1\textwidth}{!}
{
\begin{tabular}{lrlr|lrlr}
\hline \hline
\multicolumn{4}{c|}{\bf Point P5}     & \multicolumn{4}{c}{\bf Point P6}   \\
\hline
$\lambda$          & 0.026        & $m_{h_s}$                 & 462~GeV                           &            $\lambda$          & 0.001        & $m_{h_s}$                 & 76.85~GeV     \\
$\kappa$           & -0.22         & $m_{h}$                   & 124.9~GeV                             &            $\kappa$           & 0.18         & $m_{h}$                   & 125.2~GeV    \\
$\tan{\beta}$      & 42.42       & $m_{H}$                   & 2389 ~GeV                        &              $\tan{\beta}$      & 36.19       & $m_{H}$                   & 2148 ~GeV      \\
$v_s$       	    & 630.1 ~GeV       & $m_{A_s}$                 & 263.4~GeV                            &            $v_s$       	      & 200.7 ~GeV       & $m_{A_s}$                 & 282.8~GeV    \\
$\mu_{\rm tot}$    & 514.5 ~GeV    & $m_{A_H}$	              & 2389~GeV                           &             $\mu_{\rm tot}$    & 259 ~GeV    & $m_{A_H}$	                & 2148~GeV     \\
$M_1$	             & 855.2 ~GeV       & $m_{H^\pm}$	              & 2410~GeV                          &             $M_1$	            & 704.2 ~GeV       & $m_{H^\pm}$	             & 2159~GeV   \\
$M_2$ 	          & 822.6 ~GeV       & $m_{\tilde{\chi}_1^0}$    & -389.8~GeV                        &              $M_2$ 	            & 621.4 ~GeV       & $m_{\tilde{\chi}_1^0}$    & 209.1~GeV   \\
$A_t$              & 2492 ~GeV       & $m_{\tilde{\chi}_2^0}$    & 515~GeV                        &              $A_t$              & 2424 ~GeV       & $m_{\tilde{\chi}_2^0}$    & 256.7~GeV    \\
$A_\lambda$        & 1123 ~GeV  & $m_{\tilde{\chi}_3^0}$	  & -529.8~GeV                         &             $A_\lambda$        & 1501 ~GeV  & $m_{\tilde{\chi}_3^0}$	 & -270.8~GeV   \\
$m_A$              & 2000 ~GeV       & $m_{\tilde{\chi}_4^0}$ 	  & 858.4~GeV                      &                $m_A$              & 2000 ~GeV       & $m_{\tilde{\chi}_4^0}$ 	 & 661.8~GeV   \\
$m_B$              & 465.2 ~GeV       & $m_{\tilde{\chi}_5^0}$ 	  & 871.9~GeV                      &                $m_B$              & 76.24 ~GeV       & $m_{\tilde{\chi}_5^0}$ 	 & 712.3~GeV   \\
$m_N$              & -390.2 ~GeV       & $m_{\tilde{\chi}_1^\pm}$  & 520.2~GeV                        &              $m_N$              & 209.5 ~GeV       & $m_{\tilde{\chi}_1^\pm}$  & 262.1~GeV    \\
$\Omega h^2$       & 0.12         & $m_{\tilde{\chi}_2^\pm}$  & 867~GeV                        &              $\Omega h^2$       & 0.12         & $m_{\tilde{\chi}_2^\pm}$  & 662.9~GeV       \\
$\sigma^{\rm SD}_{\tilde{\chi}_1^0-n}$  & 8.24$\times 10^{-46}{\rm ~cm^2}$  &$\sigma^{\rm SI}_{\tilde{\chi}_1^0-p}$  & 1.9$\times 10^{-47}{\rm ~cm^2}$ &
$\sigma^{\rm SD}_{\tilde{\chi}_1^0-n}$  & 9.89$\times 10^{-50}{\rm ~cm^2}$  &$\sigma^{\rm SI}_{\tilde{\chi}_1^0-p}$  & 3.76$\times 10^{-48}{\rm ~cm^2}$ \\
\hline
\multicolumn{2}{l}{$V_{h_s}^{\rm S}, ~V_{h_s}^{\rm SM}, ~V_{h}^{\rm S}, ~V_{h}^{\rm SM}$}     & \multicolumn{2}{l|}{~~0.999, ~~0.022, ~-0.022,~~0.999}&
\multicolumn{2}{l}{$V_{h_s}^{\rm S}, ~V_{h_s}^{\rm SM}, ~V_{h}^{\rm S}, ~V_{h}^{\rm SM}$}     & \multicolumn{2}{l} {~-0.999, ~~0.009, ~-0.009,~-0.999} \\
\multicolumn{2}{l}{$N_{11}, ~N_{12}, ~N_{13}, ~N_{14}, ~N_{15}$}      &\multicolumn{2}{l|}{ ~0.001, -0.001,  -0.014, -0.019, ~0.999}&
 \multicolumn{2}{l}{$N_{11}, ~N_{12}, ~N_{13}, ~N_{14}, ~N_{15}$}      &\multicolumn{2}{l}{ -0.001, ~0.001,  -0.002, ~0.002, -0.999}    \\
\multicolumn{2}{l}{$N_{21}, ~N_{22}, ~N_{23}, ~N_{24}, ~N_{25}$}      &\multicolumn{2}{l|}{ -0.089, ~0.163,  -0.702, ~0.687, ~0.003}&
 \multicolumn{2}{l}{$N_{21}, ~N_{22}, ~N_{23}, ~N_{24}, ~N_{25}$}      &\multicolumn{2}{l}{ ~0.068, -0.141,  ~0.711, -0.685, -0.003}    \\
\multicolumn{2}{l}{$N_{31}, ~N_{32}, ~N_{33}, ~N_{34}, ~N_{35}$}      &\multicolumn{2}{l|}{ ~0.022, -0.040,  -0.704, -0.708, -0.023}&
 \multicolumn{2}{l}{$N_{31}, ~N_{32}, ~N_{33}, ~N_{34}, ~N_{35}$}      &\multicolumn{2}{l}{ ~0.031, -0.060,  -0.700, -0.711, -0.001}    \\
\multicolumn{2}{l}{$N_{41}, ~N_{42}, ~N_{43}, ~N_{44}, ~N_{45}$}      &\multicolumn{2}{l|}{ -0.564, -0.821,  -0.046, ~0.074, ~0.001}&
 \multicolumn{2}{l}{$N_{41}, ~N_{42}, ~N_{43}, ~N_{44}, ~N_{45}$}      &\multicolumn{2}{l}{ ~0.112, ~0.983,  ~0.056, -0.133, -0.001}    \\
\multicolumn{2}{l}{$N_{51}, ~N_{52}, ~N_{53}, ~N_{54}, ~N_{55}$}      &\multicolumn{2}{l|}{ ~0.821, -0.545,  -0.089, ~0.145, ~0.001}&
 \multicolumn{2}{l}{$N_{51}, ~N_{52}, ~N_{53}, ~N_{54}, ~N_{55}$}      &\multicolumn{2}{l}{ ~0.991, -0.099,  -0.034, ~0.085, ~0.001}    \\
\hline
\multicolumn{2}{l}{annihilations}               & \multicolumn{2}{l|}{Fractions [\%]} 			& \multicolumn{2}{l}{annihilations}               & \multicolumn{2}{l}{Fractions [\%]}            \\
\multicolumn{2}{l}{$\tilde{\chi}_1^0\tilde{\chi}_1^0 \to h_s A_s   $}  & \multicolumn{2}{l|}{99.6}& \multicolumn{2}{l}{$\tilde{\chi}_1^0\tilde{\chi}_1^0 \to h_s A_s  / h_s h_s $} 	   & \multicolumn{2}{l}{97.1/2.87}  \\
\hline
\multicolumn{2}{l}{Decays}   & \multicolumn{2}{l|}{Branching ratios [\%]} 	 &
\multicolumn{2}{l}{Decays}   & \multicolumn{2}{l} {Branching ratios [\%]}  \\
\multicolumn{2}{l}{$\tilde{\chi}^0_2 \to \tilde{\chi}^0_1 Z$}       & \multicolumn{2}{l|}{~100}  &
\multicolumn{2}{l}{$\tilde{\chi}^0_2 \to \tilde{\chi}^0_1 Z^\ast$}  & \multicolumn{2}{l} {~100}    \\
\multicolumn{2}{l}{$\tilde{\chi}^0_3 \to \tilde{\chi}^0_1 h / \tilde{\chi}^0_1 Z$} & \multicolumn{2}{l|}{~92.5/7.38} &
\multicolumn{2}{l}{$\tilde{\chi}^0_3 \to \tilde{\chi}^0_2 Z^\ast /\tilde{\chi}^\mp_1(W^\pm)^\ast $}   & \multicolumn{2}{l}{~79.5/20.4}    \\
\multicolumn{2}{l}{$\tilde{\chi}^0_4 \to \tilde{\chi}^\pm_1 W^\mp /\tilde{\chi}^0_3 Z /\tilde{\chi}^0_2 h$}         & \multicolumn{2}{l|}{~84.7/7.27/6.9}  &
\multicolumn{2}{l}{$\tilde{\chi}^0_4 \to \tilde{\chi}^\pm_1 W^\mp / \tilde{\chi}^0_3 Z / \tilde{\chi}^0_2 h$}         & \multicolumn{2}{l}{~59.6/18.3/16}    \\
\multicolumn{2}{l}{$\tilde{\chi}^0_5 \to \tilde{\chi}^0_3 Z/\tilde{\chi}^0_2 h/\tilde{\chi}^0_2 Z /\tilde{\chi}^0_3 h$}         & \multicolumn{2}{l|}{~47.9/45.1/3.12/2.39}  &
\multicolumn{2}{l}{$\tilde{\chi}^0_5 \to \tilde{\chi}^\pm_1 W^\mp / \tilde{\chi}^0_3 Z / \tilde{\chi}^0_2 h/ \tilde{\chi}^0_2 Z/\tilde{\chi}^0_3 h$}         & \multicolumn{2}{l}{~31.2/30.1/27.5/5.8/5.2}    \\
\multicolumn{2}{l}{$\tilde{\chi}^+_1 \to \tilde{\chi}^0_1 (W^+)^\ast$}         & \multicolumn{2}{l|}{~100}&
\multicolumn{2}{l}{$\tilde{\chi}^+_1 \to \tilde{\chi}^0_1 (W^+)^\ast \ / \tilde{\chi}^0_2 (W^+)^\ast$}         & \multicolumn{2}{l}{~53.2/46.8}     \\
\multicolumn{2}{l}{$\tilde{\chi}^+_2 \to  \tilde{\chi}^+_1 Z / \tilde{\chi}^0_2 W^+/\tilde{\chi}^0_3 W^+/\tilde{\chi}^+_1 h$}         & \multicolumn{2}{l|}{~25.5/25.4/25.3/23.6}  &
\multicolumn{2}{l}{$\tilde{\chi}^+_2 \to \tilde{\chi}^0_3 W^+ / \tilde{\chi}^+_1 Z / \tilde{\chi}^0_2 W^+ / \tilde{\chi}^+_1 h$}         & \multicolumn{2}{l}{~26/25.4/25/23.2}
\\
\multicolumn{2}{l}{$h_s \to W^+ W^-/hh/ZZ/b\bar{b}$}       & \multicolumn{2}{l|}{~44.5/23/20/12.3}  &
\multicolumn{2}{l}{$h_s \to b\bar{b}/gg/\tau^+ \tau^-$}       & \multicolumn{2}{l} {~73/11/8}    \\
\multicolumn{2}{l}{$h \to b\bar{b}/WW^\ast/gg/\tau^+ \tau^-/c\bar{c}$}       & \multicolumn{2}{l|}{~54.7/25.4/8.4/6.3/2.5}  &
\multicolumn{2}{l}{$h \to b\bar{b}/WW^\ast/gg/\tau^+ \tau^-$}       & \multicolumn{2}{l} {~54.2/25.9/8.4/6.3}    \\
\multicolumn{2}{l}{$H \to b\bar{b}/\tilde{\chi}^\pm_1\tilde{\chi}^\mp_2 /\tau^+ \tau^-/\tilde{\chi}^0_3\tilde{\chi}^0_5$}       & \multicolumn{2}{l|}{~55.1/18.6/12.6/5.9}  &
\multicolumn{2}{l}{$H \to b\bar{b}/\tilde{\chi}^\pm_1\tilde{\chi}^\mp_2/\tau^+ \tau^-/\tilde{\chi}^0_3\tilde{\chi}^0_4$}       & \multicolumn{2}{l} {~49.6/23.7/9.5/6}    \\
\multicolumn{2}{l}{$A_H \to b\bar{b}/\tilde{\chi}^\pm_1\tilde{\chi}^\mp_2 /\tau^+ \tau^-/\tilde{\chi}^0_2\tilde{\chi}^0_5$}       & \multicolumn{2}{l|}{~55.1/18/12.6/5.5}  &
\multicolumn{2}{l}{$A_H \to b\bar{b}/\tilde{\chi}^\pm_1\tilde{\chi}^\mp_2/\tau^+ \tau^-/\tilde{\chi}^0_2\tilde{\chi}^0_4$}       & \multicolumn{2}{l} {~49.6/23.5/9.5/5.7}    \\
\multicolumn{2}{l}{$H^+ \to t\bar{b}/\tau^+ \nu_{\tau}/\tilde{\chi}^0_4\tilde{\chi}^+_1/\tilde{\chi}^0_2 \tilde{\chi}^+_2/\tilde{\chi}^0_3\tilde{\chi}^+_2$}       & \multicolumn{2}{l|}{~54.4/13.7/12.5/9.6/9.2}  &
\multicolumn{2}{l}{$H^+ \to t\bar{b}/\tilde{\chi}^0_4 \tilde{\chi}^+_1/\tilde{\chi}^0_2 \tilde{\chi}^+_2/\tilde{\chi}^0_3 \tilde{\chi}^+_2$}       & \multicolumn{2}{l} {~49/13.5/12.5/11.7}    \\
\hline
\multicolumn{2}{l}{$R$ value: 0.10}   ~~~~~~~~~~  & \multicolumn{2}{l|}{Signal Region: 3LI in Ref.~\cite{ATLAS:2021moa}}&
\multicolumn{2}{l}{$R$ value: 0.74}               & \multicolumn{2}{l} {Signal Region: SR-WZoff-high-njd in Ref.~\cite{ATLAS:2021yqv}}\\
\hline \hline
\end{tabular}
}
\caption{Same as Table \ref{tab:6} except that the Singlino-dominated DM achieved the observed relic abundance primarily through the process $\tilde{\chi}_1^0 \tilde{\chi}_1^0 \to h_s A_s$. \label{tab:8} }
\end{table}

In Fig.~\ref{Fig5}, we project three types of samples onto the $|m_{\tilde{\chi}_1^0}|-m_{h_s}$, $|m_{\tilde{\chi}_1^0}|-m_{A_s}$, and $|m_{\tilde{\chi}_1^0}|-m_{h_s}$ planes, respectively. The color bar on the left panels indicates the value of $|\kappa|$, which is crucial in determining the relic abundance. On the right panel, the color bar shows the value of $\ln \lambda$, critical for DM-nucleon scattering cross sections. Analysis of the left panels yielded the following conclusions:
\begin{itemize}
\item Increasing $m_{h_s}$ in annihilation processes $\tilde{\chi}_1^0 \tilde{\chi}_1^0 \to h_s h_s$ and $\tilde{\chi}_1^0 \tilde{\chi}_1^0 \to h_s A_s$ or increasing $m_{A_s}$ in annihilation process $\tilde{\chi}_1^0 \tilde{\chi}_1^0 \to A_s A_s$ favors a larger $|\kappa|$ to reach the observed relic abundance at a fixed $|m_{\tilde{\chi}_1^0}|$. In particular, $|\kappa|$ reaches its maximum when the scalar mass closely matches the DM mass due to the phase space suppression in these annihilations.
\item An increase in $|m_{\tilde{\chi}_1^0}|$ correlates with a preference for a higher $|\kappa|$ to achieve the observed relic abundance, which is demonstrated by the approximations in Eqs.~(\ref{Kappa-approximation-1}), (\ref{Kappa-approximation-2}), and (\ref{Kappa-approximation-3}) for simplified cases.
\item Among the three annihilations, $\tilde{\chi}_1^0 \tilde{\chi}_1^0 \to h_s A_s$ requires the smallest $|\kappa|$ to achieve the observed relic abundance due to its s-wave nature, in contrast with the p-wave nature of the other channels. The comparison of $\tilde{\chi}_1^0 \tilde{\chi}_1^0 \to A_s A_s$ with $\tilde{\chi}_1^0 \tilde{\chi}_1^0 \to h_s h_s$ shows that the former prefers a larger $|\kappa|$, as explained by Eqs.~(\ref{Kappa-approximation-1}) and (\ref{Kappa-approximation-2}).
\item The s-wave nature of the annihilation $\tilde{\chi}_1^0 \tilde{\chi}_1^0 \to h_s A_s$ with $h_s \to b \bar{b}$ and $A_s \to b \bar{b}$ also allows for indirect DM search experiments to establish a lower bound on $|m_{\tilde{\chi}_1^0}|$ (approximately $58~{\rm GeV}$). This bound is substantially less stringent than that of the direct annihilation $\tilde{\chi}_1^0 \tilde{\chi}_1^0 \to b \bar{b}$, as the power produced through the former channel is suppressed by its relatively heavy DM mass~\cite{Berlin:2014pya,Gherghetta:2015ysa}. In contrast, no experimental limits apply to p-wave annihilations $\tilde{\chi}_1^0 \tilde{\chi}_1^0 \to h_s h_s$ and $\tilde{\chi}_1^0 \tilde{\chi}_1^0 \to A_s A_s$, permitting potentially much lighter DM candidates down to 10 GeV.
\item The complex dependence of the annihilation cross sections on the model's parameters suggests that all values of $\kappa$ ranging from 0.05 to 0.75 are capable of providing an explanation for the observed relic abundance.
\end{itemize}

The information provided in the right panels is as follows:
\begin{itemize}
\item For Type-I and -III samples, $\lambda$ is typically very small under the $h_2$ scenario due to significant singlet--doublet Higgs mixing, resulting in
$\sigma^{\rm SI}_{\tilde{\chi}_1^0-p}$ being proportional to $\lambda^2 \kappa^2$. Therefore, the results of the LZ experiment stringently limit the magnitude of $\lambda$.
However, a significant $\lambda$ may still be possible in rare cases where contributions mediated by $h_s$ cancel out those mediated by $h$~\cite{Badziak:2015exr,Zhou:2021pit}.

\item As $m_{h_s}$ and $|m_{\tilde{\chi}_1^0}|$ increase from $300~{\rm GeV}$ for Type-I and -III samples, larger $\lambda$ values may arise. In such instances, the singlet--doublet Higgs mixing is very small, resulting in the dominant contribution to $\sigma^{\rm SI}_{\tilde{\chi}_1^0-p}$ being proportional to $\lambda^4$. Moreover, the contribution mediated by $h_s$ is gradually reduced, and simultaneously, the LZ constraint is relaxed, both allowing for relatively large $\lambda$ values, particularly when  $\tilde{\chi}_1^0$ and $h_s$ are all massive.

\item In Type-II samples, the value of $\lambda$ can significantly increase when the values of $|m_{\tilde{\chi}_1^0}|$ and $m_{A_s}$ are enhanced from their starting point at $m_{\tilde{\chi}_1^0} \simeq 10~{\rm GeV}$ and $m_{A_s} \simeq 10~{\rm GeV}$. Specifically, for this type of sample, $h_s$ is always heavier than $\tilde{\chi}_1^0$, exceeding $300~{\rm GeV}$ for almost all cases, which results in a minor impact on $\sigma^{\rm SI}_{\tilde{\chi}_1^0-p}$. The behavior of $\lambda$ can be explained by two typical mass spectrum configurations: increasing $|m_{\tilde{\chi}_1^0}|$ while keeping $m_{A_s}$ constant or fixing the ratio $m_{A_s}/|m_{\tilde{\chi}_1^0}|$ and increasing $|m_{\tilde{\chi}_1^0}|$. In both cases,
    the lower limit of $m_{h_s}$ increases since $m_{h_s} > 2 m_{\tilde{\chi}_1^0} - m_{A_s}$, leading to the tendency to decrease its contribution to $\sigma^{\rm SI}_{\tilde{\chi}_1^0-p}$. Along with the relaxation of the LZ constraint, this situation allows for larger values of $\lambda$.

\item Compared to Type-I and -III samples, a larger portion of Type-II samples predict a sizable $\lambda$. The prevalence of large $\lambda$ values in Type-II samples can be attributed to the tendency for a heavy $h_s$, resulting in the dominant contribution to $\sigma^{\rm SI}_{\tilde{\chi}_1^0-p}$ being proportional to $\lambda^4$. Additionally, Type-I samples favor smaller values of $\lambda$ compared to Type-III samples largely due to a significant number of samples predicting relatively small $\mu_{\rm tot}$ values.
\end{itemize}

We have also projected three types of samples onto the $|m_{\tilde{\chi}_1^0}|- \sigma^{\rm SI}_{\tilde{\chi}_1^0-p}$ and $|m_{\tilde{\chi}_1^0}|-\mu_{\rm tot}$ planes to create Fig.~\ref{Fig6}. In this visualization, gray indicates samples excluded by the LHC constraint, while blue represents samples in agreement with all experimental constraints. Fig.~\ref{Fig6} reveals the following key observations:
\begin{itemize}
\item The LZ experiment and the LHC's search for supersymmetry complement each other in restricting the parameter space of the GNMSSM.
\item Type-II samples tend to predict a smaller $\sigma^{\rm SI}_{\tilde{\chi}_1^0-p}$ than the other sample types.
\item The LHC constraint is particularly effective in ruling out samples that predict moderately light DM and Higgsinos, notably enhancing lower limits on $|m_{\tilde{\chi}_1^0}|$ from approximately $23$, $14$, and $58~{\rm GeV}$ to approximately $70$, $47$, and $89~{\rm GeV}$ for Type-I, -II, and -III samples, respectively. Additionally, it set a lower bound of approximately $180~{\rm GeV}$ on the Higgsino mass, which is much smaller than that of the MSSM~\cite{He:2023lgi}.
\item Among three types of samples, the Type-I samples are significantly more impacted by the LHC constraint due to its inclination toward moderately light Higgsinos.
\end{itemize}

In Tables~\ref{tab:6}, \ref{tab:7}, and \ref{tab:8}, we present benchmark points for three types of samples in both $h_1$ and $h_2$ scenarios. The detailed information of these points elucidate the underlying physics of Singlino-dominated DM.

\section{Summary}\label{sec:sum}

Traditional supersymmetric models like the MSSM and $\mathbb{Z}_3$-NMSSM are facing increasingly stringent constraints as direct detection experiments for DM and searches for supersymmetry at the LHC continue to advance rapidly. This has led to significant fine-tuning of the parameters within these models to predict viable DM candidates, as evidenced by Refs.~\cite{Cao:2019qng,Cao:2018rix,Zhou:2021pit}. In response to these challenges, we undertook Bayesian analyses of the GNMSSM to thoroughly explore its DM physics.

First, we examined the theoretical structure of the GNMSSM and selected a set of physical parameters as our input criteria. We constructed a comprehensive likelihood function that incorporated current experimental and theoretical knowledge on DM physics---including relic density and direct and indirect DM searches---as well as Higgs and B physics. This likelihood function was then used to guide detailed scans of the theory's parameter space using the nested sampling algorithm. We analyzed the resulting samples using statistical measures such as the PL and marginal posterior PDF to uncover underlying physical phenomena. Our findings indicated that the DM candidates in the GNMSSM may predominantly be either Singlino- or Bino-like DM. Specifically, the Singlino-like scenario accounted for $99.3\%$ of the Bayesian evidence in the $h_1$ scenario and $64.8\%$ in the $h_2$ scenario. Given that the Bayesian evidence significantly favored the $h_1$ scenario over the $h_2$ scenario, we concluded that the GNMSSM generally preferred Singlino-dominated DM across a much broader parameter range. This preference distinctly set the GNMSSM apart from the MSSM and $\mathbb{Z}_3$-NMSSM, highlighting its unique position within the landscape of supersymmetric models.

Second, we examined the characteristics of Singlino-dominated DM and identified significant distinctions from the $\mathbb{Z}_3$-NMSSM. In the latter, DM properties were determined by four parameters: $\lambda$, $\tan \beta$, $\mu_{\rm eff}$, and $m_{\tilde{\chi}_1^0} \simeq 2 \kappa \mu_{\rm eff}/\lambda$. For a Singlino-dominated neutralino to remain the LSP, $|\kappa|$ had to be smaller than $\lambda/2$. Furthermore, the LZ constraint on $\lambda$ necessitated $|\kappa|$ to be very small. As a result, the Singlino DM in the $\mathbb{Z}_3$-NMSSM needed to co-annihilate with electroweakinos to match the observed relic abundance. However, such dynamics differed in the GNMSSM due to the introduction of $\mathbb{Z}_3$-violating terms where the DM's properties were described through five independent parameters: $\lambda$, $\kappa$, $\tan \beta$, $\mu_{tot}$, and $m_{\tilde{\chi}_1^0}$. Specifically, allowing for decorrelation of $m_{\tilde{\chi}_1^0}$ from $\mu_{tot}$ enables $|\kappa|$ to exceed $\lambda$ when predicting Singlino-dominant DM. This structural difference significantly impacted the potential creation of a secluded DM sector within singlet particles~\cite{Pospelov:2007mp}. The key characteristics of the secluded sector in our analysis included the following:
\begin{itemize}
\item Annihilation primarily occurred through the channels $\tilde{\chi}_1^0 \tilde{\chi}_1^0 \to h_s h_s$, $\tilde{\chi}_1^0 \tilde{\chi}_1^0 \to A_s A_s$, and $\tilde{\chi}_1^0 \tilde{\chi}_1^0 \to h_s A_s$ to achieve the observed relic abundance. Among the parameters affecting the annihilation cross sections, $\kappa$ played a crucial role, with $\tilde{\chi}_1^0 \tilde{\chi}_1^0 \to h_s A_s$ preferring the smallest $|\kappa|$ value compared to other channels. However, due to the complex dependence of the annihilation cross sections on the model parameters, the values of $|\kappa|$ ranging from $0.05$ to $0.75$ were all capable of explaining the observed relic densities.
\item The s-wave nature of $\tilde{\chi}_1^0 \tilde{\chi}_1^0 \to h_s A_s$ resulted in a lower limit on $m_{\tilde{\chi}_1^0}$ at approximately $60~{\rm GeV}$ from indirect DM search experiments, whereas in other channels, $m_{\tilde{\chi}_1^0}$ could decrease to as low as 10 GeV.
\item The equations governing the interactions between DM and nucleons were characterized by parameters $\lambda$, $\kappa$, $\tan \beta$, $\mu_{tot}$, and $m_{\tilde{\chi}_1^0}$, with $\lambda$ exerting the most significant influence. It was observed that $\sigma^{\rm SI}_{\tilde{\chi}_1^0 -p}$ was proportional to $\lambda^2 \kappa^2$ in the presence of substantial singlet--doublet Higgs mixing, and to $\lambda^4$ if $h_s$ was very massive. In contrast, $\sigma^{\rm SD}_{\tilde{\chi}_1^0 -n}$ consistently varied as a function of approximately $\lambda^4$. Numerical analyses indicated that for most samples in the $h_1$ scenario,  $\lambda \lesssim 0.09$; for the $h_2$ scenario, it was suggested that  $\lambda \lesssim 0.05$. Consequently, $\sigma^{\rm SI}_{\tilde{\chi}_1^0 -p}$ and $\sigma^{\rm SD}_{\tilde{\chi}_1^0 -n}$ could be as low as $10^{-49}$ and $10^{-52}\ {\rm cm^2}$, respectively. Additionally, among the considered channels, $\tilde{\chi}_1^0 \tilde{\chi}_1^0 \to h_s h_s$ in the $h_2$ scenario faced most stringent constraints from the LZ experiment due to the significant Higgs mixing and its preference for relatively light Higgsinos.
\item For Singlino-like DM in the GNMSSM, although $\sigma^{\rm SI}_{\tilde{\chi}_1^0 -p}$ could approach the exclusion limits set by the LZ experiment, $\sigma^{\rm SD}_{\tilde{\chi}_1^0 -n}$ remained well below these bounds. Conversely, both $\sigma^{\rm SI}_{\tilde{\chi}_1^0 -p}$ and $\sigma^{\rm SD}_{\tilde{\chi}_1^0 -n}$ could reach exclusion limits for Bino-like DM. This observation suggested that if SD scattering were observed in near future, the viability of Singlino-like DM would be significantly undermined.
\item In exceptional cases, Singlino-dominated DM could achieve observed relic abundance by co-annihilating with either Higgsino- (for most cases) or Wino-like electroweakinos, with permissible values for $\mu_{\rm tot}$ down to approximate $200~{\rm GeV}$. In contrast, Bino-like DM with a mass at the electroweak scale could only co-annihilate with Wino-like electroweakinos to achieve correct abundance, and the LZ experiment set a threshold for $\mu_{\rm tot}$ at approximately $380~{\rm GeV}$.
\end{itemize}

Finally, we investigated the impact of the LHC's search for supersymmetry on DM physics, which was effective in excluding scenarios that predicted moderately light DM and Higgsinos. Our findings suggested that this constraint significantly raised the lower limits for  $|m_{\tilde{\chi}_1^0}|$ from approximately $23$, $14$, and $58~{\rm GeV}$ in $\tilde{\chi}_1^0 \tilde{\chi}_1^0 \to h_s h_s$, $ \tilde{\chi}_1^0 \tilde{\chi}_1^0 \to A_s A_s$, and $\tilde{\chi}_1^0 \tilde{\chi}_1^0 \to h_s A_s$ to about $70$, $47$, and $89~{\rm GeV}$, respectively.  Moreover, a lower limit of around $180~{\rm GeV}$ was imposed on the Higgsino mass, representing a significant departure from assumptions in the MSSM~\cite{He:2023lgi}. Notably, among these channels, $\tilde{\chi}_1^0 \tilde{\chi}_1^0 \to h_s h_s$ was most influenced by the LHC constraint due to its preference for moderately light Higgsinos.

This study differed from our previous work cited in Ref.~\cite{Cao:2021ljw} in the following aspects.
\begin{itemize}
\item In contrast to Ref.~\cite{Cao:2021ljw}, which focused on the $\mu$-term-extended NMSSM and investigated the viability of Singlino-dominated DM, Bayesian analyses were conducted in this study within a general theoretical framework of the NMSSM. Specifically, we proposed a set of parameters with clear physical meanings as theoretical inputs and concluded that the theory favors Singlino-dominated DM across a much wider parameter space compared to Bino-like DM.
\item We incorporated the most recent experimental results to constrain the DM physics of the GNMSSM, including the 2023 LZ search for both SI and SD DM-nucleon scattering, various LHC searches for supersymmetry, and measurements of $h$'s couplings at the LHC with $136~{\rm fb^{-1}}$ of data. We derived mass bounds on $\tilde{\chi}_1^0$ from indirect DM search experiments and the LHC supersymmetry search.

\item We provide analytical formulas for the SI and SD cross sections of DM-nucleon scattering. These demonstrate that $\sigma^{\rm SI}_{\tilde{\chi}_1^0 -p}$ is proportional to $\lambda^2 \kappa^2$ amid substantial singlet--doublet Higgs mixing and $\lambda^4$ in heavy $h_s$ limit, whereas $\sigma^{\rm SD}_{\tilde{\chi}_1^0 -n}$ consistently scales with  $\lambda^4$. Additionally, we present analytic expressions for the annihilation cross sections of $\tilde{\chi}_1^0 \tilde{\chi}_1^0 \to h_s h_s$, $\tilde{\chi}_1^0 \tilde{\chi}_1^0 \to A_s A_s$, and $\tilde{\chi}_1^0 \tilde{\chi}_1^0 \to h_s A_s$.
\end{itemize}
Our study delivered valuable insights into the DM physics of the GNMSSM, reopening the $\mathbb{Z}_3$-NMSSM's extensive parameter space previously excluded by experiments. We eagerly anticipate further results from DM and LHC experiments that may provide more clues about the properties of DM and help pinpoint the correct theoretical framework.

\acknowledgments

J. Cao thanks Dr. Yuanfang Yue and Jinwei Lian for helpful discussions. This work was supported by the National Natural Science Foundation of China (NNSFC) under grant No. 12075076.

\bibliographystyle{CitationStyle}
\bibliography{dm}

\end{document}